\documentclass[10pt]{article}
\usepackage[margin=1in]{geometry} 
\usepackage[table]{xcolor} 
\usepackage{amsmath,amsfonts,gensymb,stmaryrd} 
\usepackage[affil-it]{authblk} 
\usepackage{abstract} 
\usepackage[numbers,sort&compress]{natbib}

\usepackage{subcaption,multirow,sidecap,listings}
\usepackage[pdftex]{graphicx}
\usepackage[font=footnotesize]{caption}
\graphicspath{{../Figures/}}
\usepackage{tikz}
\newcommand*\circled[1]{\tikz[baseline=(char.base)]{
            \node[shape=circle,draw,inner sep=2pt] (char) {#1};}} 
\newcommand*\dcircled[1]{\tikz[baseline=(char.base)]{
            \node[shape=circle,draw,inner sep=2pt,fill=black] (char) {\textcolor{white}{#1}};}} 

\definecolor{softblue}{HTML}{C5E7FB}
\definecolor{softred}{HTML}{F17877}
\definecolor{softgray}{HTML}{ACA7A7}
\definecolor{softgreen}{HTML}{81F75E}
\definecolor{newgreen}{HTML}{088A08}

\input Qcircuit

\title{Simulating the performance of a distance-3 surface code in a linear ion trap}
\author{Colin Trout$^{1}$, Muyuan Li$^{2}$, Mauricio Guti\'{e}rrez$^{1}$, Yukai Wu$^{4}$, Sheng-Tao Wang$^{4}$, Luming Duan$^{4}$, and Kenneth R. Brown$^{1,2,3}$}
\affil{\small Schools of Chemistry and Biochemistry$^{1}$, Computational Science and Engineering$^{2}$, and Physics$^{3}$,\\
 Georgia Institute of Technology, Atlanta, Georgia 30332-0400, USA\\ \vspace{5mm}
Department of Physics, University of Michigan, Ann Arbor, Michigan 48109, USA$^{4}$}
\date{}

\begin{document}

\maketitle

\vspace{-0.7in}

\begin{abstract}
\noindent We explore the feasibility of implementing a small surface code with 9 data qubits and 8 ancilla qubits, commonly referred to as  surface-17, using a linear chain of $^{171}$Yb$^{+}$ ions.  Two-qubit gates can be performed between any two ions in the chain with gate time increasing linearly with ion distance.  Measurement of the ion state by fluorescence requires that the ancilla qubits be physically separated from the data qubits to avoid errors on the data due to scattered photons. We minimize the time required to measure one round of stabilizers by optimizing the mapping of the two-dimensional surface code to the linear chain of ions.   We develop a physically motivated Pauli error model that allows for fast simulation and captures the key sources of noise in an  ion trap quantum computer including gate imperfections and ion heating.  Our simulations showed a consistent requirement of a two-qubit gate fidelity of $\ge 99.9\%$ for the logical memory to have a better fidelity than physical two-qubit operations.  Finally, we perform an analysis on the error subsets from the importance sampling method used to bound the logical error rates to gain insight into which error sources are particularly detrimental to error correction.
\end{abstract}

\section{Introduction}

A quantum computer is a device engineered to utilize the complexity of a many-particle wavefunction for the purpose of solving computational problems.  For specific problems, quantum algorithms are predicted to surpass the ability of classical information processing \cite{ShorsAlgorithm, FeynmanQC, GroversAlgorithm, AaronsonBosonSample2011,  QuantSimResources, HarrowLinearSysEqn2009} but the computational space solvable to quantum algorithms has yet to be rigorously explored due to the absence of a working physical architecture.  Experimental implementations of small quantum algorithms in systems containing under $10$ qubits have been exhibited in a variety of architectures \cite{DJ3QNMR, ShorQCNMR, DJQCatomic, DJQCphotonic, QFTQCatomic, GroversQCatomic, GroverDJQCsuperconducting, DJQCNVcenter, ShorsQubitRecycling, ShorQCatomiceff, DebnathSmallQCIons2016}.  However, realization of a large-scale algorithm consisting of hundreds or thousands of qubits will require protocols that protect the quantum states from sources of decoherence.  Quantum error correction (QEC) is a viable method for protecting of quantum states from sources of decoherence \cite{QEC1,QEC2,QEC3}.  Error correction routines embed logical qubits into subspaces of a multi-qubit Hilbert space and uses active feedback to remove entropy from the system.

An enticing selection for an error correction protocol is the surface code \cite{SCorig} which exhibits an error correction threshold in the circuit model between $0.5\% - 1 \%$ for depolarizing Pauli noise \cite{RaussendorfClusterState12007,RaussendorfClusterState22007,FowlerSurfaceCodeThresh2009,SCThresholdsStephens2014}.  This threshold represents the error rate below which logical gates and memories can be made arbitrarily good by increasing the distance of the surface code.  Here we examine the smallest surface code with nine data qubits and eight ancilla qubits, known as surface-17 \cite{Tilted13SC,Tilted17SC,TomitaLowDSC2014}. In principle only a single ancilla qubit could be used over and over, but the gains from parallelism are even apparent in studies comparing 8 ancilla qubits to 6 ancilla qubits \cite{TomitaLowDSC2014}.  With 10-20 qubits, a number of QEC codes can be implemented fault-tolerantly including the 5-qubit code \cite{Laflamme5qubit1996}, Steane [[7,1,3]] \cite{SteaneSteaneCode1996,SteaneCSScodes1999}, Bare [[7,1,3]] \cite{LiBareAnc2017},  the Bacon-Shor [[9,1,3]] \cite{BaconBaconShor2006,ShorBaconShor1995}, or the twisted surface \cite{YoderSurfaceCodeTwist2017} code.  We chose to study the surface code because the memory pseudothreshold, the error rate below which the encoded qubit outperforms the physical qubit,  is superior to the 5-qubit code, the Steane code, and the Bare code, and comparable to the Bacon-Shor and twisted surface code \cite{YoderSurfaceCodeTwist2017}. 

Atomic ions have proven to be high-fidelity qubits for quantum information processing. The internal states of the ions  are controlled by the application of electromagnetic radiation with lasers \cite{HayesEntangleOptComb2010} or microwaves \cite{OspelkausMicrowaveGates2011,WarringIndividMicroAddr2013}.  Two-qubit gates are performed by conditionally exciting the coupled motion of ions in the chain dependent on the ion's internal states  \cite{CiracZollerGate,MolmerSorensenHotIons,BellStatePrep,GatesWarmIons}.  The normal modes of motion are nonlocal allowing interactions between any ions in the chain without requiring additional overhead from moving information through local couplings \cite{SCqubitsNature} or storing qubits in auxiliary states \cite{ShorsSuperconducting}.  This arbitrary connectivity without altering the intrinsic nature of the qubit adds modularity at the hardware level which relaxes software constraints on compilation when building up high level algorithms from the hardware primitives \cite{DebnathSmallQCIons2016, Linke5qubitComp2017}.  Qubits can be encoded into either optical states \cite{RoosCaOpticalQubit1999}, Zeeman states \cite{SpectRbZeemanQubit2011,RusterCaZeemanQubit2016}, or hyperfine states of the ions \cite{BallanceControlHFQubits2016,BrownHighFidBe2011,NoekHighFidYb2013}.  For this study, information will be stored in the hyperfine ``clock" states of $^{171}$Yb$^{+}$.  While single-qubit operations in this system have displayed error rates below the surface code pseudothreshold \cite{CompPulseYb, NoekHiFidSPAMYb2013} reported from Tomita and Svore \cite{TomitaLowDSC2014}, two-qubit gate fidelities are limited by a number of factors including spontaneous Raman scattering during gates and residual entanglement between the internal state and the motional modes of the ion.  Compensation pulses have been developed with a predicted error rate due to scattering of $10^{-4}$ \cite{SpontScatErrYb} and control sequences have been implemented exhibiting single- and two-qubit gate fidelities of $99.9\%$ using the hyperfine ground states of trapped $^9$Be$^+$ \cite{BrownHighFidBe2011}  and $^{43}$Ca$^{+}$ \cite{BallanceControlHFQubits2016} ions, respectively.  However, quantum control applied to a scaled-up five-ion chain consisting of $^{171}$Yb$^{+}$ qubits currently exhibits two-qubit error rates of $2\%$ \cite{DebnathSmallQCIons2016}.  This current error rate is well above the reported pseudothreshold for the surface-17 code, but gates with 99.9\% fidelity should be achievable.

Atomic ion experiments have already demonstrated classical error correction \cite{ChiaveriniBitFlipZZXXZX2004,SchindlerRepCpde2011}, encoding logical states for quantum error correction \cite{NiggSteaneEnc2014}, and fault-tolerant quantum error detection \cite{Linke422Ions2016}.  In addition, multiple theoretical studies have examined implementation of quantum error correcting codes with trapped ions.  Architectural studies with large distance codes have looked at chains of ions connected by shuttling \cite{WinelandExpIssueIons1998,KielpinskiQCCD2002,LekitscheMicroQCCD2017} or by optical interconnects \cite{ChristophEntanglePhotonIon2003,MoehringEntangleIonQubitPhoton2007,DuanEntanglePhotonColloq2010,NickersonTopQCNoisyNetwork2013,MonroeMUSIQC2014}. Studies of smaller codes include the Steane [[7,1,3]] code in a two-dimensional shuttling architecture \cite{TomitaSteaneIonTrap2013} and more recently a linear shuttling architecture 
\cite{AbuNadaSteaneIonTrap2014,BermudezITQCwithSteane2017}. The work of Tomita and Svore \cite{TomitaLowDSC2014} used optimistic ion trap parameters to study the surface-17 code on a linear chain.  Here we consider a more realistic model based on a near-term implementation in a linear chain of Yb$^{+}$ ions.  Our model includes many additional physical details such as the necessity to physically separate measured ions from data ions and the distance dependence of two-qubit gates along the ion chain.

This study provides an assessment of the feasibility of implementing the surface-17 error correcting code on a linear trap holding a chain of $^{171}$Yb$^{+}$ ions.  Furthermore, this study will provide target fidelities for experimentalists to realize error correction with the 17-qubit surface code.  The paper is structured in the following manner.  First, a resource efficient implementation of the surface code, the 17-qubit rotated surface code, is explained.  Following that, the ion trap architecture will be defined and a map between the linear ion chain and the two-dimensional surface code is provided.  The remainder of the paper will focus on error correction.  Efficiently simulable models of ion trap error sources will be outlined followed by an examination of results from decoding methods tailored for such errors.

\section{The 17-Qubit Surface Code}
The surface code allows for high-threshold fault-tolerant quantum computation in a two dimensional architecture \cite{SCorig,SCThresholdsStephens2014,RaussendorfClusterState12007,RaussendorfClusterState22007,FowlerSurfaceCodeThresh2009}.  The surface code is constructed by a square lattice arrangement of data qubits where the faces of the lattice represent the stabilizer generators of the error correcting code with $X$ and $Z$ type  weight-4 stabilizers alternating in a checkerboard-like pattern throughout the lattice.  In this arrangement, measurements are local and logical operators are non-local operators that span the surface and terminate on one of two types of boundaries,  an $X$ and a $Z$ type, which label the type of stabilizers occupying the four terminating edges of the planar code.  There are two choices of edge operators: weight-3 triangles or weight-2 links depending on how the bulk of the surface code is oriented \cite{Tilted13SC}.  The logical $Z$ ($X$) operator spans the two $Z$ ($X$) boundaries. The code distance, the weight of the lowest weight Pauli operator that maps elements of one logical basis state to another state, has an intuitive representation as the length (in number of data qubits) of the boundaries of the square lattice arrangement of the code.

\begin{figure}[!b]
\centering
\begin{subfigure}[b]{0.25\textwidth}
\includegraphics[width=\textwidth]{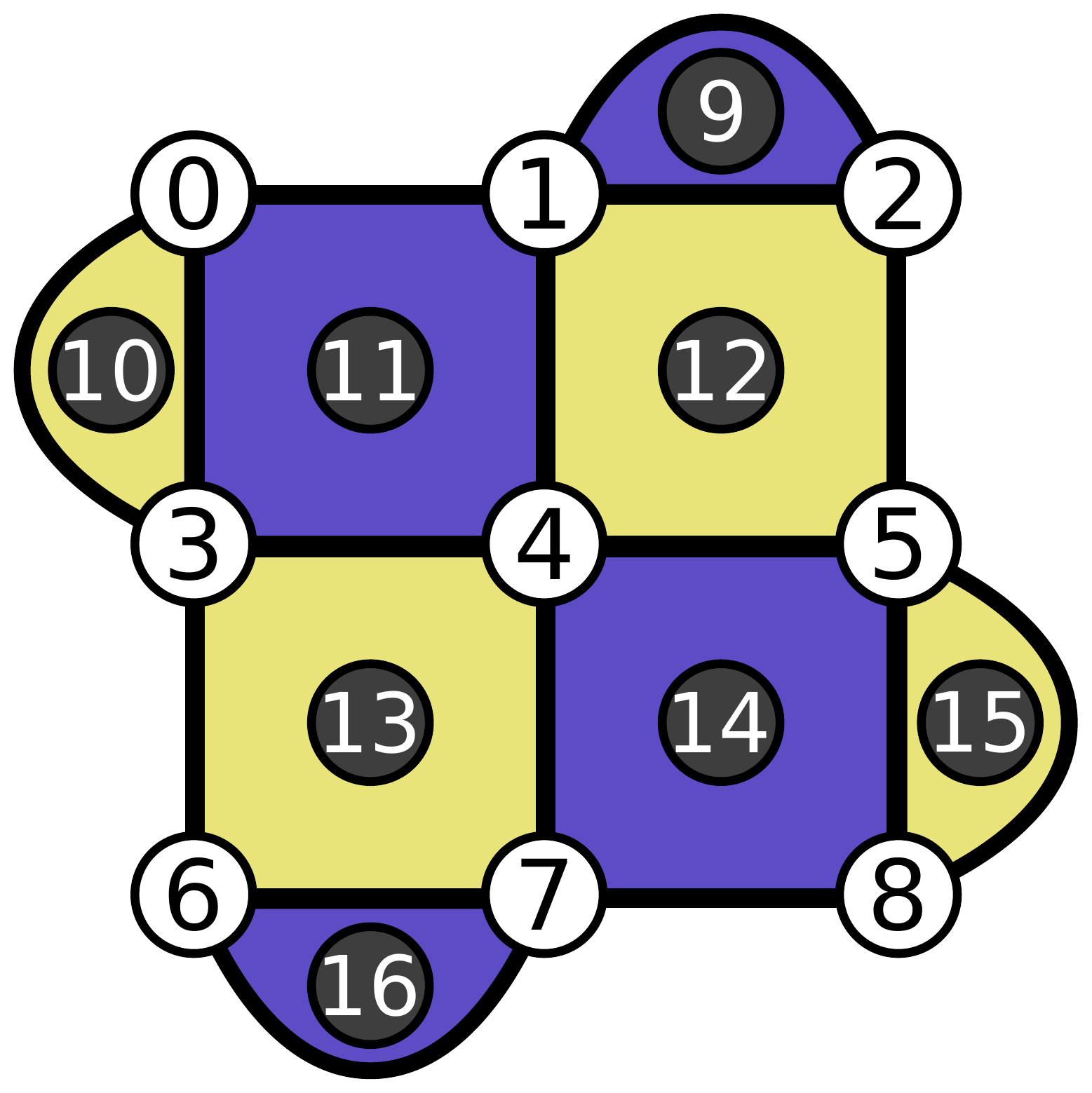}
\caption{}
\label{fig:17QSC}
\end{subfigure}
\begin{subfigure}[b]{0.45\textwidth}
\hspace{22mm}
\mbox{
\Qcircuit @C=0.75em @R=1.5em {
\lstick{\mathrm{ancilla} \; \ket{0}} & \qw & \targ & \targ & \gate{\textcolor{red}{Z}} & \targ & \targ & \meter\\
\lstick{\circled{1}} & \qw & \ctrl{-1} & \qw & \qw & \qw & \qw & \qw \\
\lstick{\circled{2}} & \qw & \qw & \qw & \qw & \ctrl{-2} & \qw & \qw & \lstick{\textcolor{red}{Z}} \\
\lstick{\circled{4}} & \qw & \qw & \ctrl{-3} & \qw & \qw & \qw & \qw \\
\lstick{\circled{5}} & \qw & \qw & \qw & \qw & \qw & \ctrl{-4} & \qw & \lstick{\textcolor{red}{Z}} \\
}
}
\caption{}
\label{fig:measerrorprop1}
\end{subfigure}
\begin{subfigure}[b]{0.25\textwidth}
\centering
\includegraphics[width=0.6\textwidth]{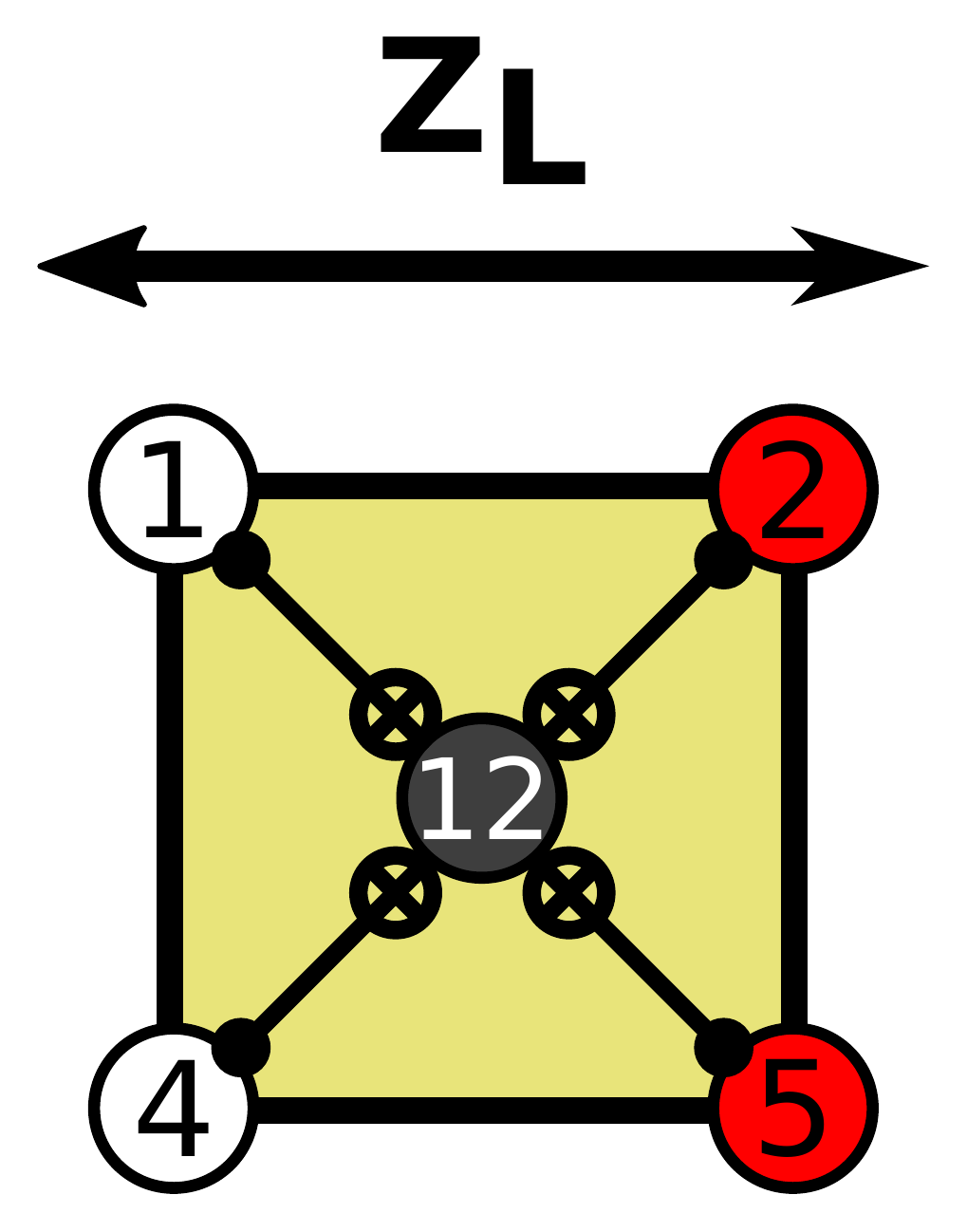}
\caption{}
\label{fig:measerrorprop2}
\end{subfigure}
\caption{a) Planar layout of the 17-qubit surface code.  White (black) circles represent data (ancilla) qubits and dark (light) faces denote $X$ ($Z$) stabilizer generators.  b) A single error on an ancilla propagating to a two-qubit error on the data; typically not fault-tolerant.  c) The ancillary error from b) propagates in a direction perpendicular to the direction of the logical operator which is equivalent to a single-qubit error \emph{from the perspective of the logical operator} retaining fault-tolerance with bare ancilla.  All images were adapted from Tomita and Svore \cite{TomitaLowDSC2014}.
}
\label{fig:17QSCfull}
\end{figure}

The 17-qubit surface code is shown in figure \ref{fig:17QSC} \cite{Tilted17SC}.  The white (black) circles represent data (ancilla) qubits and the dark (light) faces of the lattice dictate the $X$-type ($Z$-type) stabilizer generators of the code.  The 13-qubit version of the surface code is constructed by removing the ancillary qubits on the boundaries of the 17-qubit code and scheduling stabilizer measurements in a manner that each of the ancillary qubits are used to measure both a weight-2 and weight-4 stabilizer \cite{Tilted13SC}.  This work focused on the 17-qubit version because the greater circuit depth for stabilizer measurement in the 13-qubit code adversely affects error correction \cite{TomitaLowDSC2014}.

The resource win for the surface codes is the ability to use bare ancilla for fault-tolerant measurement of the stabilizer generators.  The scheduling of the two-qubit gates following an N-like pattern about the face of a weight-4 $Z$ stabilizer allows for the cases where single-ancilla qubit to two-data qubit error propagation (``hook" errors) occur in a direction perpendicular to the direction of logical $Z$ operator as shown in figures \ref{fig:measerrorprop1} and \ref{fig:measerrorprop2}.  This error is equivalent to a single-qubit error \emph{from the perspective of the $Z$ logical operator}, thus retaining fault-tolerance during syndrome measurement.  Scheduling the two-qubits gates during the measurement of an $X$ stabilizer in a Z-like pattern gives a similar result for the logical $X$ operator.  Many other error correction routines require the use of many-qubit ancillary states to ensure fault-tolerance.  Shor error correction requires many-qubit states, known as cat states, to fault-tolerantly measure stabilizers which increase the number of gates and require a number of ancilla equivalent to the sum of the operator weights of all the stabilizers to perform measurements in parallel \cite{ShorCatState}.  This would require 20 ancillary qubits to perform error correction in parallel with the surface code.  Steane \cite{SteaneSteaneEC1997} and Knill \cite{KnillKnillEC2005} error correction both require an ancillary logical state for fault-tolerance requiring 17 ancillary qubits.  Recent work has shown that the use of ``flag" ancillary qubits can reduce the number of ancillary qubits required for fault-tolerance \cite{ChaoFlagQubits12017,ChaoFlagQubits22017} but still would which corresponds to 12 ancillary qubits (in parallel) in our surface code setting.  However, a variation on the surface code, the twisted surface code, implements flag qubits and is constructed in a manner that requires only 15 total qubits with a small loss in pseudothreshold relative to the surface code \cite{YoderSurfaceCodeTwist2017}.  We chose to focus on gate fidelities and pseudothresholds; thus our choice of code.

\section{Mapping the Surface Code to an Ion Chain}
\begin{figure}
\centering
\includegraphics[width=\textwidth]{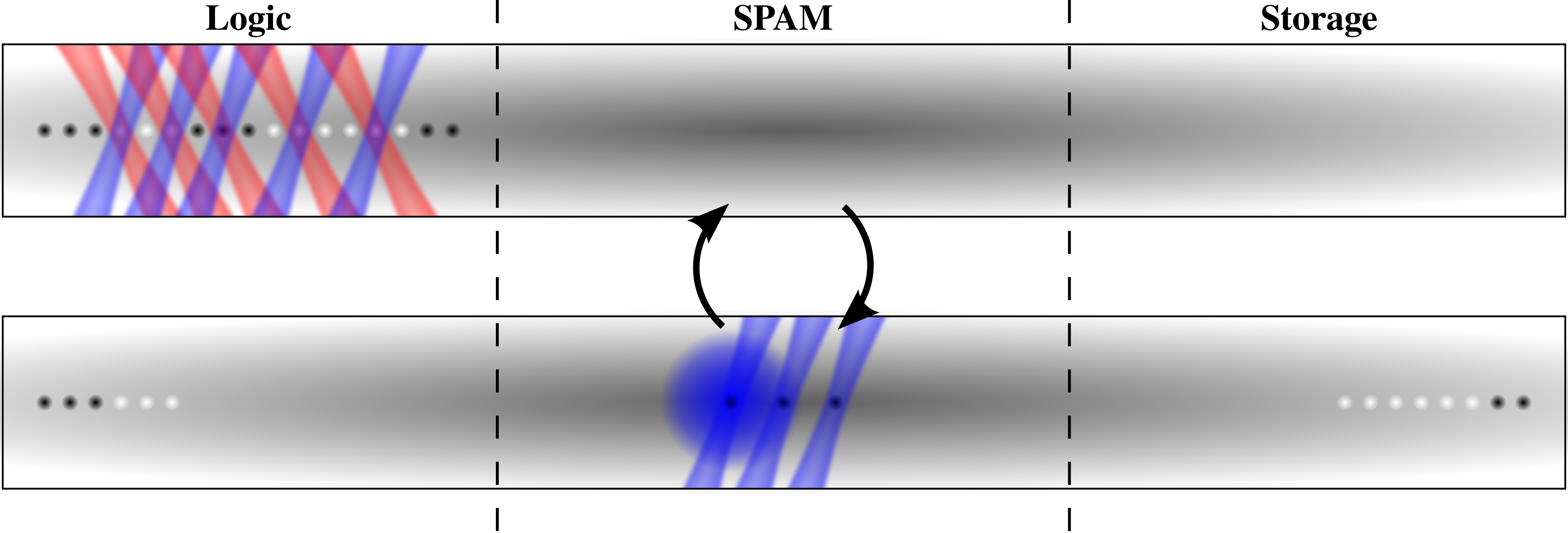}
\caption{Graphical representation of the ion trap architecture and syndrome measurement operations.  Ions are trapped in a mixed arrangement where data/ancilla qubits represented in white/black as in figure \ref{fig:17QSC}.  There are three zones: a Logic, SPAM, and Storage zone.  (Top) The Logic zone where qubit gates are applied.  (Bottom) The SPAM zone where state preparation and measurement is performed, scattering photons.  The Storage zone does not have a unique task but serves to sufficiently distance qubits from the SPAM zone.}
\label{fig:IonTrap}
\end{figure}

To perform error correction with the surface code, the required operations are single-qubit gates ($H$), two-qubit gates ($CNOT$), state initialization ($\ket{0}$ state), and measurement ($Z$-basis).  Single-qubit gates are performed by the application of laser fields \cite{HayesEntangleOptComb2010} or microwave radiation \cite{OspelkausMicrowaveGates2011,WarringIndividMicroAddr2013} to manipulate the hyperfine states of trapped $^{171}$Yb$^{+}$ ($^{2}$S$_{1/2}$$\ket{F=0;m_F=0} \leftrightarrow$  $^{2}$S$_{1/2}\ket{F=1;m_F=0}$ transition) which can drive arbitrary single-qubit rotation gates.  High fidelity (compared to other schemes), fast two-qubit gates are performed by the application of counter-propagating laser fields achieving entanglement through the coupling of the internal states with the motional modes of the ion crystal through a method known as the M\o lmer-S\o rensen gate which engineers an $XX$ entangling gate \cite{SorensenPRLMSgate1999,SorensenPRAMSgate2000,DebnathSmallQCIons2016}.  Controlled-NOT ($CNOT$) gates can be built from M\o lmer-S\o rensen gates and available single-qubit rotations \cite{MaslovCircuitCompIT2017} (see figure \ref{fig:MStoCNOT}).  State initialization and measurement are performed by applying laser beams resonant with the $^2$S$_{1/2} \leftrightarrow \, ^{2}$P$_{1/2}$ transition.  For $\ket{0}$ state preparation, qubits are optically pumped out of the $^2$S$_{1/2} \ket{F=1}$ state into the $^2$P$_{1/2} \ket{F=1}$ manifold which, with high probability, falls into the $^2$S$_{1/2} \ket{F=0}$ state \cite{WinelandExpIssueIons1998,TheoryAtomicSpec,YbstructureReadout}.  For measurement in the $Z$-basis, a $^2$S$_{1/2} \ket{F=1} \leftrightarrow \, ^2$P$_{1/2}\ket{F=0}$ cycling transition is induced where the discrepancy between scattered photon counts of the qubit states serves as readout \cite{WinelandExpIssueIons1998,TheoryAtomicSpec,YbstructureReadout}.  Note that the state preparation and measurement processes scatter photons that should not interact with surrounding ions.  This requirement introduces an additional operation, ion shuttling \cite{BlakestadXjunction2009,MoehringYjunct2011,BowlerFastShuttle2012,WaltherFastShuttle2012,WrightXtrap2013,ShuYtrap2014}, which will be used to separate qubits in memory from the scattered photons during measurement/preparation.  An alternative approach would be to use two ion species so the data ions are insensitive to the fluorescence of the measurement ions \cite{Tan2SpeciesLogicMgBe2015,HumeThesisCaBeControl2010}, but we avoided this method due to technical issues in shuttling mixed-ion crystals.

There exist many ion trap architectures containing both one-dimensional and two-dimensional ion layouts.  For a first generation implementation of a logical qubit consisting of atomic ions, a trapped linear chain of ions was favored over two-dimensional architectures due to technological challenges in the latter that result in issues such as additional ion heating through shuttling junctions in traps \cite{BlakestadXjunction2009,MoehringYjunct2011,WrightXtrap2013,ShuYtrap2014}, high idle ion heating rates \cite{Sterling2DTrap2014}, and single-ion addressing/readout issues in two-dimensional trap layouts.  The linear trap is composed of at least three zones: a Logic, State Preparation and Measurement (SPAM), and Storage zone (figure \ref{fig:IonTrap}).  Ion shuttling across the axial direction of the trap allows for the 17-ion chain to be arbitrarily split into three, separate linear chains of ions inhabiting each of the three zones.  The Logic zone is where all single- and two-qubit gates are applied.  The central SPAM zone is where state preparation and measurement operations are performed.  The Storage zone serves the purpose that its name implies and is required due to the geometric constraint of having the ions confined in a linear chain.  In addition to these three zones, one or two additional zones may be capped at the ends of the trap that hold atomic ions for sympathetic cooling of the motional modes of the qubit ions \cite{LarsonSympHg1986,KielpinskiSympTheory2000,BarrettSymCoolLogic2003}.

Now we illustrate how a round of stabilizer measurements would proceed in such an architecture.  Initially, all of the qubit ions would be prepared and cooled in the SPAM zone; an initialization step.  After initialization, all 17 ions are shuttled into the Logic zone.  In the Logic zone, the circuit implementing the measurement of the stabilizers of the surface-17 code would be implemented in, ideally, a parallel fashion.  After the application of all the gates, groups of ancillary qubits would be shuttled to the SPAM zone for measurement. During the measurement, only ancillary qubits occupy the SPAM zone and any data qubits would be stored in either the Logic or Storage zone sufficiently far away from the SPAM zone.  The ancillary qubits in the SPAM zone will be measured in parallel and in sets dictated by the data/ancilla assignment of the qubits in the ion chain.  Following readout of all ancillary qubits, all qubits will be shuttled back to the Logic zone and the process is repeated.  Such an implementation begs the question: how should the qubits in the surface code be assigned to the linear chain of ions?  We are particularly interested in configurations that minimize the gate times (errors) of the error correction circuit.  To proceed, we must first discuss two-qubit gates.

\begin{figure}[t!]
\centering
\includegraphics[width=0.8\textwidth]{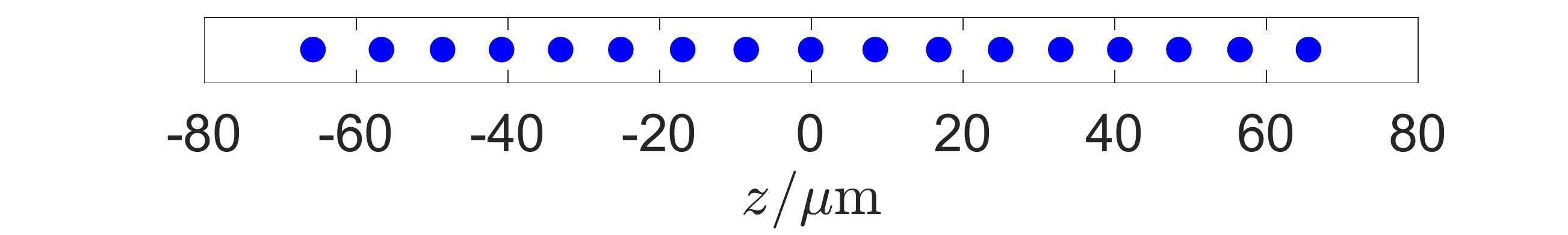}
\caption{Equilibrium positions of the 17 ions along the $z$ direction when $l_0=25\,\mu$m and $\gamma_4=0.86$.}
\label{fig:IonConfiguration}
\end{figure}

\begin{figure}[b!]
\includegraphics[width=\textwidth]{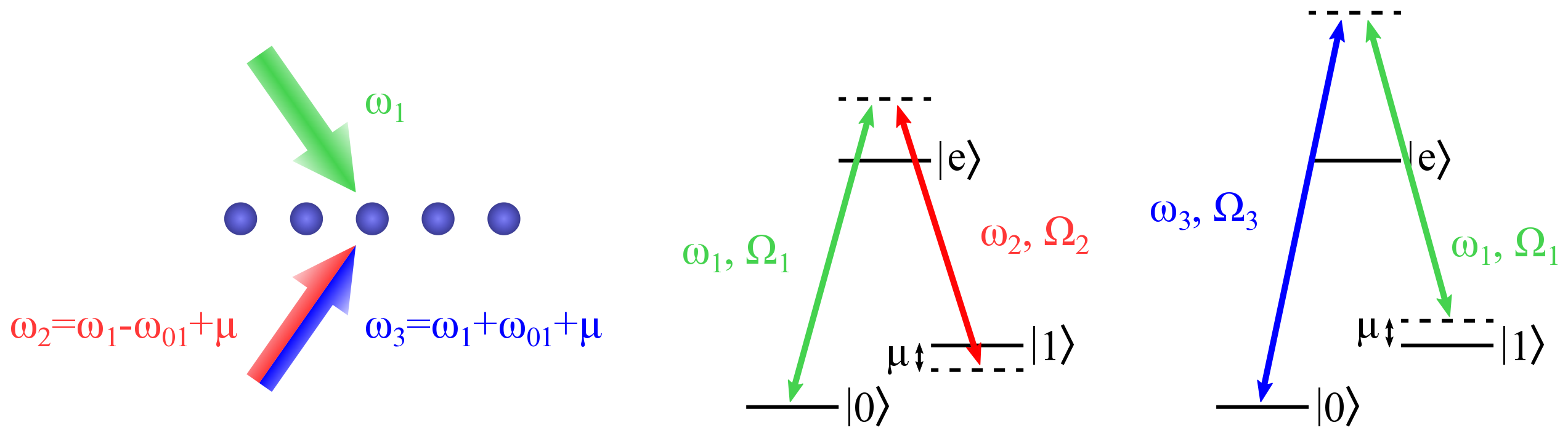}
\caption{Experimental setup (left) and energy level diagrams for the M\o lmer-S\o rensen gate implemented with counter-propagating laser fields.  In the experiment, two beams are shined on the ion with one beam in the direction of ${\bf k}_1$ and the other one containing both red and blue detuned components in the ${\bf k}_2$ direction.}
\label{fig:MSlasers}
\end{figure}

For computation of the two-qubit gate times, current gate protocols \cite{DebnathSmallQCIons2016} and motional decoupling techniques \cite{MotionDecoupling} were modeled; the latter of which contributes significantly to the distance dependence of the gate times.  The calculation of the gate time of an ion pair is outlined below.  In the weak trap limit, a Paul trap can be well approximated by a pseudo-harmonic potential (see e.g.\ Ref.~\cite{LeibfriedQDSingleIon2003}).  Here we consider ions in a linear Paul trap along the $z$ direction $\left(\omega_z \ll \omega_x,\,\omega_y\right)$.  With a harmonic trap potential, the spacing between ions in the chain will be nonuniform, which can lead to undesired transition into a zigzag shape \cite{SchifferPhaseTransAnisotropic1993,DubinTheoryCoulombCrystal1993}, as well as the difficulty in cooling many low frequency modes.  To overcome this problem, an additional quartic potential can be added to the $z$ direction \cite{LinQCAnharmIonTrap2009} giving the total potential energy:
\begin{equation}
V = \sum_i \left(-\frac{1}{2} \alpha_2 \, z_i^2 + \frac{1}{4} \alpha_4 \, z_i^4\right) + \sum_{i < j} \frac{e^2}{4 \pi \epsilon_0 \left|z_i - z_j\right|}
\end{equation}
where $\alpha_2,\,\alpha_4 > 0$ are two parameters characterizing the strength of the quadratic and the quartic potentials.  The ion configuration is then fully determined by a length unit $l_0 \equiv \left(e^2/4\pi \epsilon_0 \alpha_2 \right)^{1/3}$ and a dimensionless parameter $\gamma_4 \equiv \alpha_4 l_0^2 / \alpha_2$. For $N=17$ $^{171}$Yb$^{+}$ ions, we choose $\gamma_4=0.86$ to minimize the relative standard deviation of the ion spacings. An average ion distance of about $8.2\,\mu$m can then be realized by setting $l_0 = 25\,\mu$m. The equilibrium configuration of the 17 ions is shown in figure \ref{fig:IonConfiguration}.

The two-qubit entangling gate is implemented with a spin-dependent force on the two ions via the transverse collective modes. For example, we can use the transverse modes in the $x$ direction whose $k$-th normalized mode vector is denoted as ${\bf b}_j^k$ with a mode frequency $\omega_k$ where the index $j$ runs over all ions $\left(j=1,\,2,\cdots, N\right)$. The creation and annihilation operators corresponding to this collective mode are denoted as $\hat{a}_k^\dag$ and $\hat{a}_k$ respectively. The transverse trap frequency is set to a typical value $\omega_x = 2\pi \times 3\,$MHz and the temperature is set to $k_B T = \hbar \omega_x$ giving an average phonon number of $\bar{n} \approx 0.5$ for each transverse mode. This can be easily achieved with a Raman sideband cooling. The spin-dependent forces are generated by counter-propagating laser beams on the two ions that we choose to entangle (see figure \ref{fig:MSlasers}). The Hamiltonian, in the interaction picture, can be represented as:
\begin{equation}
\hat{H}_I = \hbar \sum_{j} \tilde\Omega_j \sum_k \eta_k {\bf b}_j^k \sin \mu t \left(\hat{a}_k e^{-i \omega_k t} + \hat{a}_k^{\dag} e^{i \omega_k t} \right) \hat{\sigma}_j^x
\end{equation}
where we further define the Lamb-Dicke parameter $\eta_k \equiv \Delta k \sqrt{\hbar/2m\omega_k}$, $\Delta k$ is the difference in the wavevectors of the counter-propagating Raman beams, $\mu$ is the two-photon detuning, and $\hat{\sigma}_j^x$ is the $\hat{\sigma}^x$ Pauli matrix on ion $j$. For the $^{171}$Yb$^{+}$ qubit transitions, the laser beams have wavelengths around $\lambda = 355\,$nm \cite{CampballUltrafastGates2010} and for counter-propagating pairs $\Delta k = 2k$, hence the Lamb-Dicke parameter $\eta_k \approx 0.111$. In the above equation, $\tilde\Omega_j$ is the effective Rabi frequency of the Raman transition pairs shown in figure \ref{fig:MSlasers} ($\tilde\Omega_j \approx \Omega_1\Omega_3/\Delta = \Omega_1\Omega_2/\Delta$ where $\Delta$ is the single-photon detuning from the excited state). From now on we will drop the tilde notation for simplicity. Note that one of the laser beams contains two frequency components and we assume that the two Raman transition pairs have the same effective Rabi frequency $\Omega_j$, opposite detunings $\pm\mu$, and opposite wavevector differences $\pm\Delta k$. This is known as the phase-insensitive geometry \cite{LeePhaseControlIon2005}.

\begin{figure}[t!]
\centering
\includegraphics[width=0.8\textwidth]{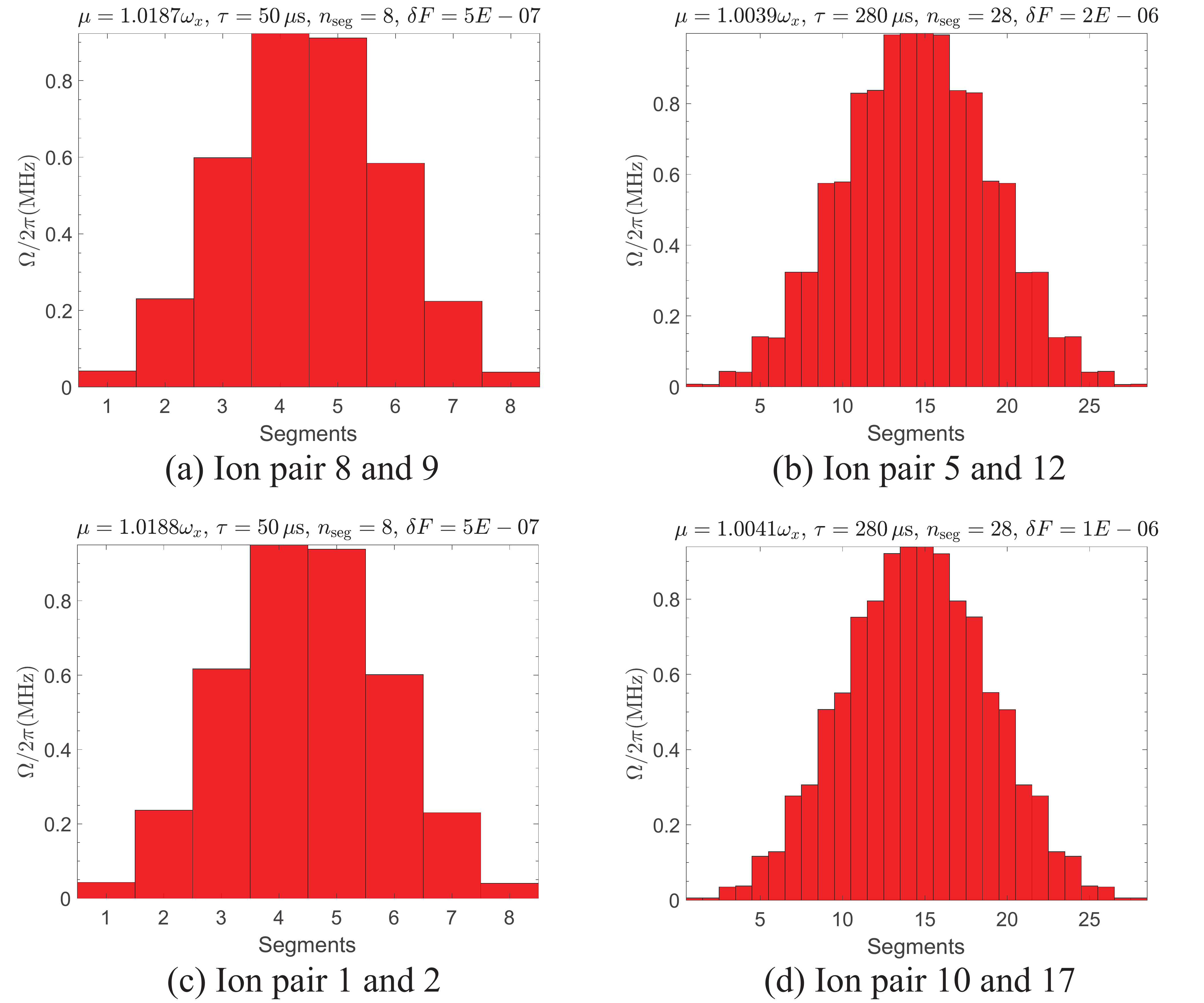}
\caption{Example pulse sequences with discrete Rabi frequencies for performing an entangling gate between two ions in the 17 ion chain: nearest neighbor ions (Pair 8 and 9, Pair 1 and 2) and ions separated by a distance of 7 ion spacings (Pair 5 and 12, Pair 10 and 17).  Thanks to the nearly uniform ion spacings, the required gate times for ion pairs at the same distance are roughly the same. In general, the ion pair at a larger distance requires a longer gate time $\tau$ due to their weaker coupling.}
\label{fig:gatetimeLuming}
\end{figure}

The time evolution under the above Hamiltonian can be written as \cite{LeePhaseControlIon2005,LinQCAnharmIonTrap2009}:
\begin{equation}
\hat{U}_I(\tau) = \exp\left(i\sum_{j} \hat{\phi}_j(\tau)\hat{\sigma}_j^x + i\sum_{i<j}\Theta_{ij}(\tau)\hat{\sigma}_i^x \hat{\sigma}_j^x\right)
\end{equation}
where $\hat{\phi}_j(\tau) = -i\sum_k [\alpha_j^k(\tau)\hat{a}_k^\dag - {\alpha_j^k}^*(\tau)\hat{a}_k]$.  The parameters $\alpha_j^k$ and $\Theta_{ij}$ are purely numbers related to the phase space displacement of the motional states after the gate and angle of the entanglement gate, respectively. For the following calculations, we assume that $\Omega_j$ is the same for both ions and we divide it into segments with equal durations; that is, a piecewise constant $\Omega(t)$ (see figure \ref{fig:gatetimeLuming}). With a suitable choice of detuning $\mu$, gate time $\tau$, and Rabi frequency $\Omega(t)$, we can suppress all the $\alpha_j^k(\tau)$ terms and realize an ideal entangling gate $e^{\pm i \pi \hat{\sigma}_i^x\hat{\sigma}_j^x/4}$ with high fidelity.  Here, we focus on the intrinsic gate infidelity caused by the residual coupling to multiple phonon modes after the entangling gate.  Other noise sources from technical control errors are not included for this calculation.

\begin{figure}[b!]
\centering
\begin{subfigure}[t]{0.45\textwidth}
\includegraphics[width=\textwidth]{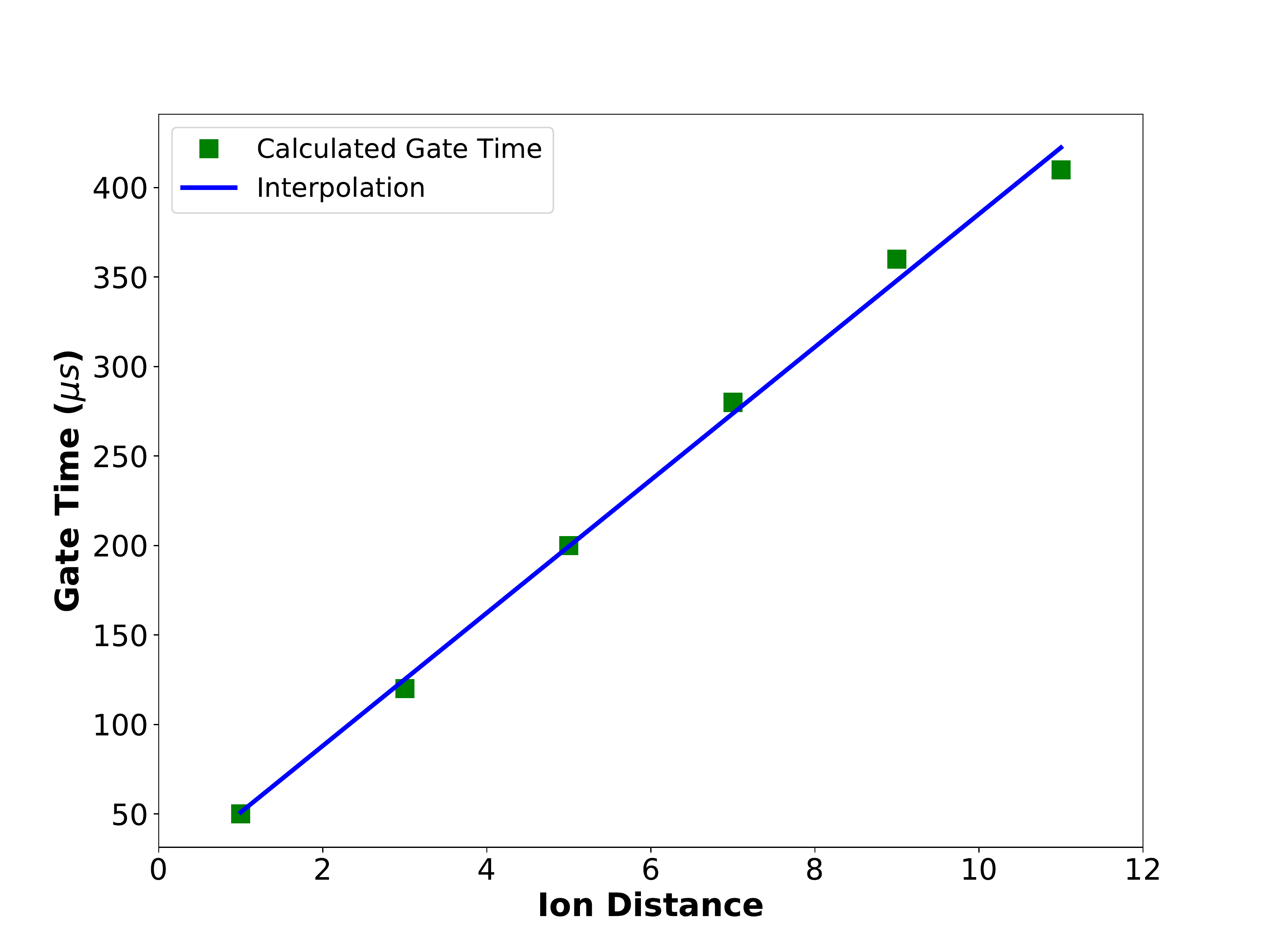}
\caption{Gate times with respect to ion distance}
\label{fig:GateTimeGraph}
\end{subfigure}
\begin{subfigure}[t]{0.5\textwidth}
\includegraphics[width=\textwidth]{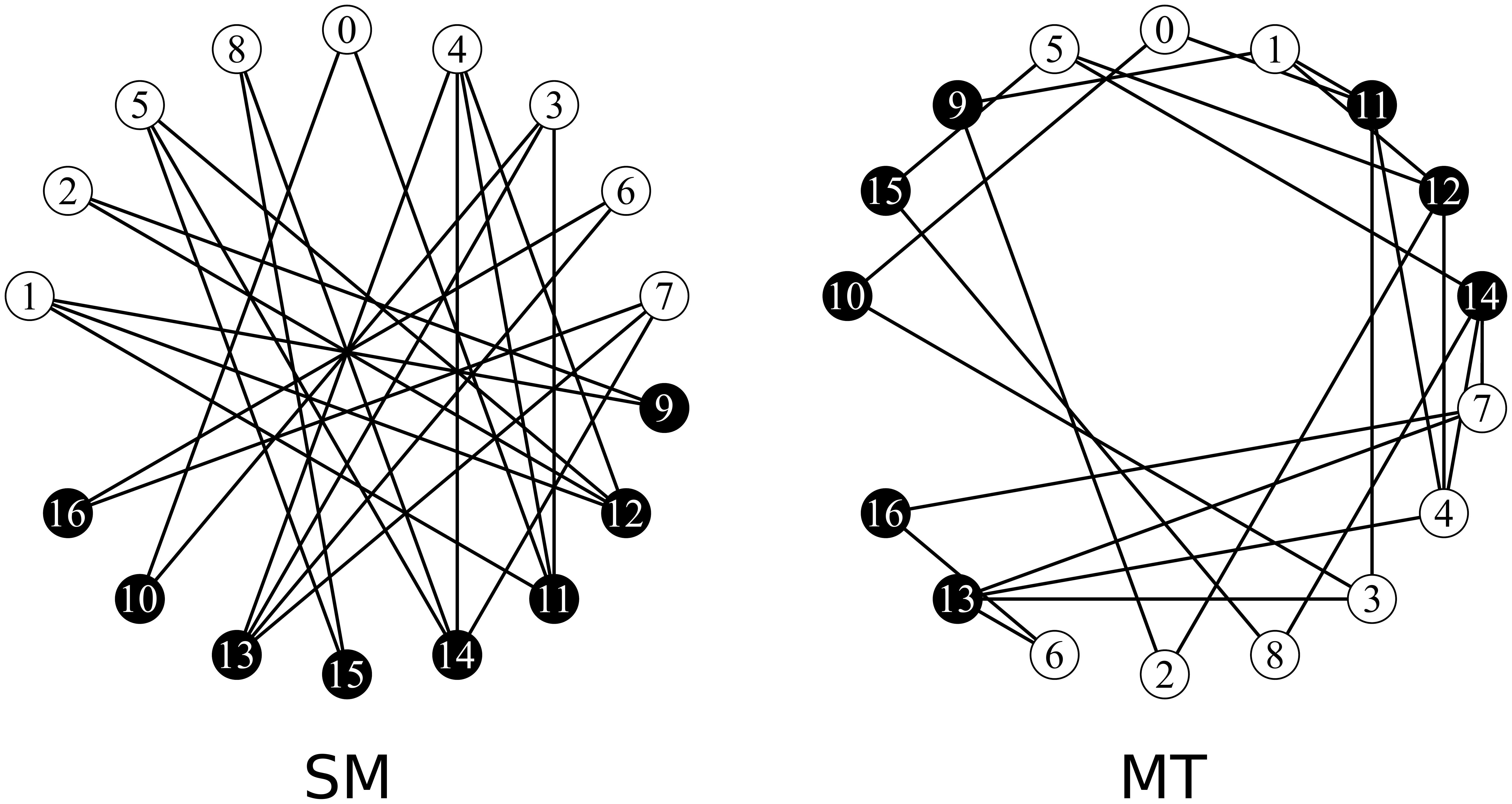}
\caption{Qubit to ion maps}
\label{fig:ionconnectivity}
\end{subfigure}
\caption{a) Two-qubit gate times calculated as a function of the ion distance in the linear chain.  Extrapolation of this data was used to calculate the time required to measure the error syndrome.  b)  Ion chains mapped to a circle.  The node labels correspond to the qubit numbers in figure \ref{fig:17QSC}.  The labels refer to different optimizations with configurations shown in figure \ref{fig:GateTimeTable}.}
\end{figure}

Figure \ref{fig:gatetimeLuming} shows example calculations of the gate sequences for different ion pairs: two nearest-neighbor pairs and another two separated by 7 ion spacings. Because the ion spacings have been configured to be nearly uniform, the gate times do not vary much for ion pairs with the same separation. In figure \ref{fig:GateTimeGraph}, we show the calculated minimal gate times for ions pairs at the distance of 1, 3, 5, 7, 9 and 11 ion spacings.  To find these ``minimal" gate times, we searched different detunings and number of segments with a step of $10\,\mu$s and screened for solutions with an infidelity below $3\times 10^{-6}$.  We further require the effective Rabi frequency $\Omega_j(t)$ to be below $2 \pi \times 1\,$MHz, which is comparable with the value in current experiments. We note that the gate time limits here are not fundamental and alternative approaches could lead to faster two-qubit gates \cite{LeungFMgates2017,SchaferFastLogicIon2017}.

The underlying connection graph of a trapped linear ion chain is a fully connected graph \cite{BrownCoDesignQC2016}.  Therefore, there are many ways to map the surface-17 code to the linear ion chain.  A natural mapping is to split the ion chain into groups of data and ancillary qubits which appears to be advantageous by minimizing ion shuttling times since all measurement can be performed in parallel and would not require the storage zone.  However, the data-ancilla distance between ions in this configuration is larger which, as we have shown above, results in slower two-qubit gates.  We fit the gate time as a result of the ion distance (figure \ref{fig:GateTimeGraph}) to a linear function yielding:
\begin{equation}
t_g = 10 + 38 d
\end{equation}
where $t_g$ is the gate time ($\mu$s) and $d$ is the ion distance.  As we can see from figure \ref{fig:gatetimeLuming}, the nearest neighbor calculated gate time (ion pair 8 and 9) corresponds to $d=1$ which gives a $t_g \approx 50\,\mu$s.  Single-qubit gates can be performed in parallel with a gate time of $10 \, \mu s$.  With this relationship between ion distance and gate times, we screened for the optimal ion chain configurations using a simulated annealing algorithm that minimized several parameters of interest.  Three parameters were minimized: the maximum ion distance between entangled ions (M), the average ion distance between entangled ions (A), and the total time for one round of syndrome measurement in parallel (T) corresponding to the second letter in the labels in figure \ref{fig:GateTimeTable}.  In addition, the optimizations were performed with constraint that the data and ancilla qubits are separate (S) and are allowed to be mixed  together (M) corresponding to the first letter in the labels in figure \ref{fig:GateTimeTable}.  The corresponding connection graphs for two optimized chains (SM and MT) are shown in figure \ref{fig:ionconnectivity}.

The ion splitting time was assigned to be $100 \, \mu$s between neighboring zones \cite{WaltherFastShuttle2012,BowlerFastShuttle2012,BermudezITQCwithSteane2017}.  Therefore, splitting ions in the chain from the logic zone and shuttling the ions to the SPAM or storage zone requires a time of $100 \, \mu$s or $200\, \mu$s, respectively.  This time is built from an assumption of a $200\,$kHz lowest axial frequency that implies that splitting/merging of subsets of ions in the ion chain can occur at a rate almost at this frequency.  For splitting the ion chain, the transport is expected to be limited to $7\,$m/s assuming a $50\,$kHz update rate in the transport waveforms \cite{ShuYtrap2014}.  Therefore, the remaining $95 \, \mu$s allows for the chains to be separated by a distance of $665 \, \mu$m which is excellent separation between the detection lasers and the data qubits.  It is assumed that operations can happen in parallel so part of an ancillary ion subchain can be shuttled to the storage zone while the other ions within the same subchain remain in the detection zone.  The rejoining of the ions shuttled to the detection zone is assumed to occur in parallel with the next splitting operation which leads to a fixed cost for rejoining the chain of also $100 \, \mu$s.  The final assumption is that a three way split requires the same amount of time as single splitting operation of the ion chain due to the parallelism which is reasonable due to the small contribution of splitting operations to the total zone to zone movement time.  The measurement time was also fixed to $100 \, \mu$s which is a lax requirement on the experimental apparatus and will allow for high fidelity state detection \cite{NoekHiFidSPAMYb2013}.

\begin{figure}
\scriptsize
\centering
\begin{tabular}{ |c | c | c | c |  c | c |}
\hline
{\bf Label} & {\bf Logic} & {\bf Shuttle} & {\bf Measure} & {\bf Total} & {\bf Ion Ordering} \\ \hline
\multirow{2}{*}{SM} & 7650  & \multirow{2}{*}{200} & \multirow{2}{*}{100} & 7950 & \multirow{2}{*}{\tiny \circled{$\;1\;$} \circled{$\;2\;$} \circled{$\;5\;$} \circled{$\;8\;$} \circled{$\;0\;$} \circled{$\;4\;$} \circled{$\;3\;$} \circled{$\;6\;$} \circled{$\;7\;$} \dcircled{$\;9\;$} \dcircled{12} \dcircled{11} \dcircled{14} \dcircled{15} \dcircled{13} \dcircled{10} \dcircled{16}}\\
& \emph{3920} &  &  & \emph{4220}  & \\ \hline
\multirow{2}{*}{SA} & 7240 & \multirow{2}{*}{200} & \multirow{2}{*}{100} & 7640 & \multirow{2}{*}{\tiny \circled{$\;0\;$} \circled{$\;2\;$} \circled{$\;6\;$} \circled{$\;8\;$} \circled{$\;1\;$} \circled{$\;4\;$} \circled{$\;3\;$} \circled{$\;7\;$} \circled{$\;5\;$} \dcircled{11} \dcircled{12} \dcircled{10} \dcircled{15} \dcircled{13} \dcircled{14} \dcircled{$\;9\;$} \dcircled{16}}\\
& \emph{4140}  &  &  & \emph{4440}  & \\ \hline
\multirow{2}{*}{MM} & 3080 & \multirow{2}{*}{1200} & \multirow{2}{*}{500} & 4780 & \multirow{2}{*}{\tiny \circled{$\;5\;$} \dcircled{15} \circled{$\;2\;$} \dcircled{12} \dcircled{14} \dcircled{$\;9\;$} \circled{$\;8\;$} \circled{$\;1\;$} \circled{$\;4\;$} \circled{$\;7\;$} \dcircled{11} \circled{$\;3\;$} \dcircled{13} \dcircled{16} \circled{$\;0\;$} \dcircled{10} \circled{$\;6\;$}}\\
& \emph{1690}  &  &  & \emph{3390}  & \\ \hline
\multirow{2}{*}{MA} & 2300 & \multirow{2}{*}{1800} & \multirow{2}{*}{800} & 4900 & \multirow{2}{*}{\tiny \circled{$\;2\;$} \dcircled{$\;9\;$} \circled{$\;1\;$} \dcircled{12} \circled{$\;5\;$} \dcircled{15} \circled{$\;8\;$} \dcircled{14} \circled{$\;4\;$} \dcircled{11} \circled{$\;0\;$} \dcircled{10} \circled{$\;3\;$} \dcircled{13} \circled{$\;7\;$} \dcircled{16} \circled{$\;6\;$}}\\
& \emph{1170}  &  &  & \emph{3770}  & \\ \hline
\multirow{2}{*}{MT} & 4300 & \multirow{2}{*}{700} & \multirow{2}{*}{300} & 5300 & \multirow{2}{*}{\tiny \dcircled{10} \dcircled{15} \dcircled{$\;9\;$} \circled{$\;5\;$} \circled{$\;0\;$} \circled{$\;1\;$} \dcircled{11} \dcircled{12} \dcircled{14} \circled{$\;7\;$} \circled{$\;4\;$} \circled{$\;3\;$} \circled{$\;8\;$} \circled{$\;2\;$} \circled{$\;6\;$} \dcircled{13} \dcircled{16}}\\
& \emph{2320}  &  &  & \emph{3320}  & \\ \hline
\end{tabular}
\caption{Trap operation times and ion arrangements optimized for an array of parameters.  The first letter of the label refers to S=separate and M=mixed arrangements of data and ancilla qubits.  The second letter of the label refers to the parameter minimized with M=maximum distance between entangled ions, A=average distance between entangled ions, and T=parallel total gate time.  All values are reported in microseconds and the numbers in roman and \emph{italics} refer to the gate time of the operations performed in serial and parallel, respectively.  Parallel operations allow for two simultaneous two-qubit gates exciting the independent $x$ and $y$ radial modes and fully parallel single-ion operations.  Single-qubit gates, parallel measurement/state preparation, and shuttling between neighboring zones require $10$ $\mu s$, $100$ $\mu s$, and $100$ $\mu s$ ($5$ $\mu s$ split and $95$ $\mu s$ shuttle time), respectively.}
\label{fig:GateTimeTable}
\end{figure}

The time required to measure the error syndrome for different optimized configurations are shown in figure \ref{fig:GateTimeTable}.  The gate times (Logic) for the chain configurations where the data and ancilla qubits are separate (labels SM and SA) are substantially longer than the mixed configurations.  The mixed configurations (MM, MA, and MT) have longer chain manipulation and measurement times.  The longer times are due to the inability to perform all measurements in parallel for a mixed arrangement; only subchains consisting of neighboring ancilla can be measured in parallel.  An example of a parallel step for the mixed configuration is shown in figure \ref{fig:IonTrap} where the ion configuration corresponds to the ion chain label MT.  Ions with labels 11, 12, 14 in the surface code are measured in parallel in this measurement step.  The neighboring data qubits on the ends of the subchain consisting of the three ancillary qubits restrict measurement on other ancillary qubits in this architecture.  The entanglement gates outlined above allow for parallel implementation.  Two simultaneous entanglement gates can be performed on two independent pairs of ions by exciting the $x$ and $y$ radial modes, respectively, for each pair.  Single-qubit operations are completely parallel for both the serial and parallel implementations.  The parallel operation times are shown in italics in figure \ref{fig:GateTimeTable}.  For the detailed calculations below, we chose the ion chain configuration that gives the minimal total syndrome measurement time (serial or parallel), MT.  Note that the gate times for the MM and MA configurations have shorter serial Logic times so these configurations will perform better than the MT configuration under the influence of our gate based error model outlined below.

\section{Modeling Ion Trap Error Sources} \label{sec:ErrorModels}
For accurate assessment of error correction in an ion trap quantum computer, appropriate error models must be developed to simulate noise sources in the physical architecture.  This section provides the components for building up such complexity.  The Kraus operator representation will be used to describe the components of the quantum error channel.  A graphical representation of the full ion trap error model is shown in figure \ref{fig:ITErrorModel}.

\subsection{Depolarizing Error Model}
The depolarizing error model is a standard error model used in simulations of quantum error correcting codes.  After the application of each gate in the quantum circuit implemented to measure the stabilizers, an element is sampled from the one-qubit (two-qubit) Pauli group and applied after each single-qubit (two-qubit) gate.  The one- and two-qubit Kraus channels are of the form:
\begin{equation}
\begin{split}
& E_{1,d} = \left\{ \sqrt{1-p} \, I, \; \sqrt{\frac{p}{3}} \, X, \; \sqrt{\frac{p}{3}} \, Y, \; \sqrt{\frac{p}{3}} \, Z \right\} \\
& E_{2,d} = \left\{ \sqrt{1-p} \, II, \; \sqrt{\frac{p}{15}} \, IX, \; \sqrt{\frac{p}{15}} \, IY, \; \sqrt{\frac{p}{15}} \, IZ, \; \sqrt{\frac{p}{15}} \, XX, ... , \; \sqrt{\frac{p}{15}} \, ZZ \right\}
\end{split}
\label{eqn:depchannel}
\end{equation}
where $p$ is the \emph{error rate} of the error channel.  Furthermore, the application of perfect two-qubit gates still allows for certain errors to propagate from single- to two-qubit errors.  For measurement of the stabilizers, the $CNOT$ (controlled-$X$) is the two-qubit gate and transforms two-qubit Pauli errors in the following manner:
\begin{equation}
\left\{ XI,\; XX, \; IZ,\; ZZ \right\} \rightarrow \left\{ XX, \; XI, \; ZZ,\; IZ \right\}
\end{equation}
where the first (second) operator is on the control (target).  The $Y$ error rules can be built from the relation $Y = i XZ$.  The stabilizer circuits in this work are built using only the $CNOT$ as the two-qubit gate.  This error model allows for errors on both the data and ancilla qubits, which translate into errors in the measurement of stabilizers during syndrome extraction.  Furthermore, preparation and measurement errors are modeled by the application of a single-qubit depolarizing error channel after preparation gates and before measurement.  This model will serve as a baseline error model for assessment of error correction.

\subsection{Coherent Over-Rotation of the M{\o}lmer-S{\o}rensen Gate}
\begin{figure}
\captionsetup{width=0.9\textwidth}
\centering
\mbox{
\Qcircuit @C=0.7em @R=2.1em {
& \ctrl{1} & \qw \\
& \targ & \qw
}}
~
\mbox{
\Qcircuit @C=0.6em @R=0.25em {
& \\
& \\
& \\
& \\
& \qw \\
& \qw \\
& \qw }
}
~
\mbox{
\Qcircuit @C=1em @R=0.8em {
&\gate{R_Y\left(\pm \frac{\pi}{2} \right)} & \multigate{1}{XX \left(s \, \frac{\pi}{4} \right)} & \gate{R_X\left(s \, \frac{\pi}{2} \right)} & \gate{R_Y\left(\mp \frac{\pi}{2} \right)} & \qw \\
& \qw & \ghost{XX \left(s \, \frac{\pi}{4} \right)} & \gate{R_X\left(\mp \frac{\pi}{2} \right)} & \qw & \qw
}}
\caption{The construction of a $CNOT$ logic gate from a M{\o}lmer-S{\o}rensen entangling gate and single-qubit gates as follows from \cite{MaslovCircuitCompIT2017}.  The quantity $s$ is ion specific and equal to the sign of the experimental interaction parameter $\chi$.}
\label{fig:MStoCNOT}
\end{figure}
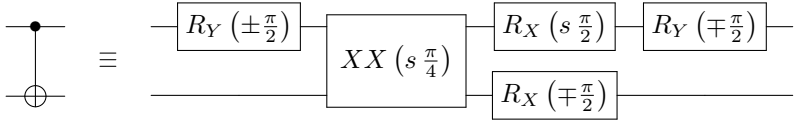

The first step in adding complexity to the error model entails compiling the two-qubit quantum logic gates in the abstract quantum circuit with experimental entangling gates.  The M{\o}lmer-S{\o}rensen (MS) entangling gate \cite{SorensenPRLMSgate1999,SorensenPRAMSgate2000} was chosen for this purpose due to its faster gate times and higher gate fidelities relative to other entangling gate schemes \cite{DebnathSmallQCIons2016}.  The MS gate uses a bichromatic laser field to induce a two-photon transition that couples $\ket{00} \leftrightarrow \ket{11}$ and $\ket{10} \leftrightarrow \ket{01}$ qubit states.  The MS gate induces a transition with a bichromatic laser tuned close to the upper and lower motional sideband of a qubit transition \cite{SorensenPRLMSgate1999,SorensenPRAMSgate2000}.  In the compuatational basis, the unitary operator associated with the M{\o}lmer-S{\o}rensen gate is:
\begin{equation}
XX \left(\chi\right) =\left( \begin{array}{c c c c}
\mathrm{cos}\left(\chi \right) & 0 & 0  & -i \, \mathrm{sin}\left(\chi \right) \\
0 & \mathrm{cos}\left(\chi \right) & -i \, \mathrm{sin}\left(\chi \right)  & 0 \\
0 & -i \, \mathrm{sin}\left(\chi \right) & \mathrm{cos}\left(\chi \right)  & 0 \\
-i \, \mathrm{sin}\left(\chi \right) & 0 & 0 & \mathrm{cos}\left(\chi \right) 
\end{array}
\right)
\end{equation}
where the parameter $\chi$ depends on the gate time applied to the specific ion pair \cite{DebnathSmallQCIons2016}.  The absolute value of the angle, $|\chi|$, may be set to any real number between $0$ and $\pi/2$ by varying the power of the laser in the experiment \cite{DebnathSmallQCIons2016}.  The sign of $\chi$ is dependent on the laser detuning which is chosen from normal modes of the ion pair \cite{DebnathSmallQCIons2016}.  The $CNOT$ gate can be achieved by assigning $\chi = \pm \pi/4$ and sandwiching the two-qubit unitary between single-qubit gates as shown in figure \ref{fig:MStoCNOT} \cite{MaslovCircuitCompIT2017}.  The M{\o}lmer-S{\o}rensen unitary implemented during the $CNOT$ can equivalently be written as:
\begin{equation}
XX\left(\chi\right) = \mathrm{exp} \left( - i \, \chi \, XX \right) = \mathrm{cos}\left(\chi\right)\, II - i \, \mathrm{sin}\left(\chi\right) \, XX
\label{eqn:MSexp}
\end{equation} 
where we attempt to assign $\chi$ as $\pi/4$ with the laser field.  However due to experimental error, a small over-rotation (with angle $\alpha$) may be applied about the $XX$ axis with the real gate applied in equation \ref{eqn:MSexp} having an angle of $\chi+\alpha$.  This error will be simulated by a probabilistic error channel of the form:
\begin{equation}
E_{2,xx} = \left\{ \sqrt{1-p_{xx}} \, II,\; \sqrt{p_{xx}} \, XX \right\}
\end{equation}
where the probability of the channel above is a function of the over-rotation angle.  For example, one possible relation between $p_{xx}$ and $\alpha$ is obtained by the Pauli twirled approximation, which results in $p_{xx} = \mathrm{sin}^2 \left(\alpha\right)$ \cite{GutierrezIncohCohNoiseEC2016}.  It is also possible to choose $p_{xx}$ such that the Pauli approximation to the over-rotation satisfies additional constraints \cite{PuzzuoliHonestApproxRealModels2014,GutierrezApproxRealError2013}.  Furthermore, the single-qubit rotation gates in the circuit (figure \ref{fig:MStoCNOT}) can also suffer over-rotations, although typically to a much less degree.  The over-rotations can be modeled in an analogous way giving three distinct gate-dependent error channels:
\begin{equation}
\begin{split}
& E_{1,x} = \left\{ \sqrt{1-p_x} \, I, \; \sqrt{p_x} \, X \right\} \\
& E_{1,y} = \left\{ \sqrt{1-p_y} \, I, \; \sqrt{p_y} \, Y \right\} \\
& E_{1,z} = \left\{ \sqrt{1-p_z} \, I, \; \sqrt{p_z} \, Z \right\}
\end{split}
\end{equation}
which are applied after every single-qubit rotation gate $R_X(\theta)$, $R_Y(\theta)$, and $R_Z(\theta)$, respectively.  For simulations, the error rates for the single-qubit gates are a factor of 10 lower than those corresponding to two-qubit gates; representing observed single- and two-qubit gate fidelities \cite{BallanceControlHFQubits2016,BrownHighFidBe2011}.

\subsection{Motional Mode Heating}

In addition to control errors, the applied field from M\o lmer-S\o rensen gate can result in motional heating of the ions, which impacts the fidelity of the two-qubit gate.  Modeling heating as a coupling of the motional states of the ions to an infinite temperature bath \cite{TurchetteIonHeatingBath2000}, Ballance et al. characterized the impact of motional heating on the error of a two-qubit entangling gate, $\epsilon_h$, giving:
\begin{equation}
\epsilon_h = \frac{\dot{\bar{n}} t_g}{2 K}
\label{eqn:EfromHeat}
\end{equation}
where $\dot{\bar{n}}$ is the average change in the thermal occupation number of the gate mode, $t_g$ is the gate time, and $K$ is the number of loops in phase space traversed by the ions during the gate \cite{BallanceControlHFQubits2016}.  We chose to study the low $K$ limit ($K = 1,2$) of equation \ref{eqn:EfromHeat} modeling heating errors with the Kraus operators which are applied after every MS gate:
\begin{equation}
E_{2,h} = \left\{ \sqrt{1-p_h} \, II,\; \sqrt{p_h} \, X X \right\}
\end{equation}
where the probabilities are ion-dependent: $p_h = \left( r_{heat} \right) \times \left( t_{MS} \right)$ where $r_h$ is the heating rate and $t_{MS}$ is the time of the M\o lmer-S\o rensen gate.  It is important to note that this model is pessimistic with respect to ion heating, even in the low $K$ limit, and the choice of coupling modes can increase $K$ by $1-2$ orders of magnitude \cite{ROzeriErrSpontScatt2007,HayesMSErrorSupp2012}.

\subsection{Background Depolarizing Noise}
For the stable ``clock" states of the hyperfine qubits, errors arise from the application of gates.  In addition to systematic over/under-rotations of the applied laser field, instabilities in the control of the qubits (laser field drifts, magnetic field fluctuations, etc.) can lead to stochastic error processes that we will model with a depolarizing error channel.  One such natural stochastic process that has shown to be a contributing source of error is scattering during the application of the gate \cite{OzeriScattErrYb2007,Gea-BanaclocheHyperRamaScatt2005}.  To model the effects of spontaneous Raman and Rayleigh photon scattering, we will apply a single-qubit depolarizing channel (equation \ref{eqn:depchannel}) after every qubit involved in a gate (single- or two-qubit gates).

\subsection{Dephasing Errors}
While the ions are located in the trap where the DC electric fields vanish, the ions may still be exposed to oscillating electric fields from blackbody radiation, laser fields, or motion around the field free point in the oscillating trap field \cite{LudlowOptClocks2014}.  The application of the oscillating electric field shifts the energy each of the states of the two-level qubit system by the AC Stark effect, which introduces dephasing errors in the applied gates.  This effect is observed for both single- and two-qubit gates.  
We choose to model these dephasing errors as a single qubit channel of the form:
\begin{equation}
E_{d} = \left\{ \sqrt{1-p_d} \, I, \; \sqrt{p_d} \, Z \right\}
\end{equation}
where each channel is applied to each qubit involved in single- and two-qubit gates and $p_d = r_d \times t_g$ for each gate where $r_d$ is the dephasing rate and $t_g$ is the time of the applied gate.  We make the approximation that single- and two-qubit dephasing errors occur at a constant rate.  This is certainly not true in that the dephasing rates will be gate dependent between two-qubit gates and will likely not be at the same rate of single-qubit dephasing but, taking that single-qubit gates have higher fidelities relative to two-qubit gates, this serves as a pessimistic approximation which is consistent with  our level of abstraction.

\begin{figure}[b!]
\includegraphics[width=\textwidth]{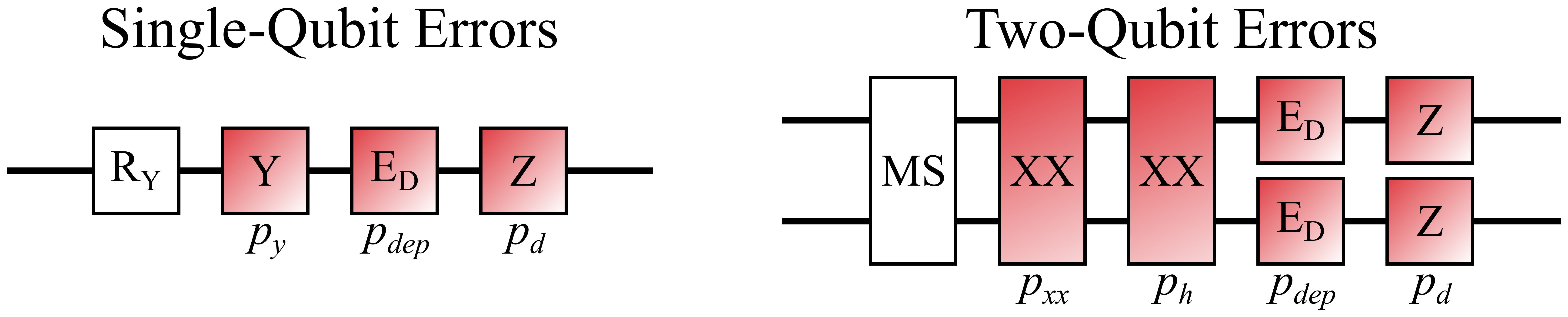}
\caption{Graphical representation of the ion trap error model implemented for the 17-qubit surface code simulations.  Rotation errors (for both single- and two-qubit gates) occur about the axis of rotation of the applied gate.  Motional mode heating errors (for two-qubit gates) manifest themselves as $XX$ errors after the applied M\o lmer-S\o rensen gate with a probability proportional to the time of the applied gate (ion/qubit dependent).  Depolarizing errors and dephasing errors are applied independently  to each qubit involved in the gate with a static probability $p_{dep}$ and probability proportional to the gate time for depolarizing and dephasing errors, respectively.}
\label{fig:ITErrorModel}
\end{figure}

\subsection{Ancilla Preparation and Measurement Errors} \label{sec:SPAM}
For the ion trap error model, measurement errors were modeled by a single-qubit depolarizing channel applied before the measurement with a probability equivalent to that of the single-qubit over-rotation errors of the single-qubit gates.  Preparation errors were modeled with a single-qubit depolarizing channel applied immediately after the preparation of the state but with a probability equivalent to the background depolarizing channel.  All states are prepared and measured in the $+Z$ basis, which can be performed with high-fidelity \cite{HartyMagClock2014}.  Note that this implementation is not ideal given that both state preparation and state readout rely on the same scattering processes.  However, the preparation and measurement errors should not be the dominant source of failure in the simulations consistent with single-qubit gate, preparation, and readout fidelities of $\ge 99.9\, \%$ \cite{HartyMagClock2014}.  Furthermore, state preparation/measurement is a high-fidelity operation (relative to two-qubit gates) so the inflated state preparation errors will give a pessimistic simulation of the fault-tolerance of the surface code on ion traps relative to the physical architecture.  These claims are reinforced in section \ref{sec:SingleErrorThresh}.

\section{Error Correction for Ion Trap Errors}

To perform error correction on the surface code, classical decoding algorithms have been developed to determine the most appropriate correction operation to perform given the limited information about the encoded state from the syndrome.  Various decoders are available that trade-off classical efficiency and observed error threshold.  We apply a few decoders for error correction on the surface code below and discuss their performance.  For all simulations, we implemented a Monte Carlo simulation of the surface code using an importance sampling method described in Ref. \cite{LiBareAnc2017}.

\subsection{Integration into Ion Trap Hardware}
When choosing a decoding method to integrate into a physical architecture, there is much to consider that extends beyond the (pseudo)threshold.  Processing, memory, and runtime requirements of the decoder play a role in the feasibility of implementing error correction with an experimental control system.

\subsubsection{Lookup Table Decoder}
The simplest decoder is a lookup table that maps a syndrome configuration to the lowest weight Pauli error corresponding to the syndrome.  We may represent an error configuration ${\bf e}$ as a binary (row) vector $\mathbb{F}_2^{18}$ where the first/last 9 elements of the vector correspond to $X$-type/$Z$-type errors on the data qubits; for instance:
\begin{equation}
{\bf e}(2563)= \left[0\;0\;0\;0\;0\;0\;1\;0\;1\;0\;0\;0\;0\;0\;0\;0\;1\;1 \right] = X_6 Z_7 Y_8
\end{equation}
Given two matrices, $H$ and $G^T$, that correspond to the binary representation of the $X$-type and $Z$-type stabilizers, respectively; one may define a mapping matrix $T$ between error configurations ${\bf e}$ and binary syndrome (column) vectors ${\bf s}$:
\begin{equation}
T =
\left(
\begin{array}{cc}
H & 0\\
0 & G^T
\end{array} \right)
= \left(
\begin{array}{ccccccccc ccccccccc}
0 & 1 & 1 & 0 & 0 & 0 & 0 & 0 & 0 & 0 & 0 & 0 & 0 & 0 & 0 & 0 & 0 & 0 \\
1 & 1 & 0 & 1 & 1 & 0 & 0 & 0 & 0 & 0 & 0 & 0 & 0 & 0 & 0 & 0 & 0 & 0 \\
0 & 0 & 0 & 0 & 1 & 1 & 0 & 1 & 1 & 0 & 0 & 0 & 0 & 0 & 0 & 0 & 0 & 0 \\
0 & 0 & 0 & 0 & 0 & 0 & 1 & 1 & 0 & 0 & 0 & 0 & 0 & 0 & 0 & 0 & 0 & 0 \\
0 & 0 & 0 & 0 & 0 & 0 & 0 & 0 & 0 & 1 & 0 & 0 & 1 & 0 & 0 & 0 & 0 & 0 \\
0 & 0 & 0 & 0 & 0 & 0 & 0 & 0 & 0 & 0 & 1 & 1 & 0 & 1 & 1 & 0 & 0 & 0 \\
0 & 0 & 0 & 0 & 0 & 0 & 0 & 0 & 0 & 0 & 0 & 0 & 1 & 1 & 0 & 1 & 1 & 0 \\
0 & 0 & 0 & 0 & 0 & 0 & 0 & 0 & 0 & 0 & 0 & 0 & 0 & 0 & 1 & 0 & 0 & 1 \\
\end{array} 
\right)
\end{equation}
Iterating over all elements of $\mathbb{F}_2^{18}$ and applying $T$, we constructed a lookup table Tab$\left[s\right] =\lfloor e \rfloor$ where $\lfloor e \rfloor = \mathrm{min}_s \left(|e|\right)$ is the minimum weight error configuration corresponding to the syndrome string $s$.  With a slight abuse of notation, we denote $|\cdot|$ as the hamming weight of the error string $e$ with the caveat that $Y$-type errors are evaluated as the same weight as $X$ and $Z$-type errors.  Those familiar with the CSS construction of quantum error correcting codes will recognize $H$ and $G^T$ as the parity check matrices of $\mathcal{C}$ and $\mathcal{C}^{\perp}$ of a classical linear error correcting code used to construct the 17-qubit surface code \cite{MikeNIke}.  All of the rules of the full lookup table (Tab$\left[s\right]$) can be constructed with two 16-element tables, each with keys corresponding to the $X$-type and $Z$-type stabilizer measurements, respectively.

For circuit-level noise, the lookup table above is not sufficient for fault-tolerance.  A set of syndrome processing rules must be imposed to ensure that measurement errors do not result in faulty corrections that introduce errors onto the data qubits.  An example of a typical set of rules is shown below ($a$, $b$, and $c$ are syndrome outcome strings):
\begin{figure}[h!]
\centering
\begin{minipage}{0.5\textwidth}
\mbox{
\Qcircuit @C=0.7em @R=2.1em {
& \gate{a} & \qw & \gate{b} & \qw & \gate{a\neq b:c} \ar@{--}[]+<2.8em,1em>;[]+<2.8em,-1em> & \qw & \gate{a} & \qw & \gate{b} & \qw & \gate{a\neq b:c} & \qw & \hdots}} \\ \vspace{2mm}
\end{minipage}
\vspace{-3mm}
\end{figure}
\\
where two rounds of stabilizer measurement are performed and, if the first two measurement outcomes disagree, a third round of stabilizer measurement is performed.  Correction is applied based upon the final measurement performed.  We chose to employ a different set of fault-tolerant syndrome processing rules that can, on average, reduce the depth of the circuit required to perform a fault-tolerant correction by one round of stabilizer measurement.  The routine:
\begin{figure}[h!]
\centering
\begin{minipage}{0.40\textwidth}
\mbox{
\Qcircuit @C=0.7em @R=2.1em {
& \gate{a} & \qw & \gate{a\neq 0: b} \ar@{--}[]+<2.8em,1em>;[]+<2.8em,-1em> & \qw & \gate{a} & \qw & \gate{a \neq 0: b} & \qw & \hdots}}
\end{minipage}
\vspace{-3mm}
\end{figure}
\\
performs one round of stabilizer measurements and performs a correction based on the following round of stabilizer measurements ($b$) only if the first round was non-trivial $\left(a \neq 0 \right)$.  These two sets of rules yield equivalent results for the 17-qubit surface code under circuit-level depolarizing noise.

\subsubsection{Minimum Weight Perfect Matching}
For topological codes, minimum weight matching algorithms have been shown to be a useful heuristic technique for performing error correction \cite{FowlerMatchingSC2012, EdmondsPaths1965, EdmondsMatroid1965}.  For the distance 3 surface code, the minimum weight perfect matching rules can be encoded into a lookup table that presents a correction operation based on three rounds of syndrome measurement (for circuit-level depolarizing noise) \cite{TomitaLowDSC2014}.

\begin{figure}[t!]
\centering
\begin{tabular}{|c | c | c |}
\hline
{\bf Decoder} & {\bf Level-1 Pseudothreshold} & {\bf Computational Time ($s$)}\\ \hline
Lookup Table & $3.0 \times 10^{-3}$ & $1.1 \times 10^{-7}$ \\ \hline
Matching (table) & $5.5 \times 10^{-3}$ & $1.43 \times 10^{-6}$  \\ \hline
\end{tabular}
\caption{Performance of the two lookup table style decoders considered for implementation into a near-term quantum error correction experiment.  Lookup table style decoders were chosen due to their easy integration into the control software of an ion trap system.}
\label{fig:DecoderPerf}
\end{figure}

\begin{figure}
\centering
\footnotesize
\begin{subfigure}[t]{0.25\textwidth}
\hspace{3mm} \mbox{
\Qcircuit @C=0.8em @R=.94em {
\lstick{0} & \multigate{8}{MS(X)} & \gate{R_X(-\frac{\pi}{2})} &  \qw \\
\lstick{1} & \ghost{MS(X)} & \qw & \qw \\
\lstick{2} & \ghost{MS(X)} & \gate{R_X(+\frac{\pi}{2})} & \qw \\
\lstick{3} & \ghost{MS(X)} & \gate{R_X(-\frac{\pi}{2})} &  \qw \\
\lstick{4} & \ghost{MS(X)} & \qw & \qw \\
\lstick{5} & \ghost{MS(X)} & \gate{R_X(+\frac{\pi}{2})} & \qw \\
\lstick{6} & \ghost{MS(X)} & \gate{R_X(-\frac{\pi}{2})} & \qw \\
\lstick{7} & \ghost{MS(X)} & \qw & \qw \\
\lstick{8} & \ghost{MS(X)} & \gate{R_X(+\frac{\pi}{2})} & \qw \\
}}
\caption{$X$-type stabilizers}
\end{subfigure}
\hspace{10mm}
\begin{subfigure}[t]{0.45\textwidth}
\hspace{3mm} \mbox{
\Qcircuit @C=0.8em @R=0.65em {
\lstick{0} & \gate{R_Y(+\frac{\pi}{2})} & \multigate{8}{MS(Z)} & \gate{R_X(+\frac{\pi}{2})} & \gate{R_Y(-\frac{\pi}{2})} & \qw \\
\lstick{1} & \gate{R_Y(+\frac{\pi}{2})} & \ghost{MS(Z)} & \gate{R_X(+\frac{\pi}{2})} & \gate{R_Y(-\frac{\pi}{2})} & \qw \\
\lstick{2} & \gate{R_Y(+\frac{\pi}{2})} & \ghost{MS(Z)} & \gate{R_X(+\frac{\pi}{2})} & \gate{R_Y(-\frac{\pi}{2})} & \qw \\
\lstick{3} & \gate{R_Y(+\frac{\pi}{2})} & \ghost{MS(Z)} & \qw & \gate{R_Y(-\frac{\pi}{2})} & \qw \\
\lstick{4} & \gate{R_Y(+\frac{\pi}{2})} & \ghost{MS(Z)} & \qw & \gate{R_Y(-\frac{\pi}{2})} & \qw \\
\lstick{5} & \gate{R_Y(+\frac{\pi}{2})} & \ghost{MS(Z)} & \qw & \gate{R_Y(-\frac{\pi}{2})} & \qw \\
\lstick{6} & \gate{R_Y(+\frac{\pi}{2})} & \ghost{MS(Z)} & \gate{R_X(-\frac{\pi}{2})} & \gate{R_Y(-\frac{\pi}{2})} & \qw \\
\lstick{7} & \gate{R_Y(+\frac{\pi}{2})} & \ghost{MS(Z)} & \gate{R_X(-\frac{\pi}{2})} & \gate{R_Y(-\frac{\pi}{2})} & \qw \\
\lstick{8} & \gate{R_Y(+\frac{\pi}{2})} & \ghost{MS(Z)} & \gate{R_X(-\frac{\pi}{2})} & \gate{R_Y(-\frac{\pi}{2})} & \qw \\
}}
\caption{$Z$-type stabilizers}
\end{subfigure}
\vspace{2mm}

\begin{subfigure}[t]{0.6\textwidth}
\centering
\begin{lstlisting}[backgroundcolor = \color{softblue}]
       MS(X)                           MS(Z)         
 GATE  s  ID1  ID2                GATE  s  ID1  ID2
 PREP     16                      PREP     10
 XX    +   6   16                 XX    -   0   10
 XX    -   7   16                 XX    +   3   10                   
 MEAS     16                      MEAS     10  

 PREP     11                      PREP     12
 XX    +   0   11                 XX    -   1   12
 XX    -   1   11                 XX    -   4   12
 XX    +   3   11                 XX    +   2   12
 XX    -   4   11                 XX    +   5   12 
 MEAS     11                      MEAS     12

 PREP     14                      PREP     13
 XX    +   4   14                 XX    -   3   13
 XX    -   5   14                 XX    -   6   13
 XX    +   7   14                 XX    +   4   13
 XX    -   8   14                 XX    +   7   13
 MEAS     14                      MEAS     13

 PREP      9                      PREP     15
 XX    +   1   9                  XX    -   5   15
 XX    -   2   9                  XX    +   8   15  
 MEAS      9                      MEAS     15             
\end{lstlisting}
\end{subfigure}
\caption{(Top) The syndrome extraction circuit for the 17-qubit surface code compiled with M\o lmer-S\o rensen entangling gates and single-qubit ion trap operations where the ancillary qubit wires have been suppressed.  The number of single-qubit gates required for the circuit is 30, which is a substantial reduction relative to the naive implementation.  (Bottom) The primitive gate operations compiling the $MS(X)$ and $MS(Z)$ gates above.  The values ID1 and ID2 correspond to the qubit indices to which the gate is applied defined in figure \ref{fig:17QSC}.  The PREP gate projects ancillary qubits into the $\ket{0}$ state and all MEAS gates are $Z$-basis measurements (see section \ref{sec:SPAM}).  The XX gates are M\o lmer-S\o rensen gates.  The parameter $s$ which is dictated by the sign experimental interaction parameter was taken as a free parameter during compilation.  The assignment of $s$ for each gate is shown explicitly.  Note that this is an intuitive representation of the stabilizer measurements and does not indicate the order of operations in our architecture (recall that all entangling gates are performed and then preparation/measurement gates).}
\label{fig:MSsyndromecirc}
\end{figure}

\subsubsection{Decoder Performances}
Figure \ref{fig:DecoderPerf} shows the performance of the two lookup table style decoders, standard lookup and matching rule derived lookup, considered for implementation in a near-term experimental quantum error correction routine.  Lookup table decoders were chosen for their easy integration into existing ion trap experimental controls which have restricted logic/memory available versus other techniques, such as maximum likelihood \cite{BravyiMLD2014,HeimCircuitLevelMLD2016} or deeper memory step matching algorithms \cite{FowlerMatchingSC2012, EdmondsPaths1965, EdmondsMatroid1965} for example, which would require additional processing power to implement/integrate into an experiment.  The lookup table decoder was favored over the matching table because of its requirement for one less round of stabilizer measurement to perform fault-tolerant error correction with a comparable level-1 pseudothreshold to the matching table (figure \ref{fig:DecoderPerf}).  Because current estimates of the syndrome extraction indicate it is relatively slow (figure \ref{fig:GateTimeTable}), the ability to choose a correction fault-tolerantly from a minimal number of experimental operations is important to maintain coherence of the encoded information.  The lookup table was implemented in all further simulations because of ease of integration into ion trap controls while requiring at most two syndrome measurements to fault-tolerantly perform error correction.

\subsection{Error Correction on Ion Traps}
Now that a fast, light memory, high-performance decoder has been identified, we will switch attention to using such a method to apply error correction on the 17-qubit surface code under the influence of ion trap errors.  First, we must map the abstract quantum circuit used for error correction in the surface code to a circuit that implements gates that would be available in an ion trap quantum computer; specifically single-qubit rotations and M\o lmer-S\o rensen gates.  Next, we will discuss the influence of the individual ion trap error sources (outlined in section \ref{sec:ErrorModels}) on the fault-tolerance of the surface code mapped to a linear ion chain highlighting the experimental parameter regimes which would allow for fault-tolerance for the surface code implementation.  Finally, we analyze the error subset probabilities from the importance sampling simulations to understand the roles of the competing error sources and gain insight into the error sources that are most influential/detrimental to the error correcting properties of the code.

\subsection{Surface-17 Syndrome Extraction Circuit Gate Compilation}
As shown in figure \ref{fig:MStoCNOT}, the two-qubit gates in the syndrome extraction circuit for the 17-qubit surface code must be decomposed into single-qubit rotation gates and two-qubit M\o lmer-S\o rensen gates.  In addition, Hadamard gates are required during the measurement of the $X$-type stabilizers which can be decomposed into rotation gates in two equivalent ways: $H \equiv R_Y\left(-\frac{\pi}{2} \right) R_X\left(\pi \right)$ or $H \equiv R_X\left(-\pi \right) R_Y\left(\frac{\pi}{2} \right)$.  Note that the implementation of the rotation gates constructing the $CNOT$ gate allows for some freedom in the direction of the rotation which can be used to reduce the number of primitive gates (an outline of the ion trap compilation techniques can be found in \cite{MaslovCircuitCompIT2017}).  The parameter $s \in \left\{+1,\, -1 \right\}$ in the circuit dictated by the sign of the interaction parameter $\chi$ between two ions which is determined by the experimental apparatus.  At our layer of abstraction, the value of $s$ is left as a free parameter.  Applying such a compilation method allowed for the reduction of the number of single-qubit gates from 48 in the naive implementation to 30 in the compiled circuit; the number of entangling gates cannot be reduced in the error correction routine leaving 24 M\o lmer-S\o rensen gates as well.  A representation of the compiled syndrome extraction circuit is shown in figure \ref{fig:MSsyndromecirc} where the ancillary wires have been suppressed.  This circuit was used for all further results.

\subsection{Single Error Source Dominant Effects} \label{sec:SingleErrorThresh}
\begin{figure}[t!]
\begin{subfigure}[t]{0.32\textwidth}
\includegraphics[width=\textwidth]{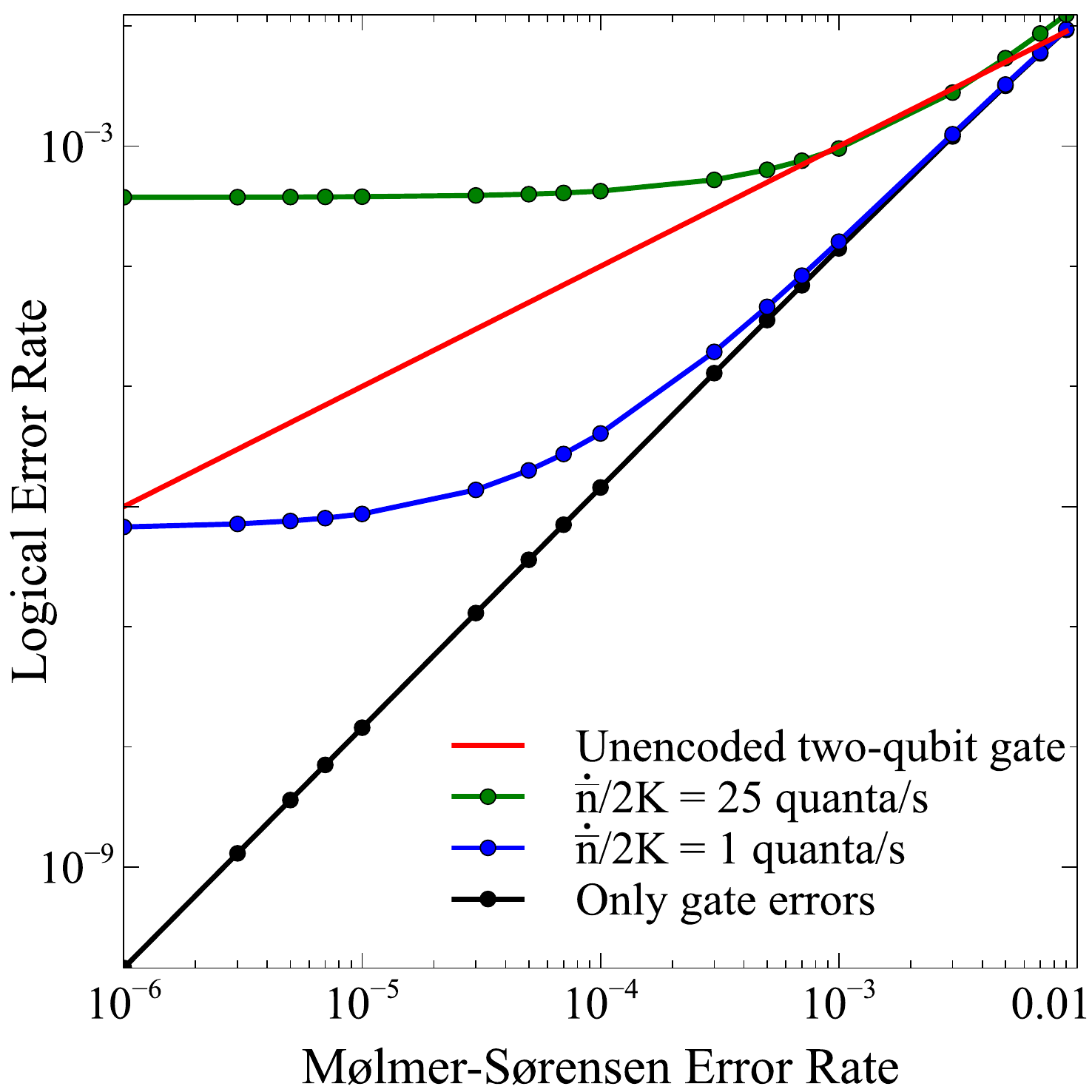}
\caption{Ion heating}
\label{fig:singlesourceheating}
\end{subfigure}
\begin{subfigure}[t]{0.32\textwidth}
\includegraphics[width=\textwidth]{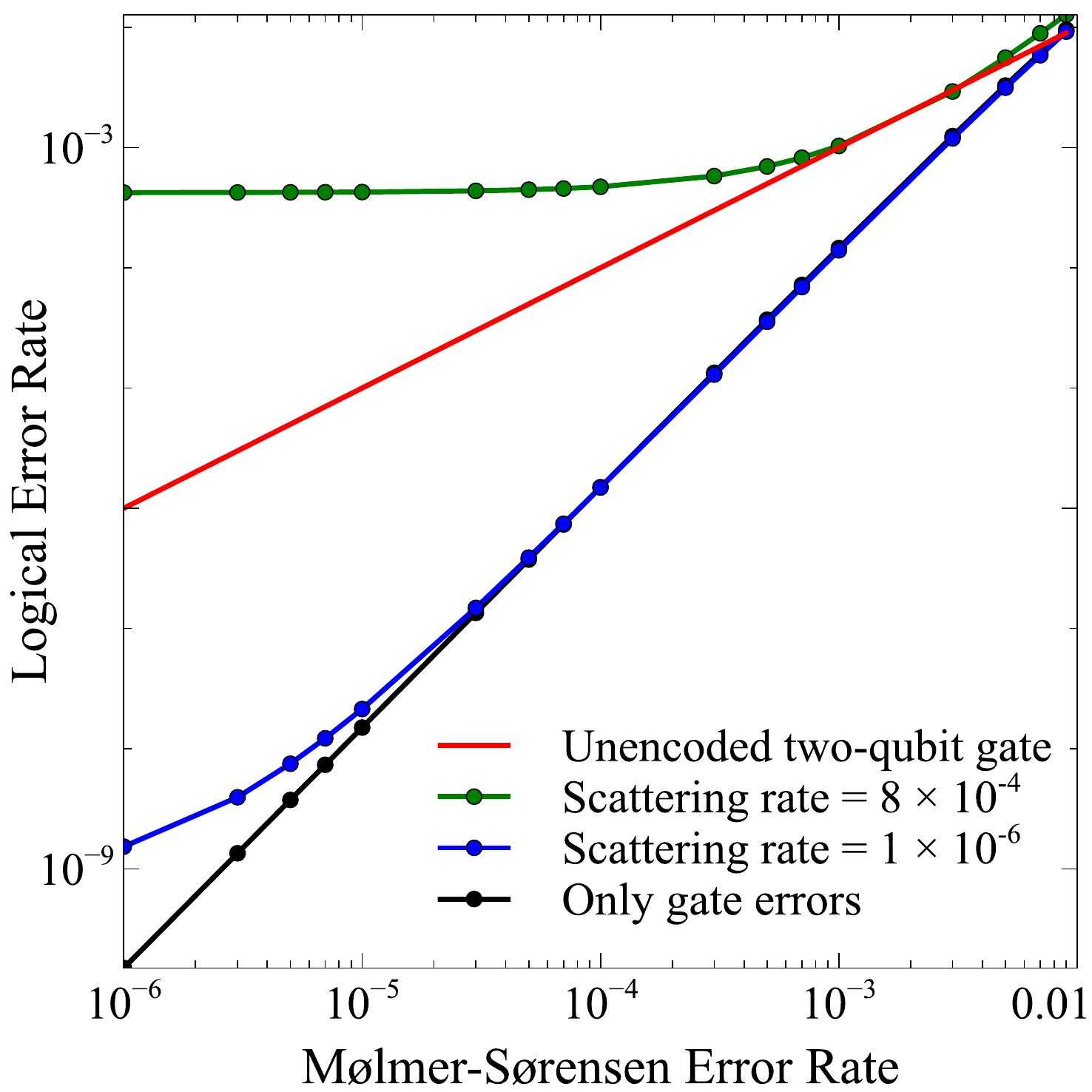}
\caption{Background Depolarizing Noise}
\label{fig:singlesourcedep}
\end{subfigure}
\begin{subfigure}[t]{0.32\textwidth}
\includegraphics[width=\textwidth]{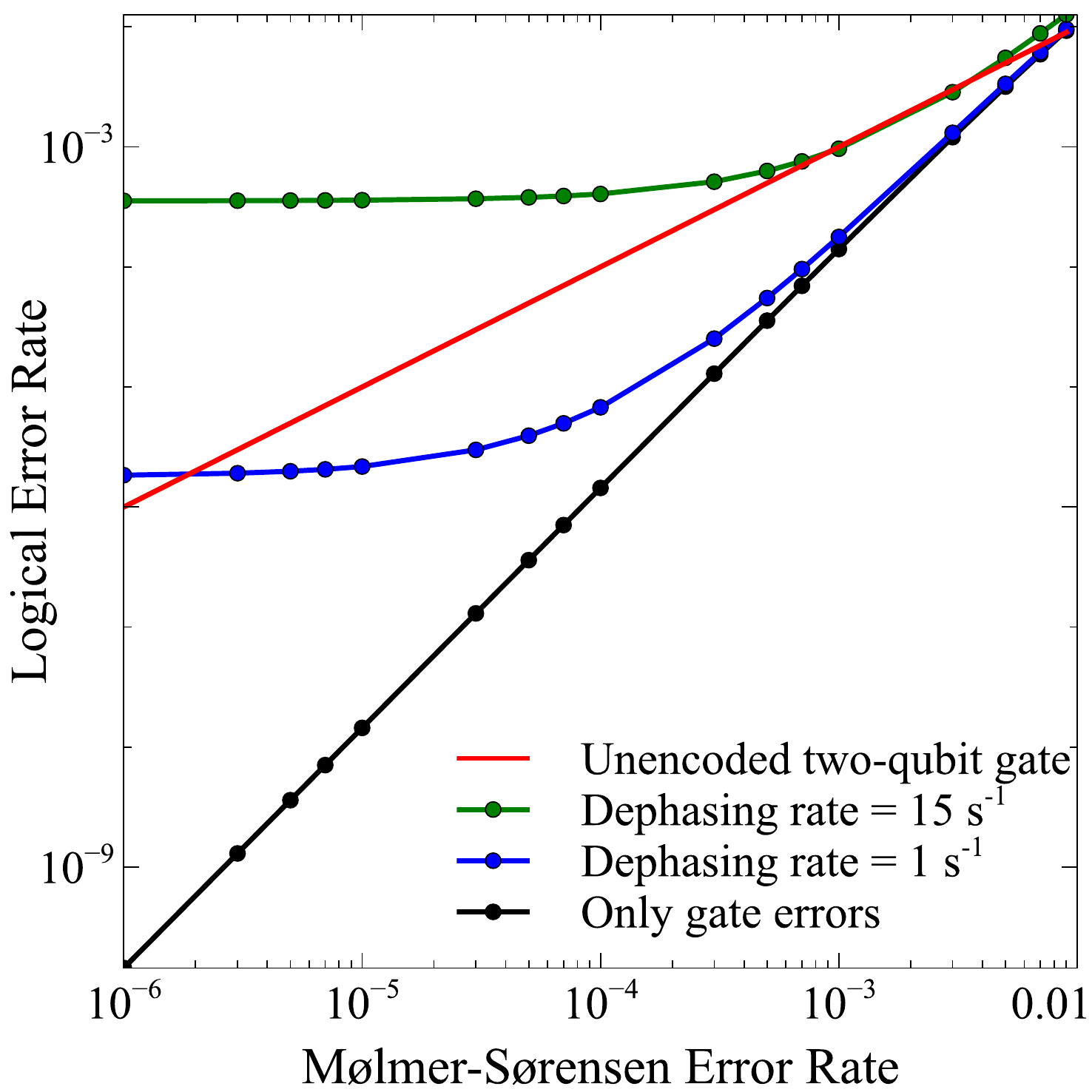}
\caption{Spin dephasing}
\label{fig:singlesourcedephase}
\end{subfigure}
\caption{The influence of ion trap error sources on the fault-tolerance of the 17-qubit surface code.  For each plot, only the labeled error source was introduced in the simulation in addition to gate errors.  To achieve fault-tolerance, a two-qubit gate fidelity of $\ge 99.9\%$ and an error in the gate from the specific error source below a critical value (green curves) is required.}
\label{fig:singlesourcethresh}
\end{figure}

In this section, we characterize the influence of the error sources in the limit where each error type is the dominant source of the error.  Therefore the simulations that generate the following pseudothresholds will have varying single- and two-qubit error rates (remember that $p_x = p_y = p_z = p_{xx}/10 $) and constant heating, depolarizing, or spin dephasing error rates during simulations.  Our goal is to find a parameter range under which, again in this limit of a dominant error source, fault-tolerant retention of the encoded information would be possible.  In all instances, a two-qubit gate fidelity of $\ge 99.9\%$ and an error source error rate below a critical rate is necessary to allow for fault-tolerance (see figure \ref{fig:singlesourcethresh}).  We discuss those critical rates for each error source below.

Ion heating was characterized by a parameterized representation of the heating rate $\dot{\bar{n}}/2K$ where $\dot{\bar{n}}$ is the heating (in quanta$/s$) of the gate motional mode and $K$ is the number of loops in phase space traversed by the M\o lmer-S\o rensen gate. As shown in figure \ref{fig:singlesourceheating}, fault-tolerance is not achieved at a heating rate above $25$ quanta$/s$ which corresponds to a motional mode heating rate during the gate of $100$ and $200$ quanta$/s$ for $K=2$ and $K=4$, respectively.  A heating rate ($\dot{\bar{n}}$) of about $58$ quanta$/s$ has been observed for a single $^{9}$Be$^+$ ion on a room temperature surface trap \cite{HiteArSurfaceTrap2012} and a silicon based trap in a cryogenic environment used to trap individual $^{40}$Ca$^{+}$ ions exhibited heating rates as low as $0.33$ quanta$/s$ ($0.6(2)$ quanta$/s$ on average) \cite{NiedermayrCryoTrap2014}.  Note that macroscopic traps exhibit significantly lower heating rates relative to surface traps; for instance a single trapped $^{111}$Cd$^{+}$ ion exhibited a heating rate of $2.48$ quant$/s$ for a room temperature macroscopic trap \cite{DeslauriersMacroCdHeating2004}.  However, additional difficulties arise for macroscopic traps in engineering a system that allows for ion separation, addressing, and detection required for an error correction protocol.  Also, the use of sympathetic cooling ions has been shown to reduce motional mode heating effects on $T_2^*$ \cite{Wang10minT22017}; a method which could reduce the heating rates of the idle computational qubits during the error correction routine.

The depolarizing error channel was applied to simulate stochastic error processes.  One such process of interest is spontaneous Raman and Rayleigh scattering which results in single- and two-qubit gate errors.  Figure \ref{fig:singlesourcedep} displays an upper limit on the scattering rate (per-gate) of $8 \times 10^{-4}$ to allow for fault-tolerance when scattering errors dominate.  Ozeri et al. have shown that gate errors due to Raman scattering to occur at a rate less than $10^{-4}$ for single-qubit gates but two-qubit gates have scattering rates on the order of $10^{-2}$ for their experimental setup for various species of trapped ions \cite{OzeriScattErrYb2007}.  These achieved scattering rates are still above the theoretical lower bound on the scattering rates for single- and two-qubit gates for $^{171}$Yb$^{+}$ by 3 and 7 orders of magnitude, respectively \cite{OzeriScattErrYb2007}, showing potential for improvement especially in the two-qubit scattering case.  Rayleigh scattering errors are less substantial, resulting in error rates per gate orders of magnitude below the Raman scattering error rates for heavy ions such as $^{171}$Yb$^{+}$ \cite{OzeriScattErrYb2007}.
  
Spin dephasing was modeled using a model that assumed a constant dephasing rate that scaled linearly with the time of the applied gate.  The upper bound on the error rate (figure \ref{fig:singlesourcedephase}) corresponds to a dephasing rate of $15 \, s^{-1}$. These values are related to $T_2^*$ \cite{Wang10minT22017}.  Note that the use of magnetic clock transitions \cite{LangerMagClock2005,HartyMagClock2014}, decoherence free subspaces \cite{Schmidt-KalerDFSCa2003}, or sympathetic cooling ions \cite{Wang10minT22017} during computation has been observed to increase the $T_2$ coherence times of the qubits to the order of seconds.  A 5-qubit system that has implemented small quantum algorithms \cite{DebnathSmallQCIons2016} and the $\llbracket4,2,2\rrbracket$ error detection code \cite{Linke422Ions2016} with hyperfine qubits exhibits a $T_2^*$ of $\approx0.5\, s$ \cite{Linke5qubitComp2017}, but further magnetic field stabilization could improve this as shown in \cite{Wang10minT22017} which exhibits over a 10 minute coherence time for trapped $^{171}$Yb$^{+}$ ions.

\subsection{Competing Error Sources: Dominant Errors}
To characterize the dominant error sources contributing to the logical error rate in the 17-qubit surface code in the case where multiple error sources are competing, we take advantage of the importance sampling technique.  We will briefly outline the importance sampling method; highlighting the use of error subsets that will be independently analyzed to gain insight into the effect of the error source on the logical error rate of the encoded state.  This will then be followed by an analysis of the statistically significant error subsets, which will be used to characterize the most malignant errors contributing to the failure rate of the error correcting circuit.

\subsubsection{Importance Sampling} \label{sec:importsample}

This method is an adaptation of the method from \cite{LiBareAnc2017} but extended to the case where multiple error sources are available during the simulation.  The method relies on approximating the logical error rate as a sum of statistically weighted logical error rates of error subsets.  For low enough physical error rates, few subsets need to be sampled in order to obtain an accurate approximation, which makes the approach considerably more efficient than the standard direct Monte Carlo sampling.  The subsets are indexed according to the number of errors present in the circuit.  For instance for the standard depolarizing error model, the subsets would be indexed according to the number of single- and two-qubit errors present in the circuit.  Sampling error configurations corresponding to the number of errors for this subset and calculating the fraction of configurations resulting in a logical error gives an effective subset error rate $A_{s,t}$.  Multiplying this subset logical error rate by the total statistical weight of the error subset will provide the subset's contribution to the total logical error rate.  Computing the statistical weight is done as so:
\begin{equation}
W_{s,t} = \left(
\begin{array}{c}
n_s \\
s
\end{array}
\right) p_s^{|e|}(1-p_s)^{n_s - |e|}
\left(
\begin{array}{c}
n_t \\
t
\end{array}
\right) p_t^{|e|}(1-p_t)^{n_t - |e|}
\end{equation}
where $s$ and $t$ are the number of single- and two-qubit errors in the circuit being considered, respectively.  These are also the indices of the subset.  The values $n_s$ and $n_t$ are the number of single- and two-qubit fault-points in the circuit, respectively.  The values $p_s$ and $p_t$ are the single- and two-qubit error channel probabilities and $|e|$ denotes the weight of the error.  Estimating the logical error rate then constitutes calculating the following sum:
\begin{equation}
\displaystyle p_L \approx \sum_{(s,t)}^{W_{s,t}<\lfloor W \rfloor} W_{s,t} \, A_{s,t}
\label{eqn:pLapprox}
\end{equation}
where subsets with statistical weights below a chosen cutoff value, $\lfloor W \rfloor$, are omitted from the sum.  Note that, with this method, the sampling of each error subset only needs to be performed once to generate a logical error curve.

\begin{figure*}[t!]
\centering
\mbox{
\Qcircuit @C=1.em @R=0.8em {
\lstick{} & \qw & \multigate{1}{a} & \multigate{1}{b} & \multigate{1}{c} & \qw & \qw  \\
\lstick{} & \qw & \ghost{a} & \ghost{b} & \ghost{c} & \qw & \qw \\
}}
\caption{Circuit containing three two-qubit gates, labeled $a$, $b$, and $c$, with error rates $p_a$, $p_b$, and $p_c$, respectively.}
\label{fig:samplecircexample}
\end{figure*}
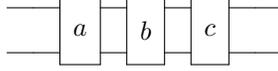

We altered the method above to handle situations where errors of equivalent types have different error rates; such is the situation for our ion heating and dephasing error models with ion dependent gate times, which influence the error rate per qubit.  To motivate this point, consider the quantum circuit in figure \ref{fig:samplecircexample}.
This circuit contains three two-qubit gates, $a$, $b$, and $c$, with different error rates $p_a$, $p_b$, and $p_c$, respectively.  The weight of the $(0,2)$ subset would then be:
\begin{equation}
p_a p_b (1-p_c) + p_a p_c (1-p_b) + p_b p_c (1-p_a)
\end{equation}
so the two-qubit subset calculation requires one more ingredient: we need to sum over all $n$-tuple error configurations ($f_n$) during the subset weight calculation:
\begin{equation}
\displaystyle W_n = \sum_{n = |e|}^{e\in f_n}
\prod_{k\in e} p_k \prod_{j \not\in e} (1-p_j)
\end{equation}
When we adapt this approach to heating errors in an ion trap circuit, we get the following calculation of the subset weight:
\begin{equation}
\left(
\begin{array}{c}
n_s \\
s
\end{array}
\right) p_s^{|e|}(1-p_s)^{n_s - |e|}
\left(
\begin{array}{c}
n_t \\
t
\end{array}
\right) p_t^{|e|}(1-p_t)^{n_t - |e|} \;
\sum_{n = |e|}^{e\in f_n} 
\prod_{h\in e} p_h \prod_{\not h\not\in e}  (1-p_{\not h})
\label{eqn:subetweightexample}
\end{equation}
where $p_h$ and $p_{\not h}$ are the individual error rates of the heating error channels for each two-qubit configuration on which an entangling gate is and is not applied in the simulation, respectively.  We have taken into account the influence of the different rates for the calculation of the subset weights, but this also has an influence on the sampled subset logical error rates as certain error configurations will be more probable than others.  Because the heating error rates are linearly proportional to the gate times in our error model, we have chosen to sample heating error configurations from a gate time weighted distribution of error configurations giving a corresponding logical error rate of $A_{s,t,h}$.  With the new subset weights and subset logical error rates, the estimation of the total logical error rate naturally extends from equation \ref{eqn:pLapprox}.  Note that heating adds an extra subset label: ($s$,$t$,$h$).  The indices $s$, $t$, and $h$ represent the number of single-qubit gate, two-qubit gate, and heating errors sampled, respectively.
 
Recall that the ion trap error model from section \ref{sec:ErrorModels} contains 5 distinct error sources.  Therefore, we extended the concepts from equations \ref{eqn:pLapprox} and \ref{eqn:subetweightexample} to calculate the logical error rate of the 17-qubit surface code under the influence of  single-qubit gate, two-qubit gate, ion heating, background depolarization, and dephasing errors.  The analysis below will include 5 index subsets ordered with the indices listing the number of single-qubit gate, two-qubit gate, heating, background depolarization, and dephasing errors sampled in the circuit, in that order.

\begin{figure}[t!]

\begin{subfigure}[t!]{0.5\textwidth}
\includegraphics[width=\textwidth]{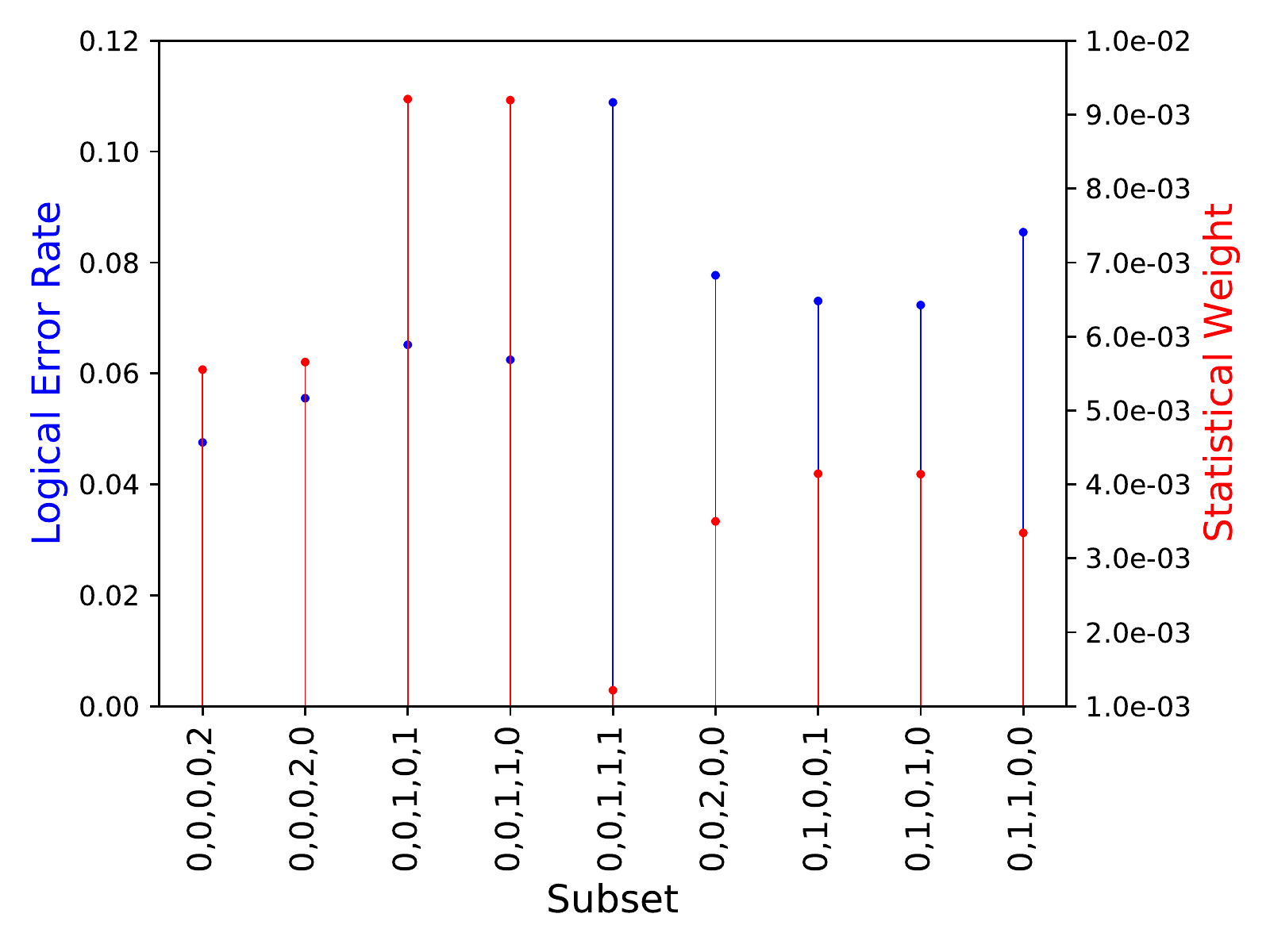}
\caption{$10^{-3} \leq W < 10^{-2}$}
\label{fig:subsetweightsa}
\end{subfigure}
\begin{subfigure}{0.5\textwidth}
\includegraphics[width=\textwidth]{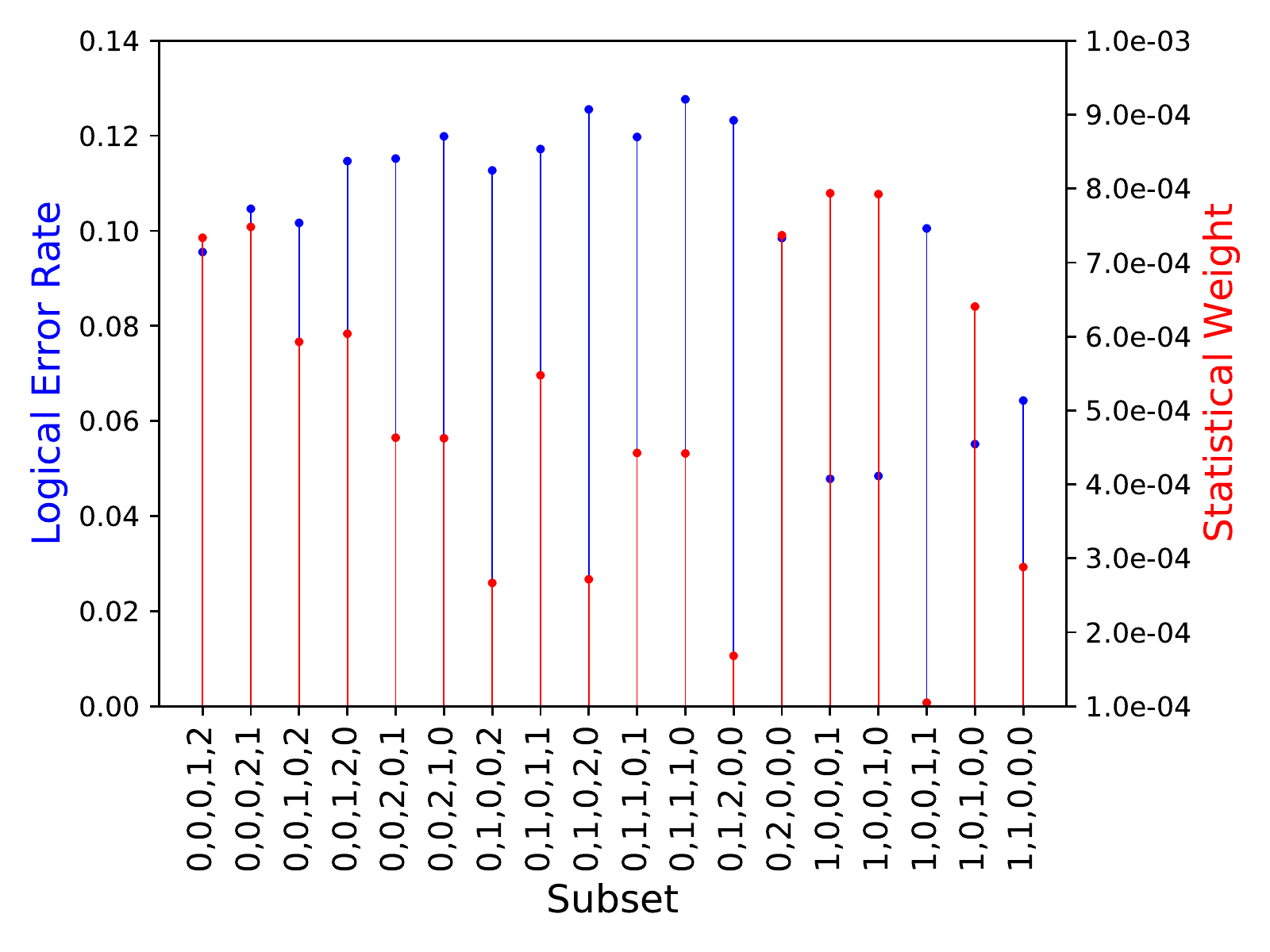}
\caption{$10^{-4} \leq W < 10^{-3}$}
\label{fig:subsetweightsb}
\end{subfigure}

\begin{subfigure}[t!]{0.5\textwidth}
\includegraphics[width=\textwidth]{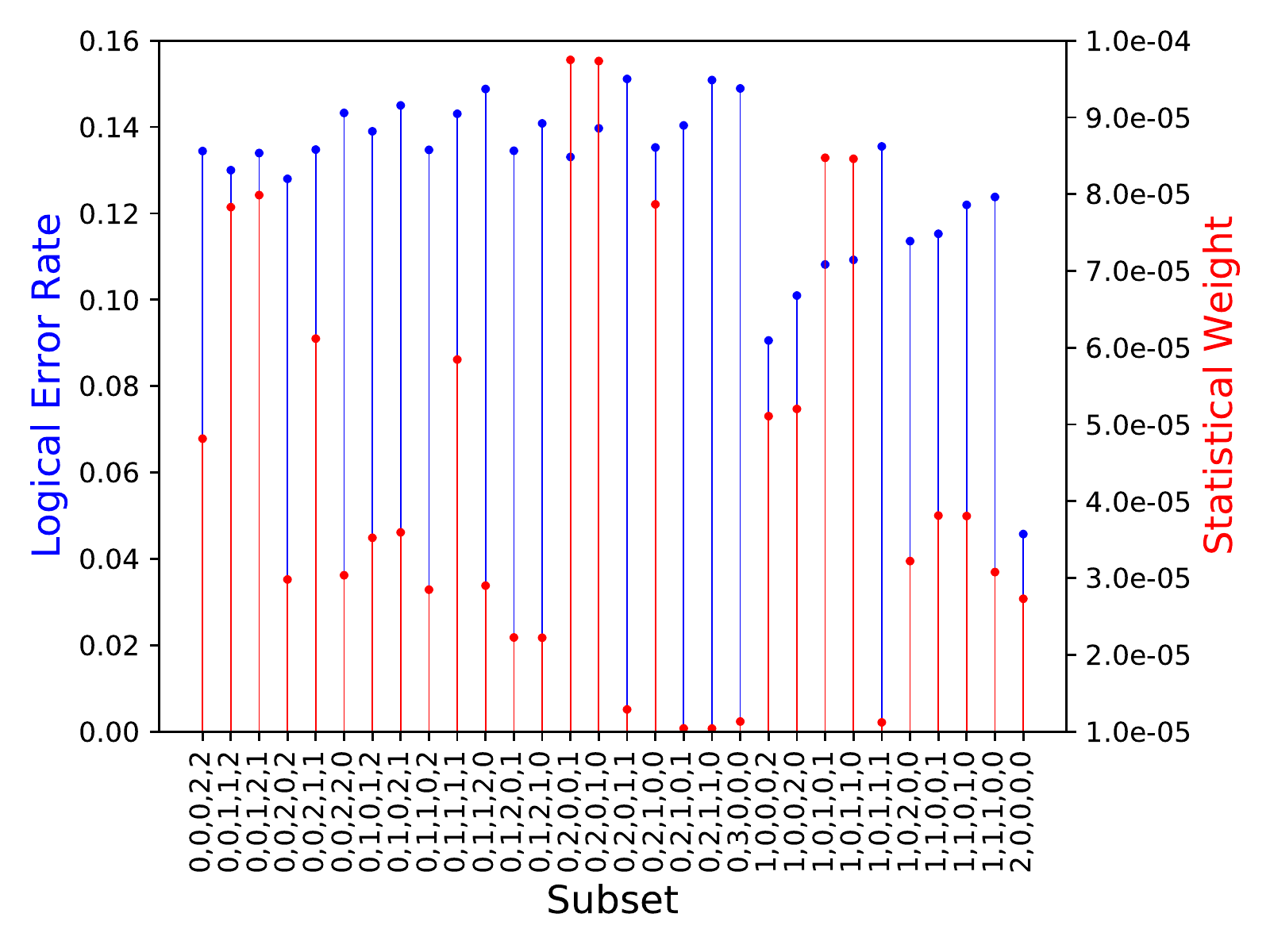}
\caption{$10^{-5} \leq W < 10^{-4}$}
\label{fig:subsetweightsc}
\end{subfigure}
\begin{subfigure}{0.5\textwidth}
\includegraphics[width=\textwidth]{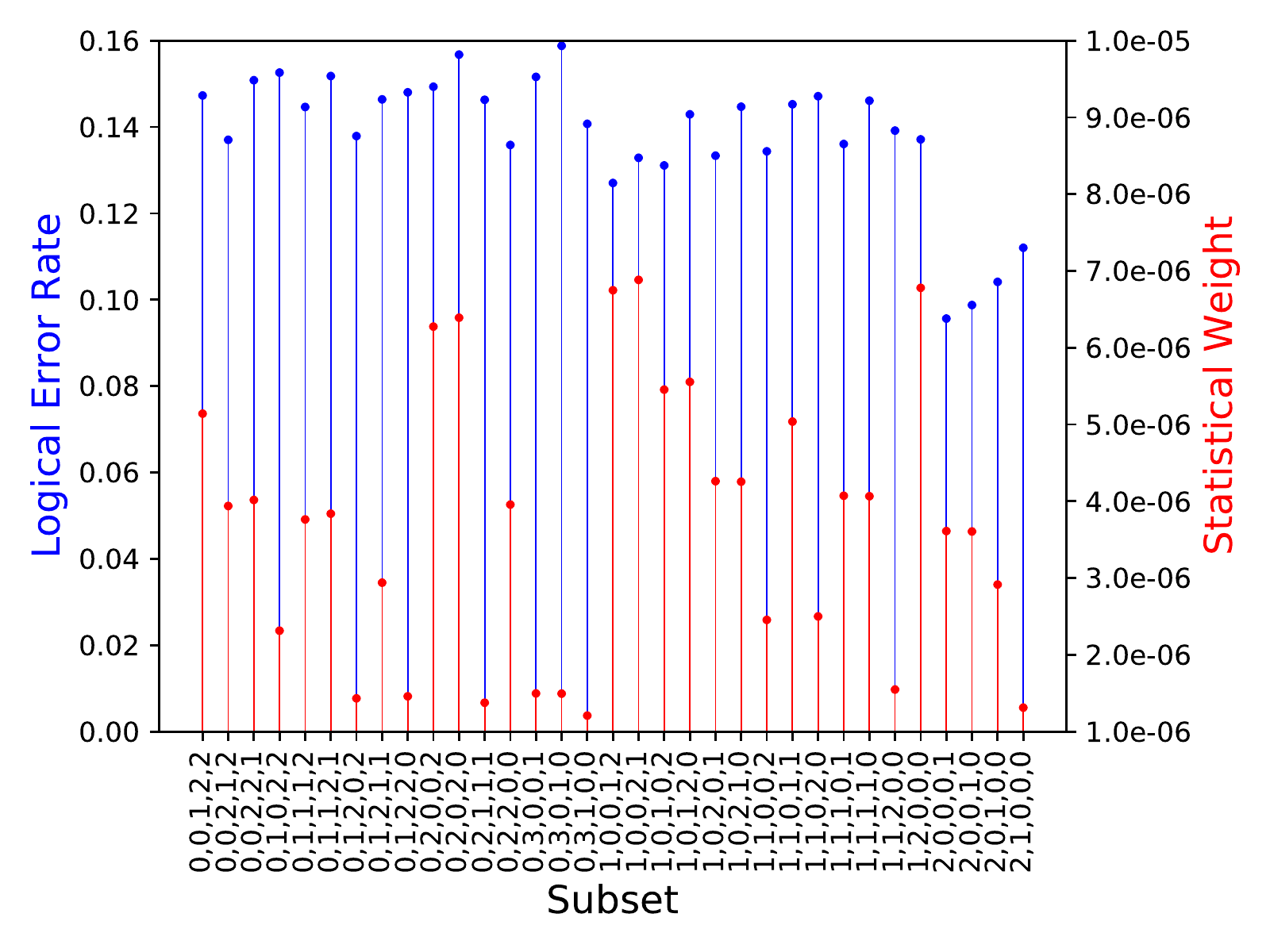}
\caption{$10^{-6} \leq W < 10^{-5}$}
\label{fig:subsetweightsd}
\end{subfigure}

\caption{The subset logical error rates and subset statistical weights above the cutoff of $10^{-6}$ corresponding to events expected to be sampled from a random distribution at least once out of a million samples.  The data is separated into four plots according to the order of magnitude of the subset statistical weights, which are plotted in red.  The logical error rates for the subsets are plotted in blue.  Note that the product of the subset weight and its corresponding logical error rate dictates the subset's contribution to the total logical error rate of the code.  For calculation of the statistical weights of the subsets, the single-qubit gate error rate $\left(p_y = p_x = p_z \right)$, two-qubit gate error rate $\left(p_{xx}\right)$, rate of heating $\left(r_{heat}\right)$, background depolarizing noise error rate $\left(p_{dep}\right)$, and rate of dephasing $\left( r_d \right)$ were $10^{-4}$, $10^{-3}$, $25$ quanta$/s$, $8 \times 10^{-4}$, and $15$ $s^{-1}$, respectively, which correspond to the parameters allowing for the logical error rate equivalent to the unencoded two-qubit gate error rate in section \ref{sec:SingleErrorThresh}.  A plot containing similar subset information for data beyond the subset cutoff is shown in figure \ref{fig:allsubsets}.}

\label{fig:SubsetTab}
\end{figure}

\begin{figure}[b!]
\begin{subfigure}{\textwidth}
\centering
\mbox{
\Qcircuit @C=0.8em @R=.94em {
\lstick{\ket{\psi}} & \multigate{1}{MS} &  \qw \\
\lstick{\ket{0}} & \ghost{MS} &  \qw
}}
\parbox{0.05\textwidth}{
\centering
$\Rightarrow$\\ \vspace{-8mm}
}
\mbox{
\Qcircuit @C=0.8em @R=.94em {
\lstick{} & \multigate{1}{\textcolor{red}{E_2}} &  \qw \\
\lstick{} & \ghost{\textcolor{red}{E_2}} &  \qw
}}
\parbox{0.05\textwidth}{
\centering
$\Rightarrow$\\ \vspace{-8mm}
}
\mbox{
\Qcircuit @C=0.8em @R=.94em {
\lstick{} & \gate{R_X(\pm\frac{\pi}{2})} & \gate{R_Y(\pm \frac{\pi}{2})} & \qw \\
\lstick{} & \qw &  \qw & \qw
}}
\parbox{0.05\textwidth}{
\centering
$\Rightarrow$\\ \vspace{-8mm}
}
\mbox{
\Qcircuit @C=0.8em @R=.94em {
\lstick{} & \multigate{1}{\textcolor{red}{E_2'}} &  \qw \\
\lstick{} & \ghost{\textcolor{red}{E_2'}} &  \qw
}}
\caption{Faults existing after a M\o lmer-S\o rensen gate are trans-\\formed by single-qubit rotations gates and their errors.}
\end{subfigure}
\vspace{3mm}

\begin{subfigure}{0.5\textwidth}
\centering
$\begin{array}{c c c}
\underline{E_2} &  & \underline{E_2'} \\
II & \rightarrow & \textcolor{newgreen}{ZI} \\
XI & \rightarrow & \textcolor{blue}{II} \\
YI & \rightarrow & \textcolor{newgreen}{YI} \\
ZI & \rightarrow & \textcolor{newgreen}{XI} \\
XX & \rightarrow & \textcolor{red}{IX} \\
XY & \rightarrow & \textcolor{red}{IY} \\
XZ & \rightarrow & \textcolor{blue}{IZ} \\
YX & \rightarrow & \textcolor{orange}{YX} \\
\end{array}$
~
$\begin{array}{c c c}
\underline{E_2} &  & \underline{E_2'} \\
YY & \rightarrow & \textcolor{orange}{YY} \\
YZ & \rightarrow & \textcolor{newgreen}{YZ} \\
ZX & \rightarrow & \textcolor{orange}{XX} \\
ZY & \rightarrow & \textcolor{orange}{XY} \\
ZZ & \rightarrow & \textcolor{newgreen}{XZ} \\
IX & \rightarrow & \textcolor{orange}{ZX} \\
IY & \rightarrow & \textcolor{orange}{ZY} \\
IZ & \rightarrow & \textcolor{newgreen}{ZZ} \\
\end{array}$
\caption{$R_X \left(\pm \frac{\pi}{2} \right)$ Gate Error: $X$}
\label{fig:RXerrors}
\end{subfigure}
\begin{subfigure}{0.5\textwidth}
\centering
$\begin{array}{c c c}
\underline{E_2} &  & \underline{E_2'} \\
II & \rightarrow & \textcolor{newgreen}{YI} \\
XI & \rightarrow & \textcolor{newgreen}{XI} \\
YI & \rightarrow & \textcolor{newgreen}{ZI} \\
ZI & \rightarrow & \textcolor{blue}{II} \\
XX & \rightarrow & \textcolor{orange}{XX} \\
XY & \rightarrow & \textcolor{orange}{XY} \\
XZ & \rightarrow & \textcolor{newgreen}{XZ} \\
YX & \rightarrow & \textcolor{orange}{ZX} \\
\end{array}$
~
$\begin{array}{c c c}
\underline{E_2} &  & \underline{E_2'} \\
YY & \rightarrow & \textcolor{orange}{ZY} \\
YZ & \rightarrow & \textcolor{newgreen}{ZZ} \\
ZX & \rightarrow & \textcolor{red}{IX} \\
ZY & \rightarrow & \textcolor{red}{IY} \\
ZZ & \rightarrow & \textcolor{blue}{IZ} \\
IX & \rightarrow & \textcolor{orange}{YX} \\
IY & \rightarrow & \textcolor{orange}{YY} \\
IZ & \rightarrow & \textcolor{newgreen}{YZ} \\
\end{array}$
\caption{$R_Y \left(\pm \frac{\pi}{2} \right)$ Gate Error: $Y$}
\label{fig:RYerrors}
\end{subfigure}

\vspace{2mm}

\begin{subfigure}{\textwidth}
\centering
\textcolor{blue}{$\bullet$ Corrected} \hspace{2mm} \textcolor{newgreen}{$\bullet$ Single Data Error} \hspace{2mm} $\bullet$ \textcolor{orange}{Flagged Error} \hspace{2mm} \textcolor{red}{$\bullet$ Meaurement Error}
\end{subfigure}

\caption{The errors in rotation gates will transform the faults that exist after a two-qubit gate.  The transformation of an existing two-qubit Pauli error (ignoring the phase) after a single-qubit gate error on wires that contain an $R_X \left(\pm \frac{\pi}{2} \right)$ and an $R_Y \left(\pm \frac{\pi}{2} \right)$ gate is shown in b.) and c.).  The first and second element of the Pauli error correspond to the error on the data and ancilla qubit, respectively.  There are two types of single-qubit over-rotation errors, $X$ and $Y$, which transform Pauli errors according to b.) and c.), respectively.  Measurement errors can introduce of errors into the code through faulty correction when errors occur in the following round of stabilizer measurement.  Flagged errors are detectable errors that are not always corrected properly but the error stays local to the qubit given that the following syndrome measurement is accurate.  Single data errors are favorable because the error stays local to the data qubit and can be more easily corrected in the next round of measurement because no faulty information was sent to the decoder.  Self correction may occur in as well for specific errors.  Applying the transformation $II \leftrightarrow ZI$ on the $E_2'$ values in c.) give the resulting error on wires containing only $R_Y \left(\pm \frac{\pi}{2} \right)$ gate errors.
}
\label{fig:sqgateerrors}
\end{figure}

\subsubsection{Competing Error Sources: Sampling Subset Analysis} \label{sec:subsetanal}

For the importance sampling simulations of the 17-qubit surface code, a subset weight cutoff of $\lfloor W \rfloor$  $>$ $10^{-6}$ was used and $30,000$ samples were collected for the calculation of each subset's logical error rate $A_{s,t,h,dep,z}$.  This weight cutoff corresponds to events expected to be sampled at least once out of a million randomly sampled events, which is sufficient for near-term error correction experiments.  To calculate the statistical weights of the subsets, a single-qubit gate error rate $\left(p_y = p_x = p_z \right)$, two-qubit gate error rate $\left(p_{xx}\right)$, rate of heating $\left(r_{heat}\right)$, background depolarizing noise error rate $\left(p_{dep}\right)$, and rate of dephasing $\left( r_d \right)$ of $10^{-4}$, $10^{-3}$, $25$ quanta$/s$, $8 \times 10^{-4}$, and $15$ $s^{-1}$ was chosen, respectively, which corresponds to the error rates that exhibit a logical error rate equal to the two-qubit gate error rate (see the green curves in figure \ref{fig:singlesourcethresh}).  The logical error rates and statistical error weights calculated for each subset are presented in figure \ref{fig:SubsetTab}.  An expanded number of subsets beyond this cutoff were run and are shown in figure \ref{fig:allsubsets} (See Supplementary Material).  The goal of this analysis is to parse out situations where certain error sources show dominant contribution to the failure rate of the quantum error correcting circuit.

The logical error rates for each of the subsets sampled are shown in blue in figure \ref{fig:SubsetTab}.  The error subsets containing two-qubit gate or heating errors tend to have higher logical error rates than other subsets containing comparable number of errors.  This occurs due to the ability of errors of this type to generate measurement faults in the circuit.  The M\o lmer-S\o rensen gate transforms single-qubit data errors in the following manner: $ZI \leftrightarrow YX$ and $YI \leftrightarrow -ZX$ where the data and ancilla qubit errors are the first and second elements, respectively.  A two-qubit gate or heating error makes preexisting errors undetectable which is a particularly malignant case.  The tendancy towards measurement errors in the ion trap error model indicates that implementing a decoder that makes a correction based on more syndrome measurement rounds may show an above average performance boost in error correction.  Error subsets containing single-qubit gate errors tend to have lower logical error rates that other subsets with comparable number of errors.  To understand why this is the case, we explore the effect the errors have on error correction.  Figure \ref{fig:sqgateerrors} shows how single-qubit gate errors transform preexisting errors in the circuit.  The particularly malignant case in when there is a measurement error which can introduce errors into the code.  For each single-qubit fault point, there are only two elements of the two-qubit Pauli group that are transformed in a manner that would result in a meaurement error.  Actually, half of the elements of this group result in single-qubit errors (or no errors) on data that can be readily decoded in the following step of stabilizer measurement (see figure \ref{fig:sqgateerrors}).  The remaining errors are detectable but not necessarily corrected properly (this depends on the location that the fault occurs).  However, these errors do alert the decoder to the location of an error on the code which is favorable and the faulty correction on these qubits will not propagate errors in a malignant manner given the next round of stabilizer measurement is correct.  Take note that one of the malignant errors transformed in figure \ref{fig:RXerrors} ($R_X(\pm \frac{\pi}{2})$ gate error) is $XX$ which is the form of the two-qubit gate and heating errors.  Therefore, compiling out the single-qubit gates $R_X (\pm \frac{\pi}{2})$ gates seems to have also boosted the efficiency of the decoder to decode two-qubit gate and heating errors in addition to the obvious performance boost from less general fault points in the compiled circuit.  Another alarming malignant configuration in figure \ref{fig:RYerrors} is the $ZX$ error which is the result of the M\o lmer-S\o rensen transformation of $YI$ (a single-qubit data error).  However, this fault requires two single-qubit $Y$ errors which have a low statistical weight of occurance (see figure \ref{fig:SubsetTab}).

How else can we use this information to improve error correction?  One obvious extension is to use such statistics to develop decoders targeted for such errors.  For instance, this information about the failure rates \emph{at the logical level} may be used to bias transition matrix elements of a maximum likelihood decoder to include information about the influence of error cosets on the code's performance instead of only considering the statistical weights of the error cosets \cite{TuckettSCBiasNoise2017}.  Code considerations when optimizing the ion chain layout could serve to bound the effects of the gate-time dependent error sources.  Specifically, optimizing the ion chain to assign the most distant qubits (with the longest gate times) to weight-2 stabilizers can reduce the influence of anisotropic error sources.  Consider a scenario where two-qubit gate and heating errors dominate, then assigning the faultiest gates to the weight-2 $X$-type stabilizers would bound the influence of the most probable heating errors to single-qubit $X$-errors on the data.  While this does not apply to our current error model where dephasing ($Z$) and heating ($XX$) errors both have the same dependency on gate time, this is probably not the case experimentally and, the greater the anisotropy of the errors, the greater one can reduce their effect.

The subset statistical weights (probabilites of occurence) are shown in red in figure \ref{fig:SubsetTab}.  These statistical weights give insight into the likelihood of sampling particular error events.  Recall that the subset's contribution to the total logical error rate of the code (used to generate the pseudothreshold plots in \ref{fig:singlesourcethresh}) is the product of the subset weight and subset logical error rate (see section \ref{sec:importsample}).  Only ten points above the subset weight of $10^{-3}$ (figure \ref{fig:subsetweightsa}) have significant contribution to the total logical error rate; that is, this small collection of subsets can be used to completely recreate the pseudothreshold plots in \ref{fig:singlesourcethresh}.  Actually, the two subsets, $(0,0,1,0,1)$ and $(0,0,1,1,0)$, have the largest contribution to the encoded logical error rate and bound the logical error rate to $p_L \approx \sum A_{s,t,h,dep,z} \times W \approx 1 \times 10^{-3}$, which corresponds to a two-qubit gate fidelity of $99.9 \%$ (recall that the gate error rate for calculation of the subset weights was $10^{-3}$).  This essentially recreates our calculation of a $99.9 \%$ two-qubit gate fidelity for fault-tolerance that used more subsets.  Therefore, changes in the statistics of the dominating subsets have significant influence on the observed pseudothreshold of the quantum error correcting code and can be considered when implementing a decoding algorithm.  This also illustrates the concern that a success metric such as the (pseudo)threshold only represents the mean statistics of an underlying error model \cite{BarnesFailureDistr2017}.

\section{Conclusions}

We studied the feasibility of implementing the quantum error correction with the 17-qubit surface code on a linear chain of atomic qubits.  Optimization of the ion chain showed a preference for mixed data/ancilla configurations to reduce the gate times for syndrome extraction.  We showed that the 17-qubit surface code contained enough structure to allow for the use two 16 key lookup tables for error correction with a respectably high pseudothreshold of $3 \times 10^{-3}$ for circuit-level depolarizing noise.  The lookup table decoder is easily integrated into the logic of the ion trap controls and decodes at a rate much faster ($\mathcal{O}\left(ns\right)$) than any physical operation on the qubits.  When modeling ion trap error sources, it was shown that a two-qubit gate fidelity of $\ge 99.9 \, \%$ is required in the cases where ion heating, scattering, or spin dephasing are the dominant error sources.  Furthermore, the parameter regimes that allow for fault-tolerant error correction are not outlandish for such error sources.  Finally, we took advantage of the error subset data required for our simulations to parse out trends that occur when multiple error sources occur during error correction.  We found that two-qubit gate and heating errors are the most malignant error sources and single-qubit gate errors are manageable in our ion trap error model.  We also speculate on how this subset information can be used to reduce the influence of malignant error sources on error correction.

Similar calculations have recently been done for the Steane [[7,1,3]]
code in a linear chain of ions that also allows for the rotation of ion
crystals.  The calculations presented in \cite{BermudezITQCwithSteane2017} use a different
Pauli error model for ion trap errors that emphasizes memory errors.
The key difference in approach is that our model includes enough
ancillae such that a full syndrome measurement takes place during a single
measurement time. The serialization of measurements in \cite{BermudezITQCwithSteane2017}, when
combined with intrinsic memory errors, requires a lower physical qubit
error rate in order to achieve a break even logical error rate.

To better assess the performance of quantum error correcting codes in
real systems, more detailed physical error models are warranted \cite{GutierrezIncohCohNoiseEC2016}.  A promising approach is to use realistic error channels
with quantum trajectories to avoid  simulating the entire density
matrix.  As recently shown for superconducting systems and the
surface-17 code, a smart choice in circuit representation allows the
entire 17 qubit system to be modeled with only 10 active qubits \cite{OBrienSurface17DensityMatrix2017}.
In the future, we plan to apply this technique to experimentally
derived error models for ion traps to help assess which coherent
errors have the most deleterious effect.

\section{Acknowledgements}
We thank Chris Monroe, Jungsang Kim, and Norbert Linke for useful discussions. This work was supported by the Office of the Director of National Intelligence - Intelligence Advanced Research Projects Activity through ARO contract W911NF-10-1-0231 and the National Science Foundation grant PHY-1415461.

\bibliographystyle{apsrev}
\bibliography{Surface17wIons}

\newpage
\section{Supplemental Information}

For the importance sampling method, error subsets were labeled according to the number of errors due to single-qubit gates, two-qubit gates, ion heating, depolarizing, and dephasing processes which were modeled with Pauli errors in a manner shown in figure \ref{fig:ITErrorModel}.  For each subset, no more than 6 errors of one type were introduced into the circuit due to the total subset weight cutoff introduced.  A natural method of partitioning such a data set is to represent each error subset as a base-7 integer in the following manner:
\begin{gather*}
n_{s},\, n_{t},\, n_{heat},\, n_{dep},\, n_{z}\\
7^0,\,7^1,\,7^2,\,7^3,\,7^4
\end{gather*}
where digits are read from left to right.  For instance, the subset $(1,\,5,\,2,\,1,\,2) = 5279$.  This representation clusters the subset data into a readable format shown in the plots below.

\begin{figure}[b!]
\caption{An expanded collection of error subsets ran for the 17-qubit surface code under the influence of the ion trap ion model described in section \ref{sec:ErrorModels}.  The data is grouped according to the base-7 representation explained above and two plots are shown per group of data: a plot for the logical error rate and a plot for the subset weight.  The subsets with weights above the cutoff of $10^{-6}$ (horizontal dashed line) are shown in figure \ref{fig:SubsetTab}.}
\label{fig:allsubsets}
\begin{subfigure}[t]{0.5\textwidth}
\includegraphics[width=\textwidth]{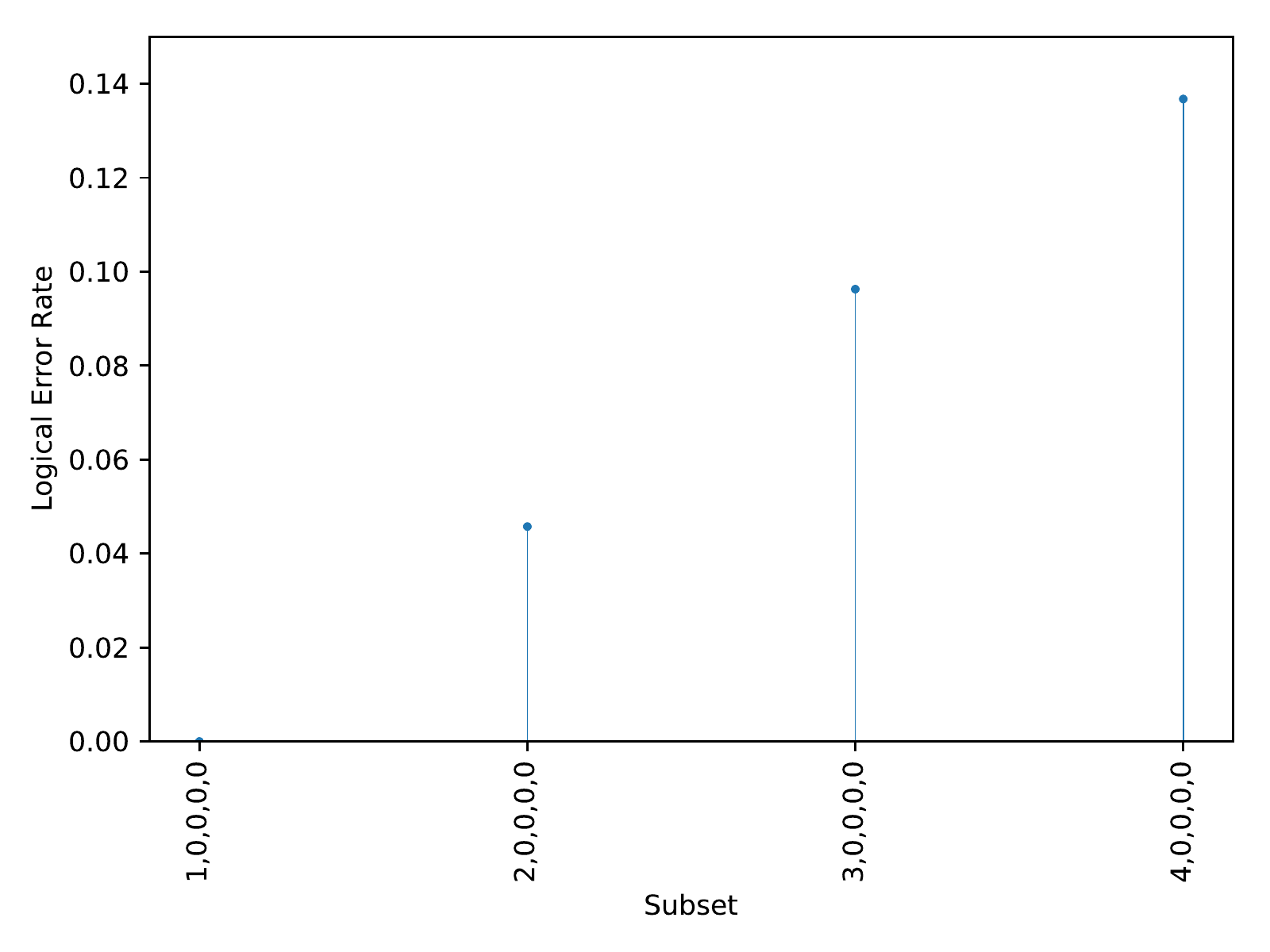}
\end{subfigure}
\begin{subfigure}[t]{0.5\textwidth}
\includegraphics[width=\textwidth]{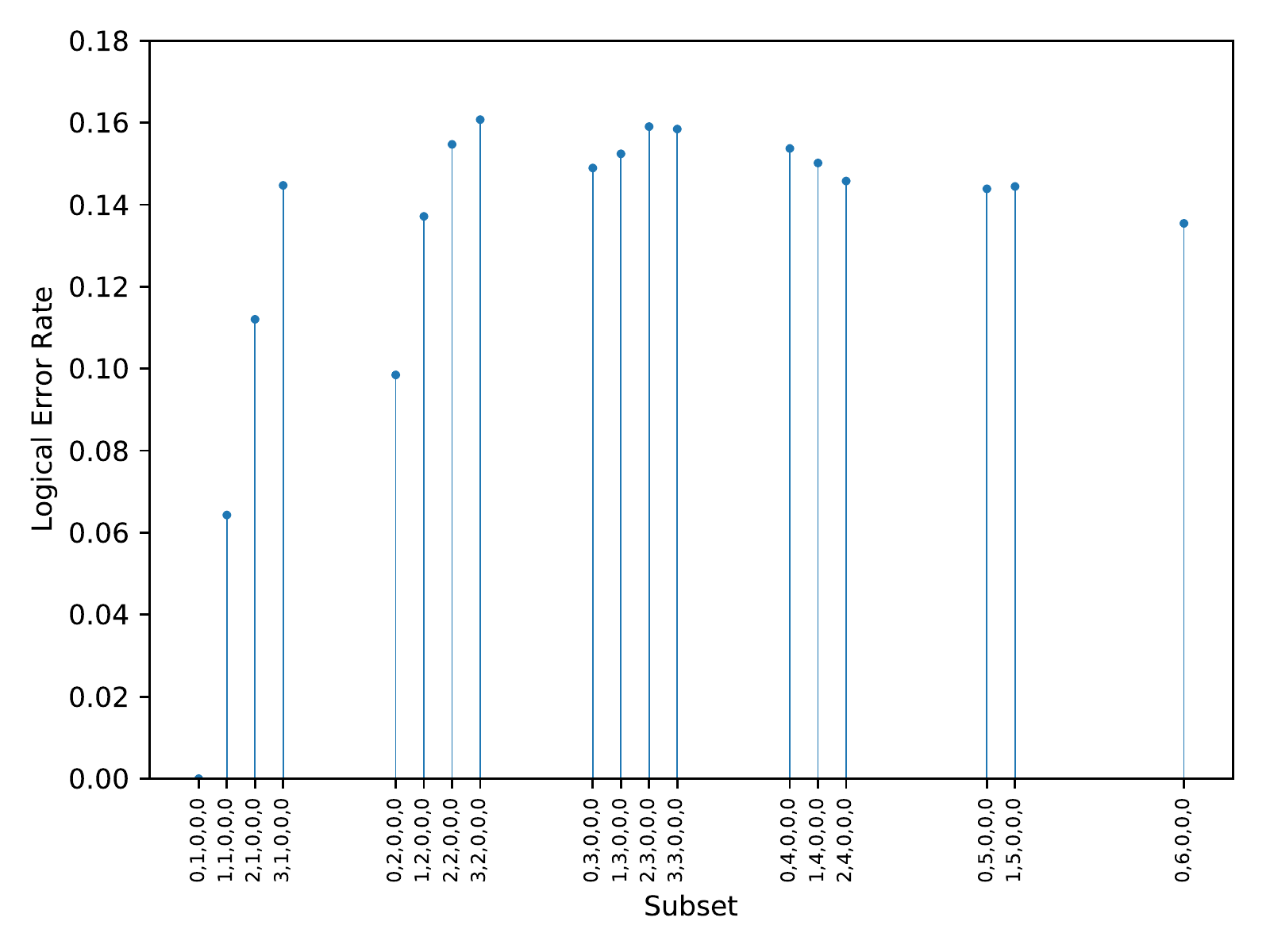}
\end{subfigure}

\begin{subfigure}[t]{0.5\textwidth}
\includegraphics[width=\textwidth]{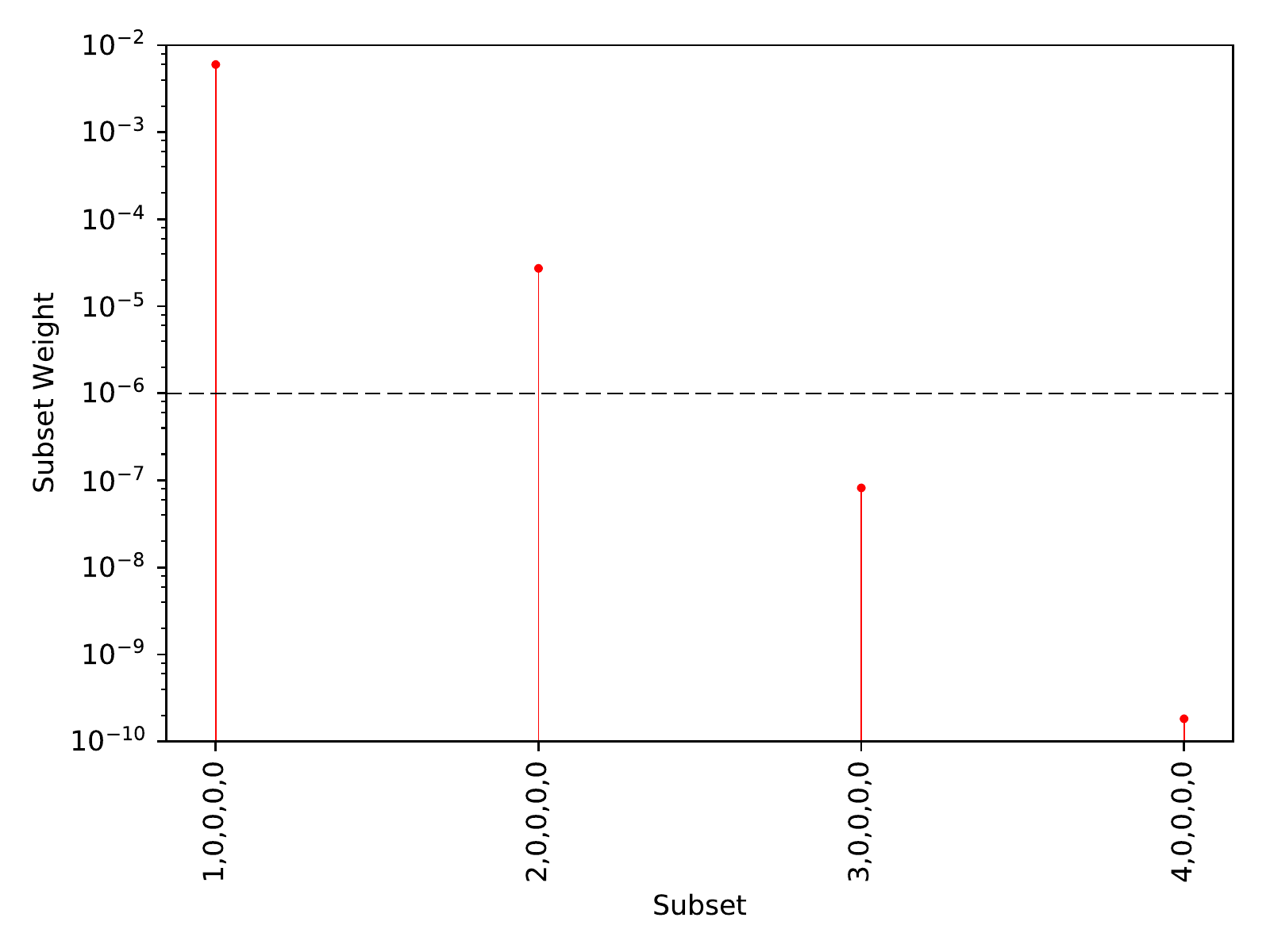}
\end{subfigure}
\begin{subfigure}[t]{0.5\textwidth}
\includegraphics[width=\textwidth]{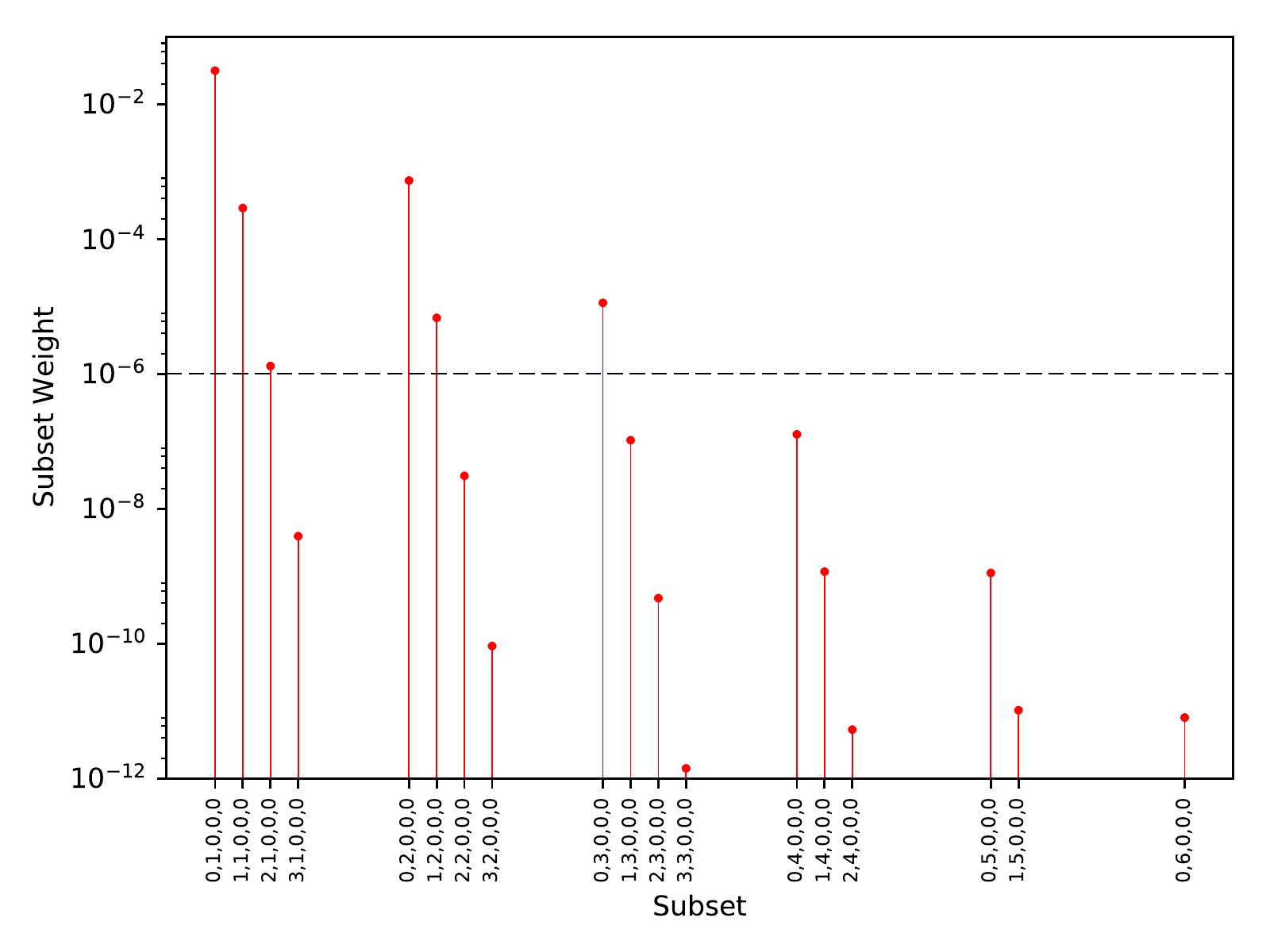}
\end{subfigure}
\end{figure}

\begin{figure}[b!]\ContinuedFloat
\begin{subfigure}[t]{\textwidth}
\includegraphics[width=\textwidth,trim={0 0.5cm 0 0.5cm}]{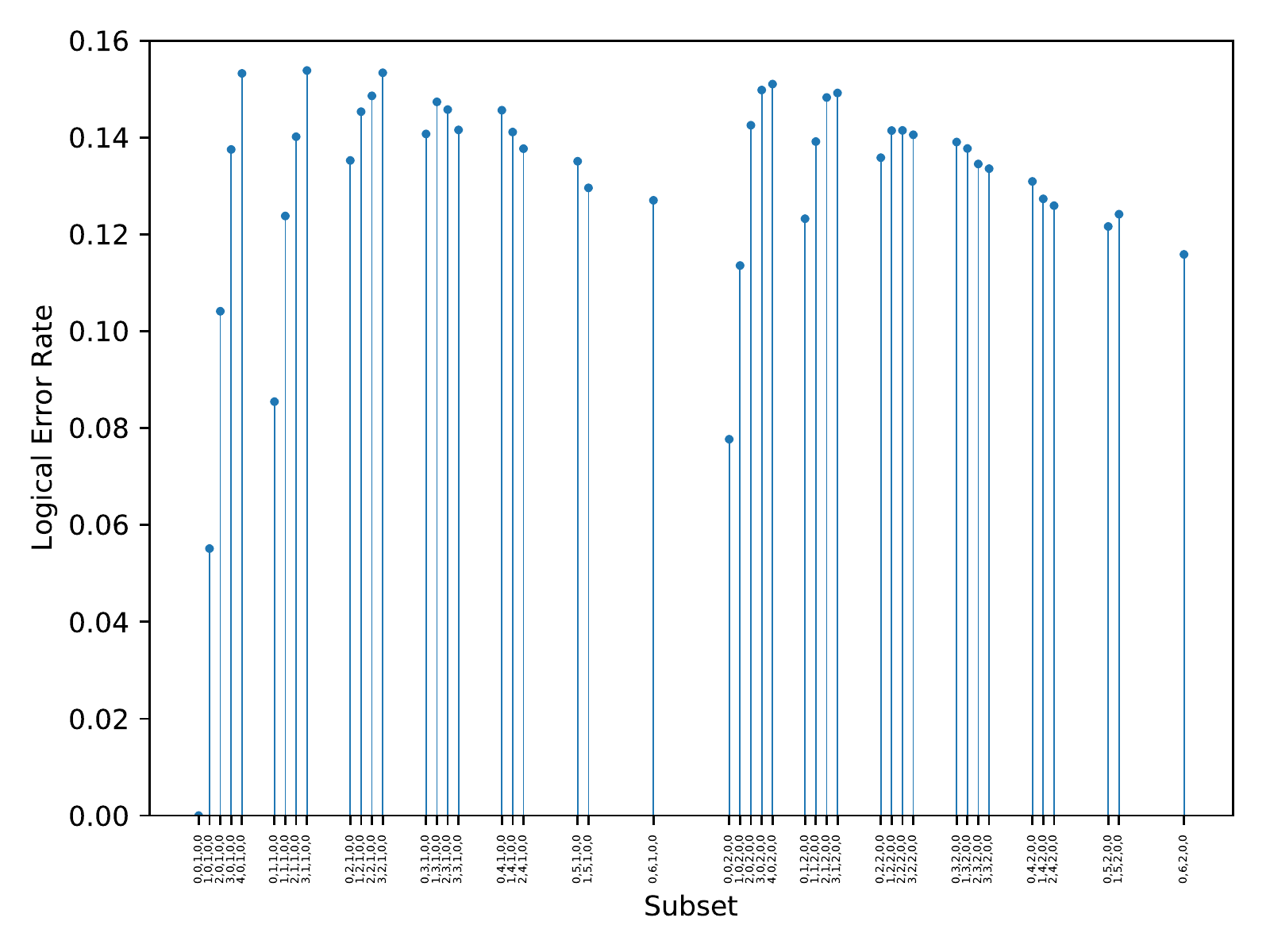}
\end{subfigure}

\begin{subfigure}[t]{\textwidth}
\includegraphics[width=\textwidth]{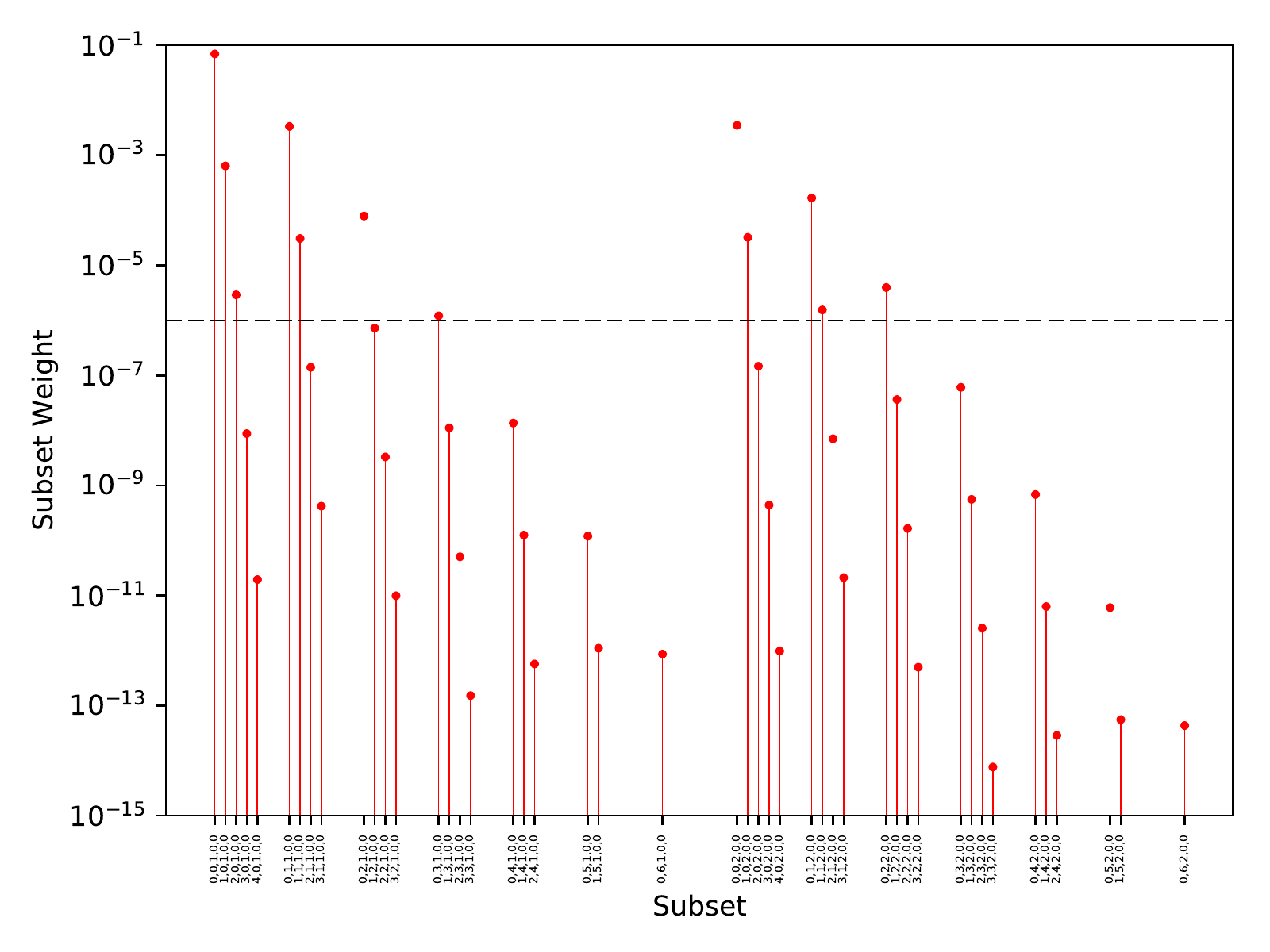}
\end{subfigure}
\end{figure}

\begin{figure}[b!]\ContinuedFloat
\begin{subfigure}[t]{\textwidth}
\includegraphics[width=\textwidth,trim={0 0.5cm 0 0.5cm}]{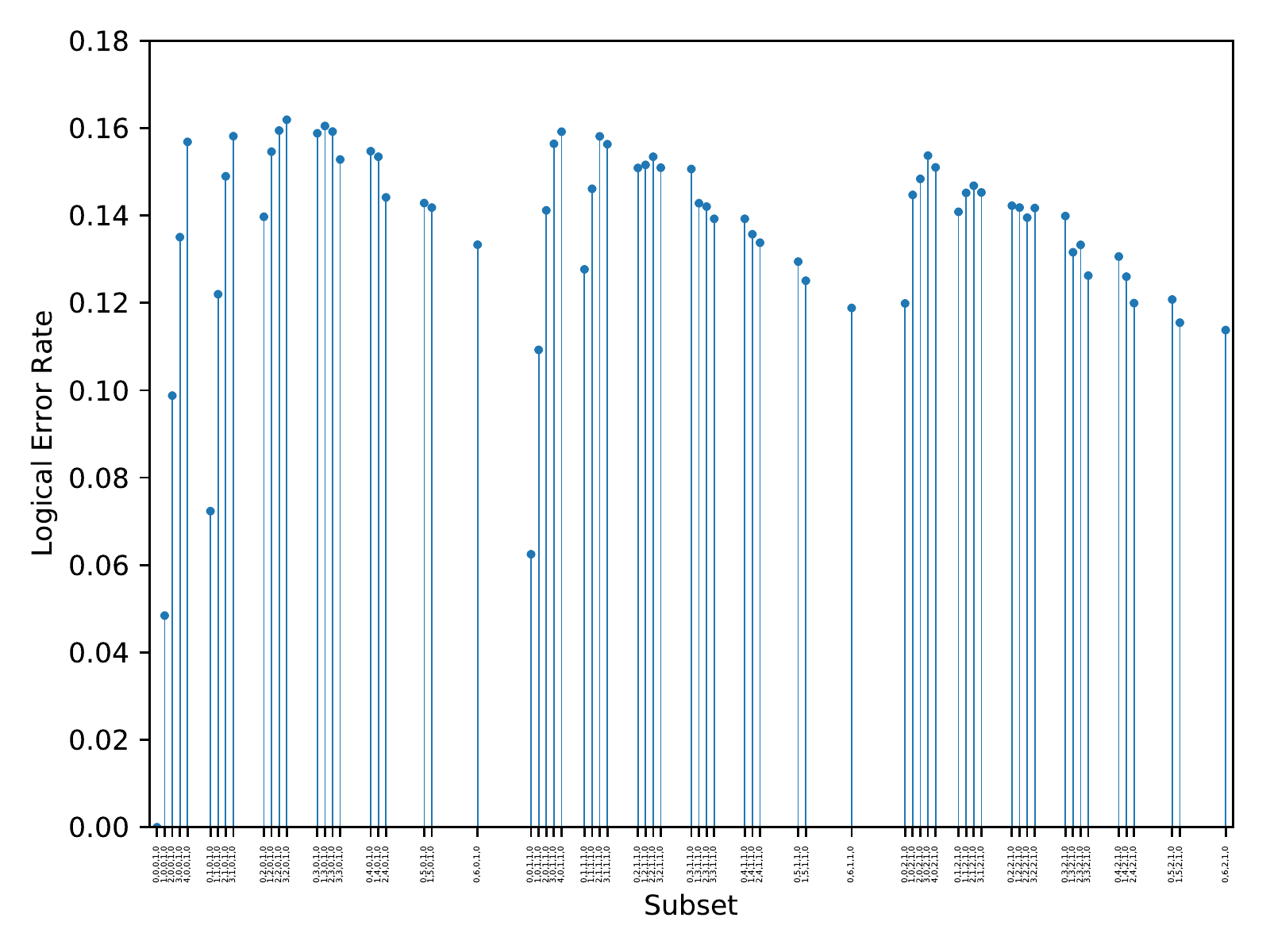}
\end{subfigure}

\begin{subfigure}[t]{\textwidth}
\includegraphics[width=\textwidth]{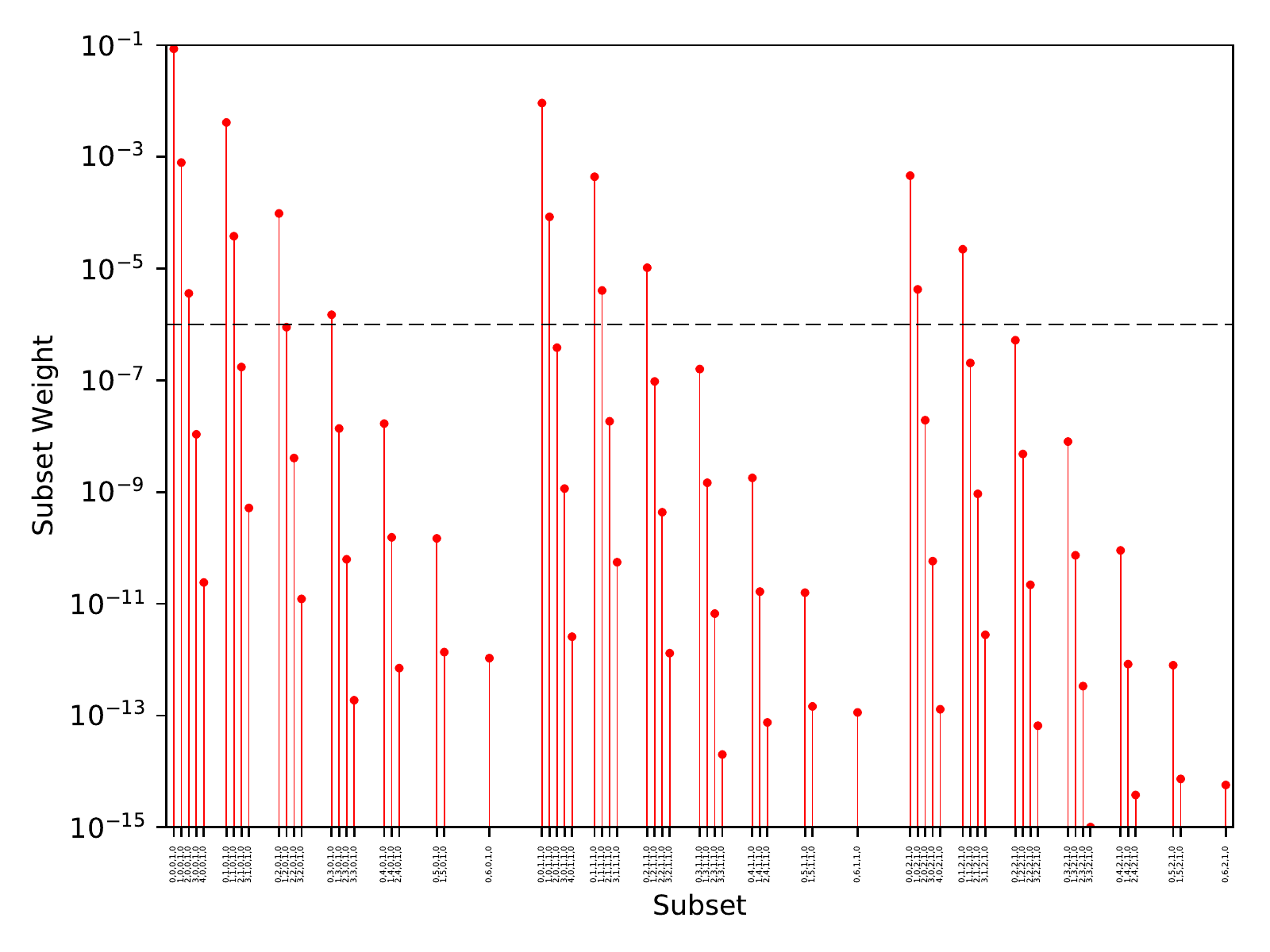}
\end{subfigure}
\end{figure}

\begin{figure}[b!]\ContinuedFloat
\begin{subfigure}[t]{\textwidth}
\includegraphics[width=\textwidth,trim={0 0.5cm 0 0.5cm}]{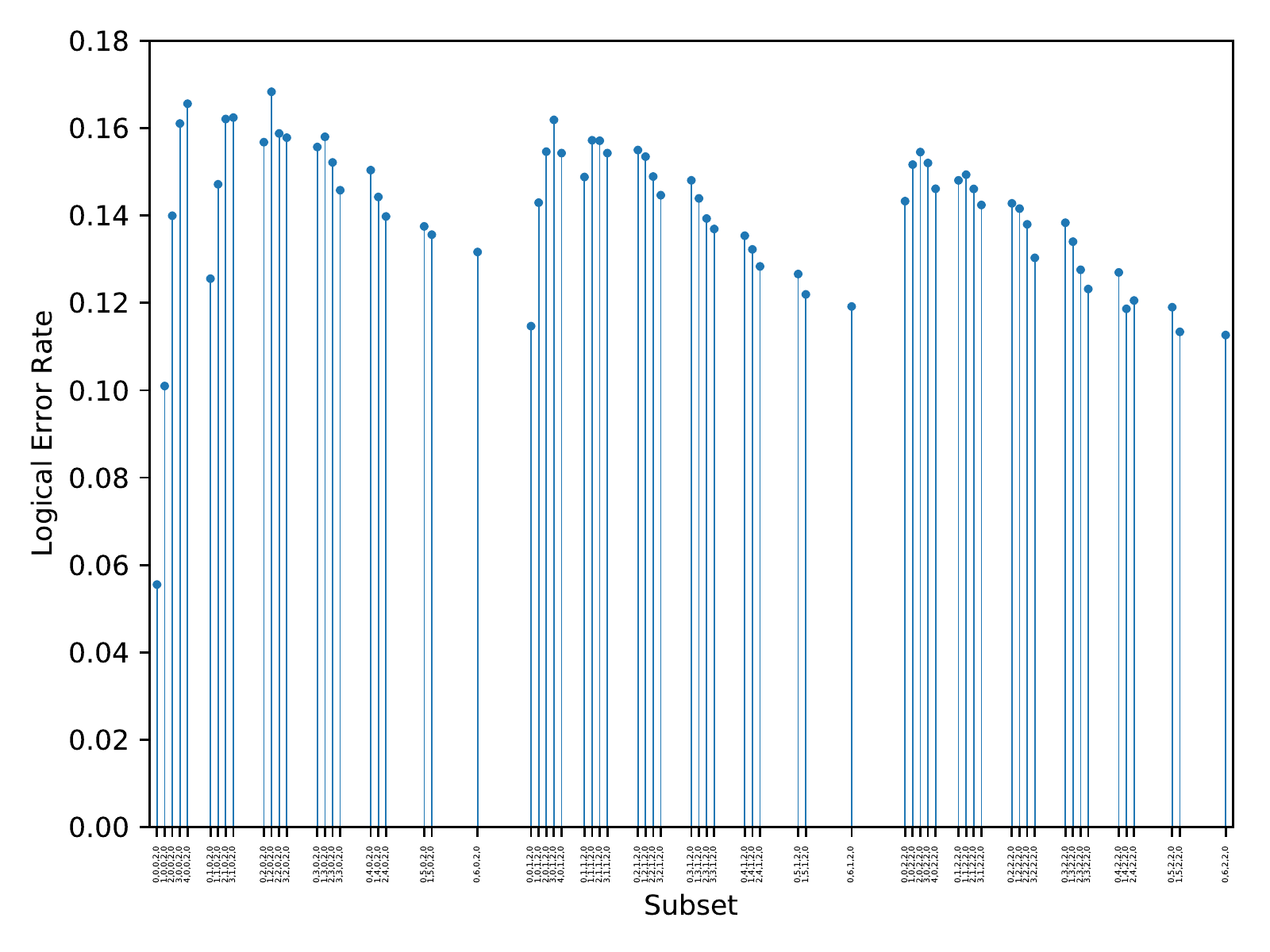}
\end{subfigure}

\begin{subfigure}[t]{\textwidth}
\includegraphics[width=\textwidth]{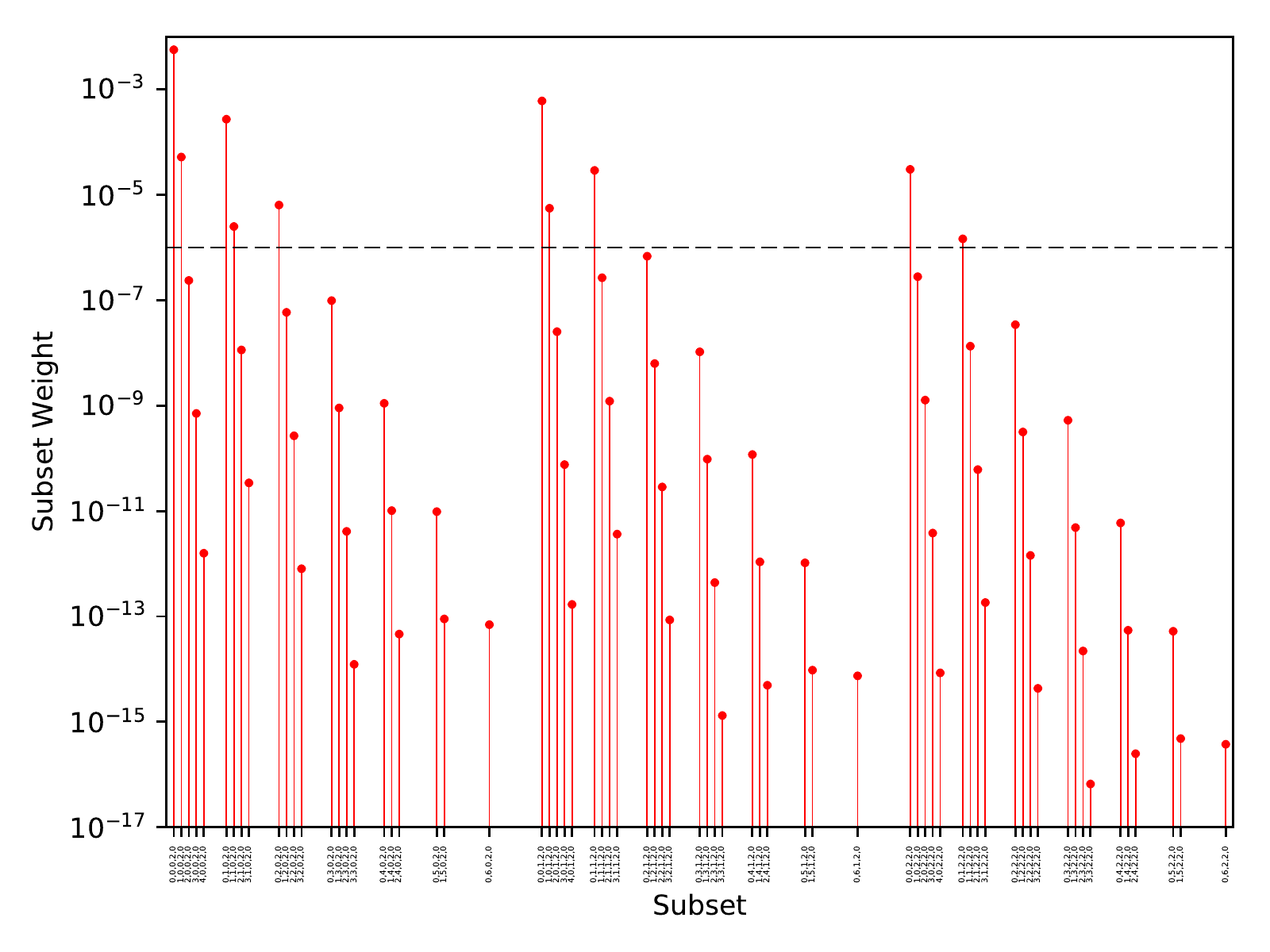}
\end{subfigure}
\end{figure}

\begin{figure}[b!]\ContinuedFloat
\begin{subfigure}[t]{\textwidth}
\includegraphics[width=\textwidth,trim={0 0.5cm 0 0.5cm}]{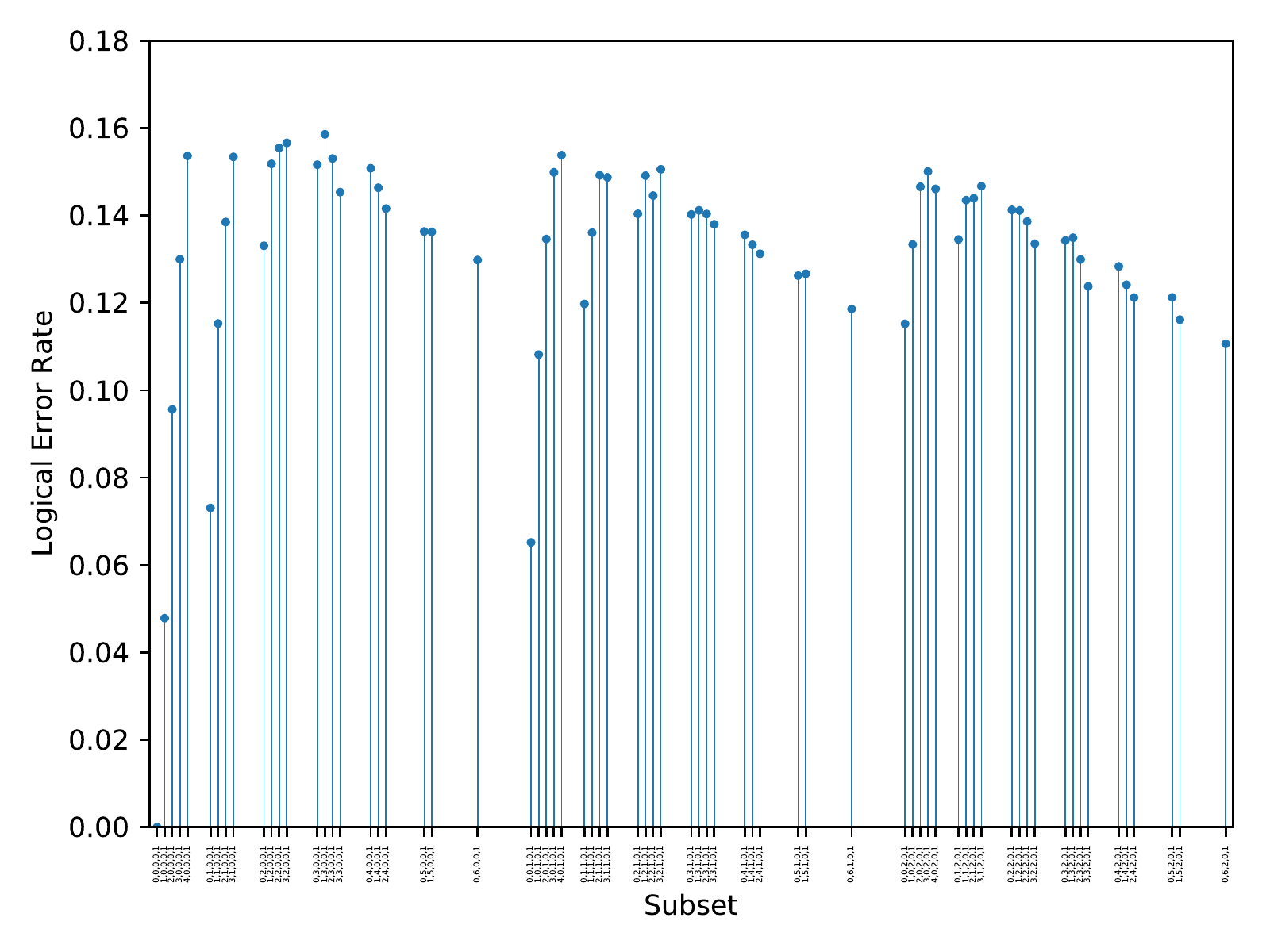}
\end{subfigure}

\begin{subfigure}[t]{\textwidth}
\includegraphics[width=\textwidth]{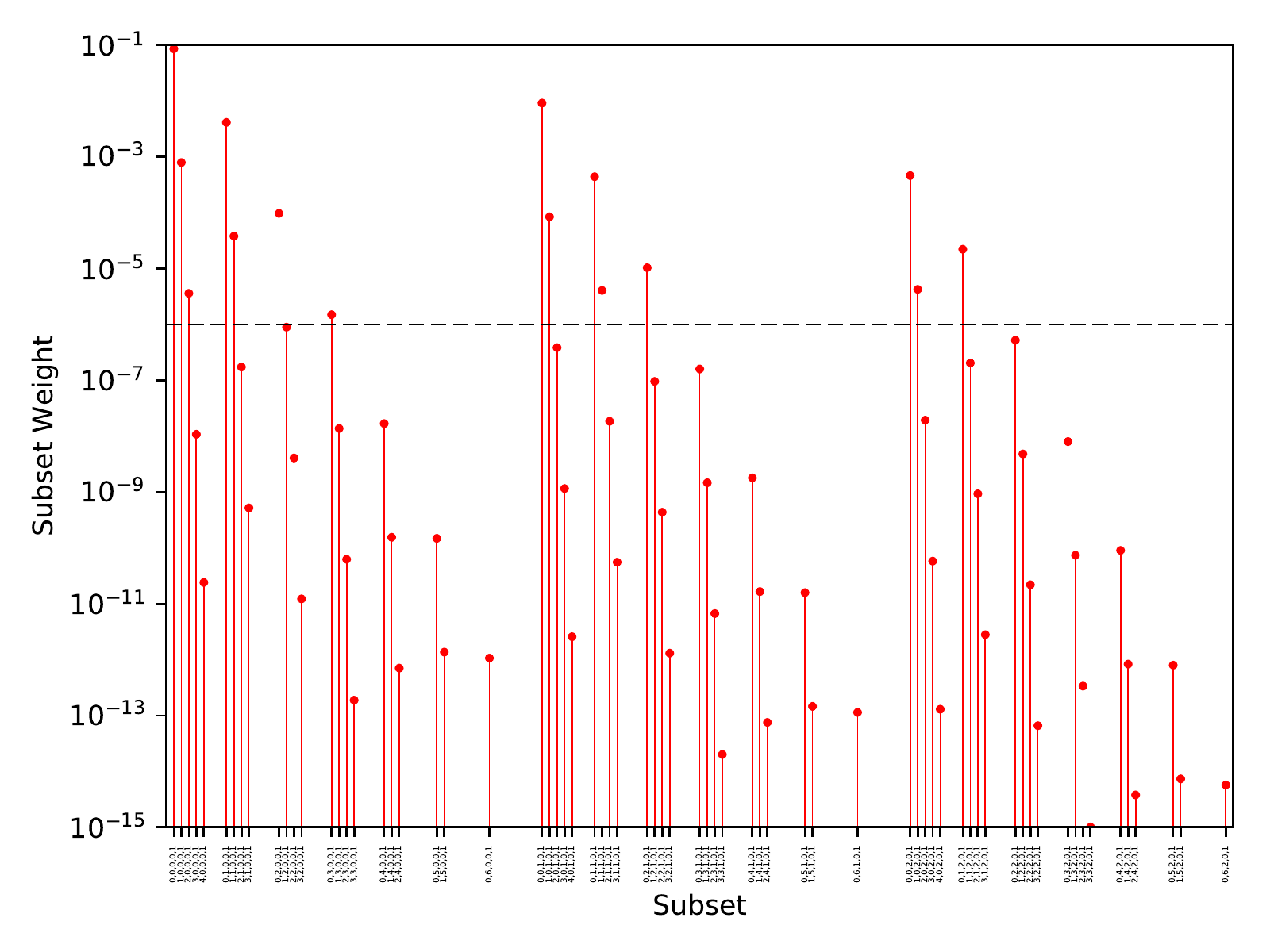}
\end{subfigure}
\end{figure}

\begin{figure}[b!]\ContinuedFloat
\begin{subfigure}[t]{\textwidth}
\includegraphics[width=\textwidth,trim={0 0.5cm 0 0.5cm}]{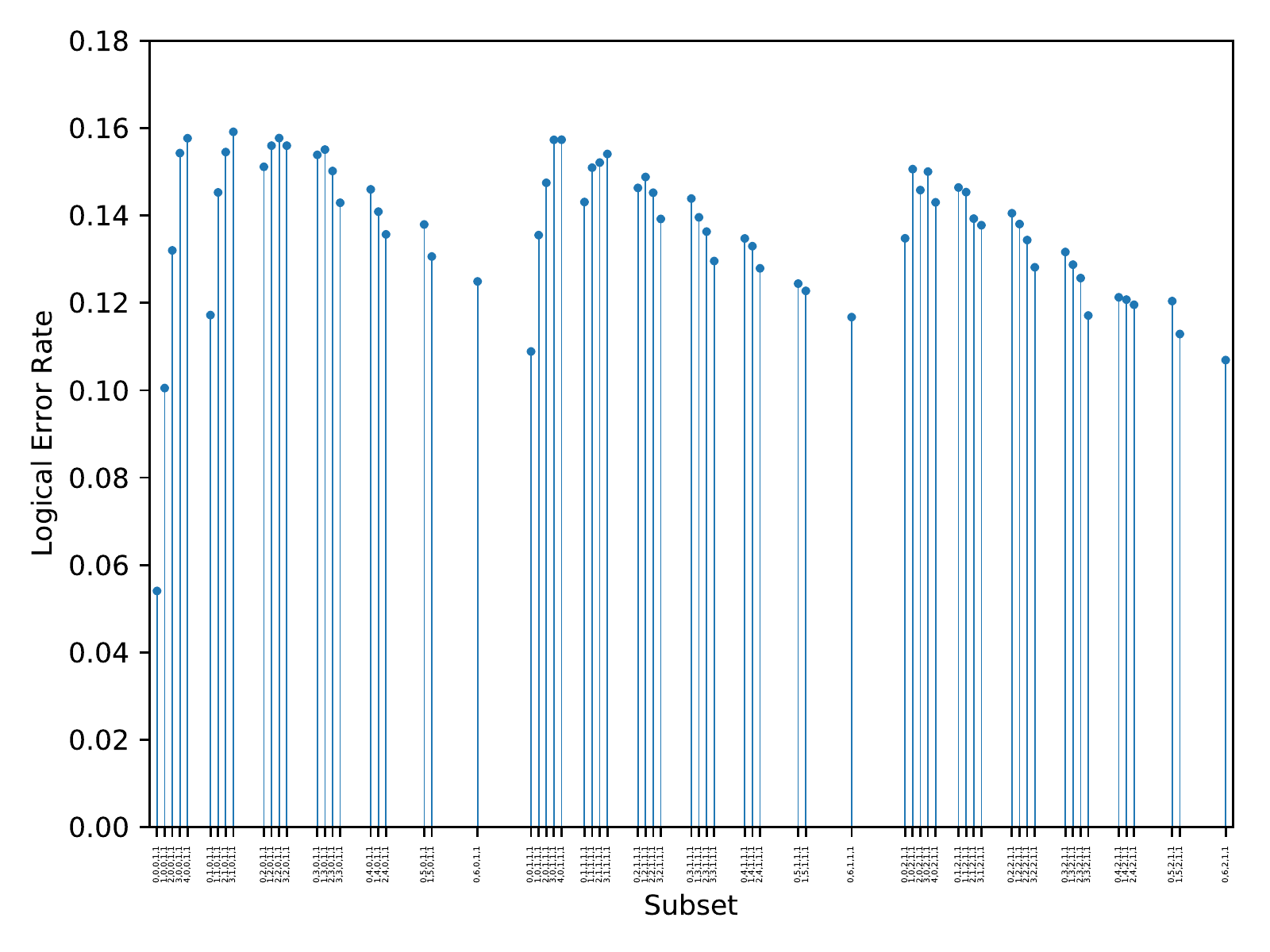}
\end{subfigure}

\begin{subfigure}[t]{\textwidth}
\includegraphics[width=\textwidth]{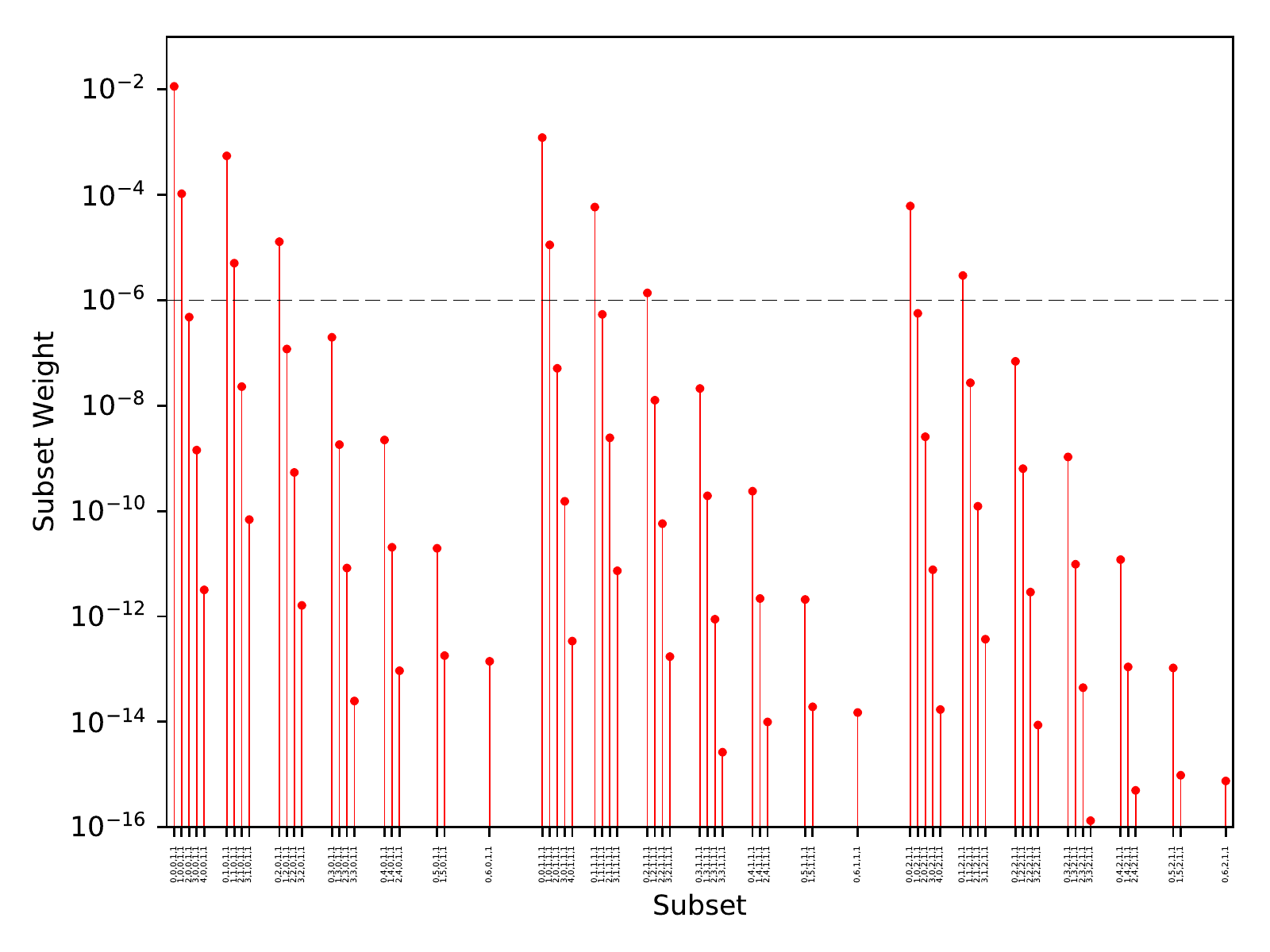}
\end{subfigure}
\end{figure}

\begin{figure}[b!]\ContinuedFloat
\begin{subfigure}[t]{\textwidth}
\includegraphics[width=\textwidth,trim={0 0.5cm 0 0.5cm}]{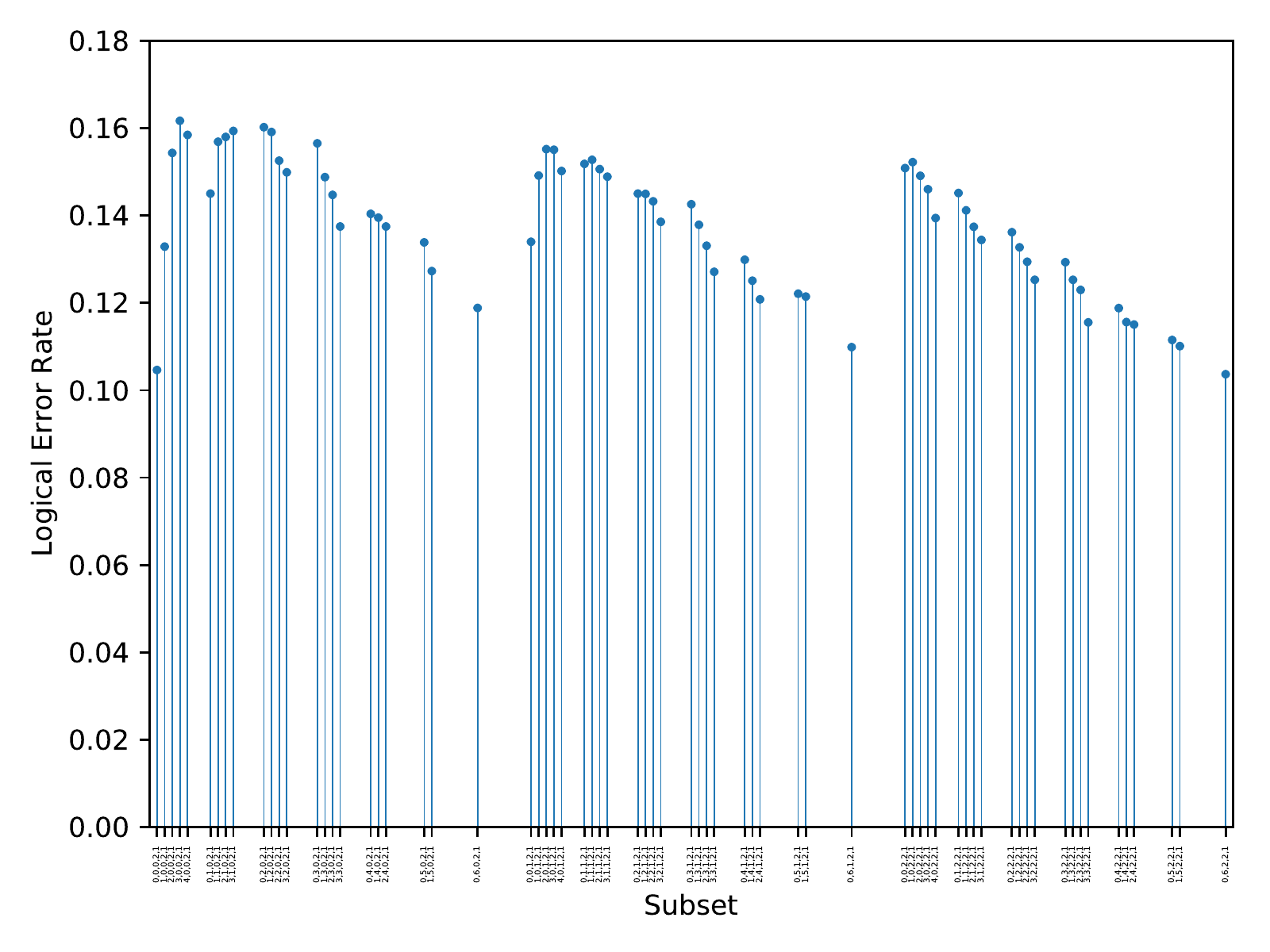}
\end{subfigure}

\begin{subfigure}[t]{\textwidth}
\includegraphics[width=\textwidth]{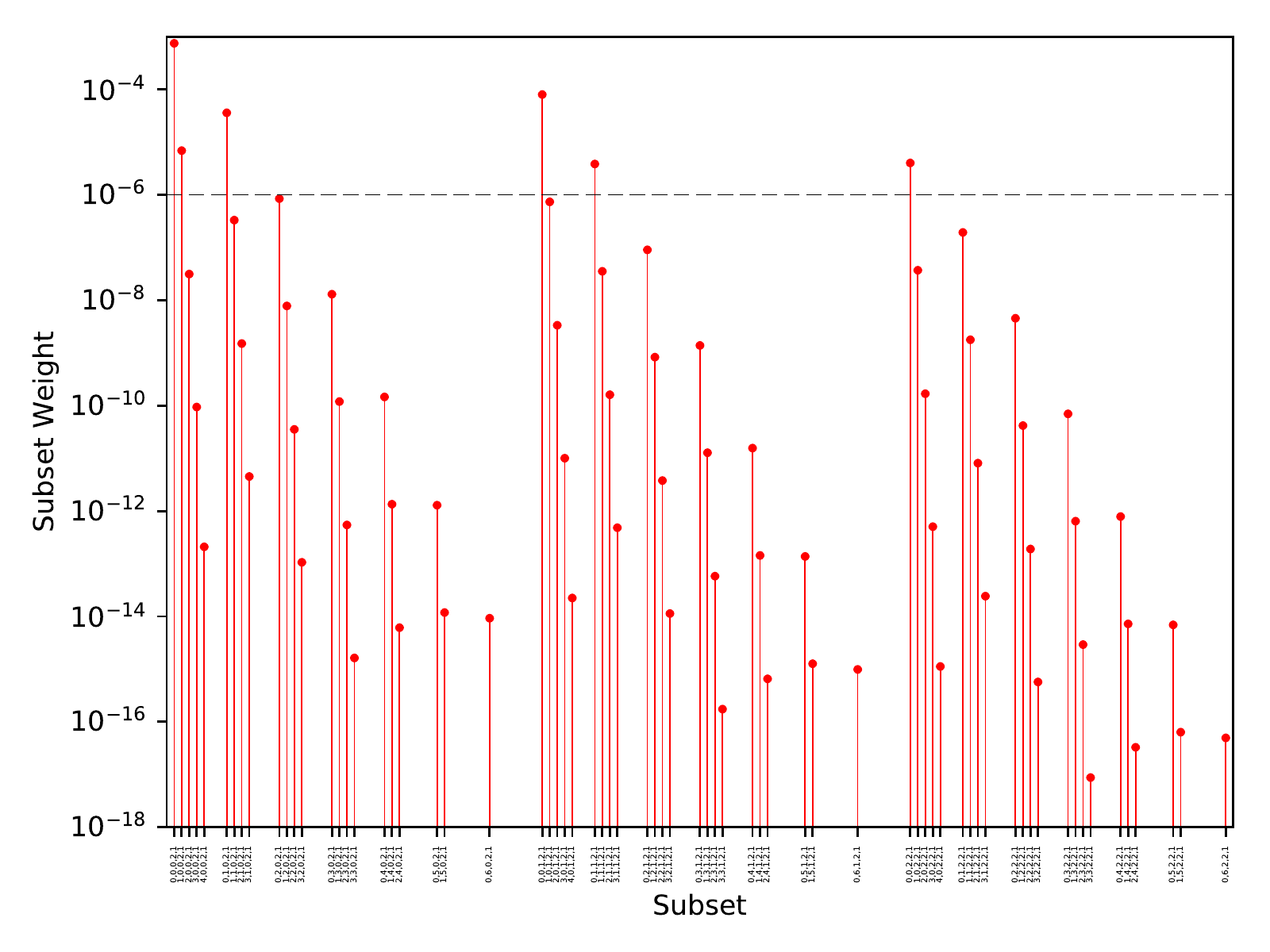}
\end{subfigure}
\end{figure}

\begin{figure}[b!]\ContinuedFloat
\begin{subfigure}[t]{\textwidth}
\includegraphics[width=\textwidth,trim={0 0.5cm 0 0.5cm}]{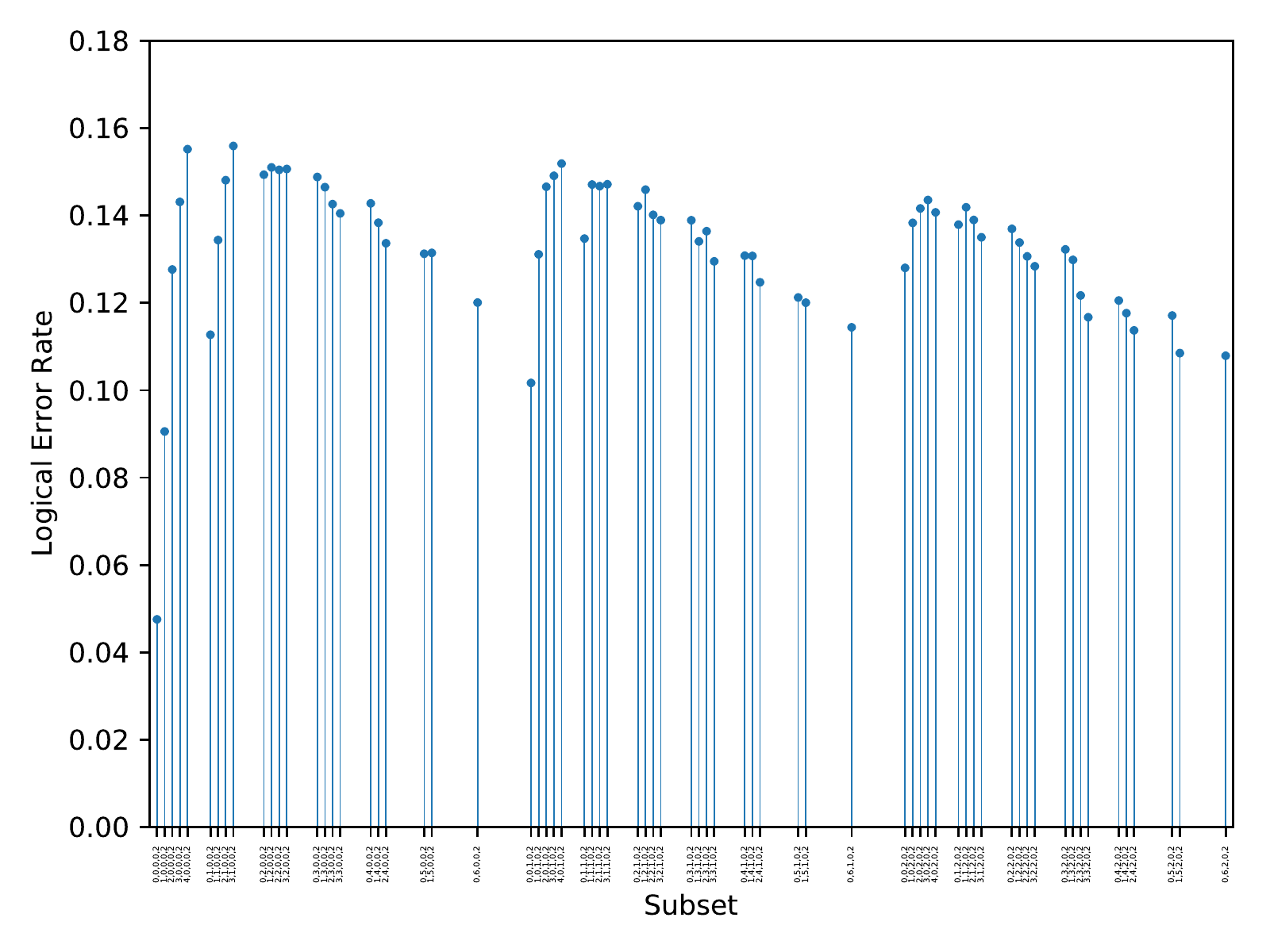}
\end{subfigure}

\begin{subfigure}[t]{\textwidth}
\includegraphics[width=\textwidth]{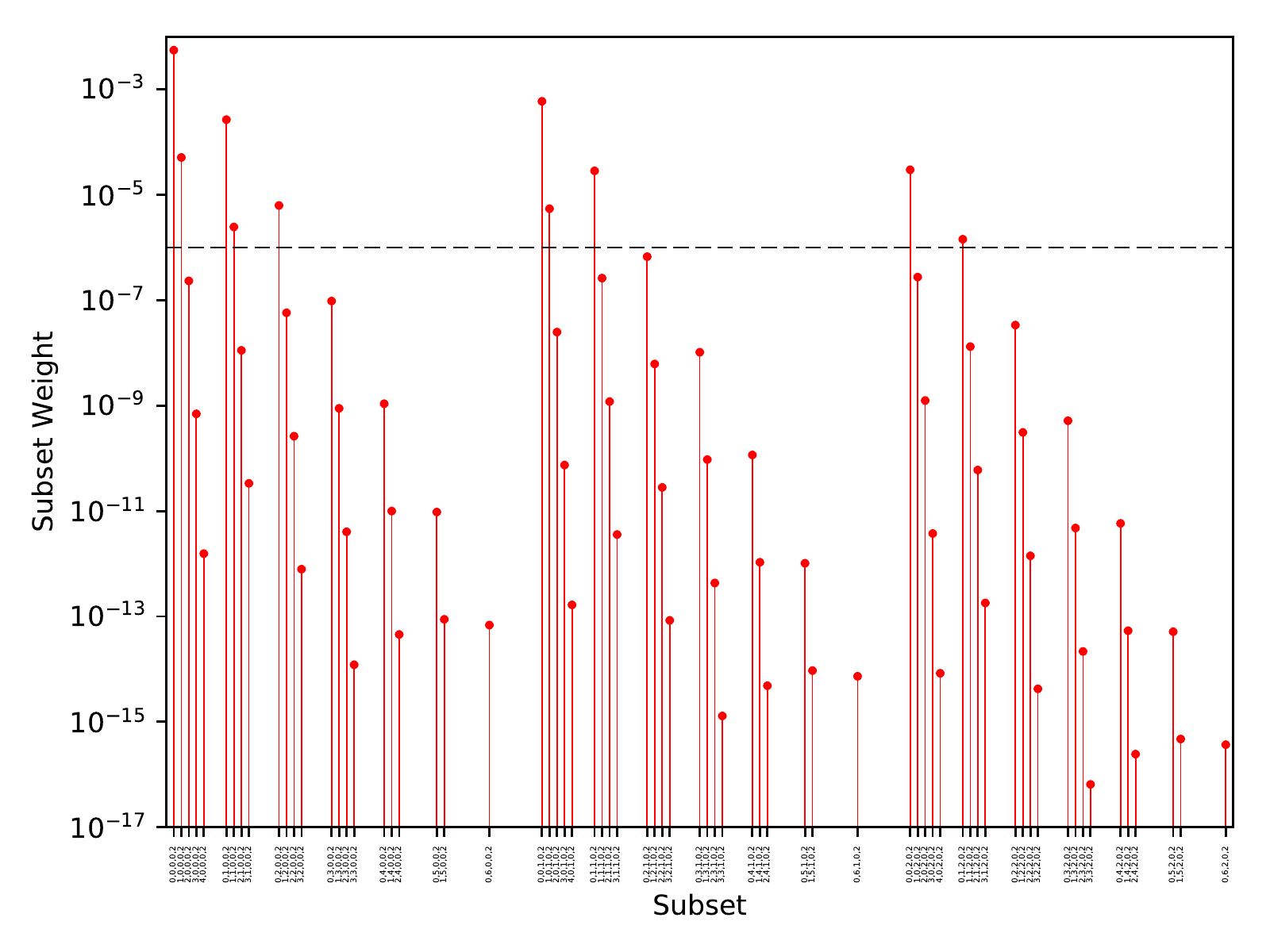}
\end{subfigure}
\end{figure}

\begin{figure}[b!]\ContinuedFloat
\begin{subfigure}[t]{\textwidth}
\includegraphics[width=\textwidth,trim={0 0.5cm 0 0.5cm}]{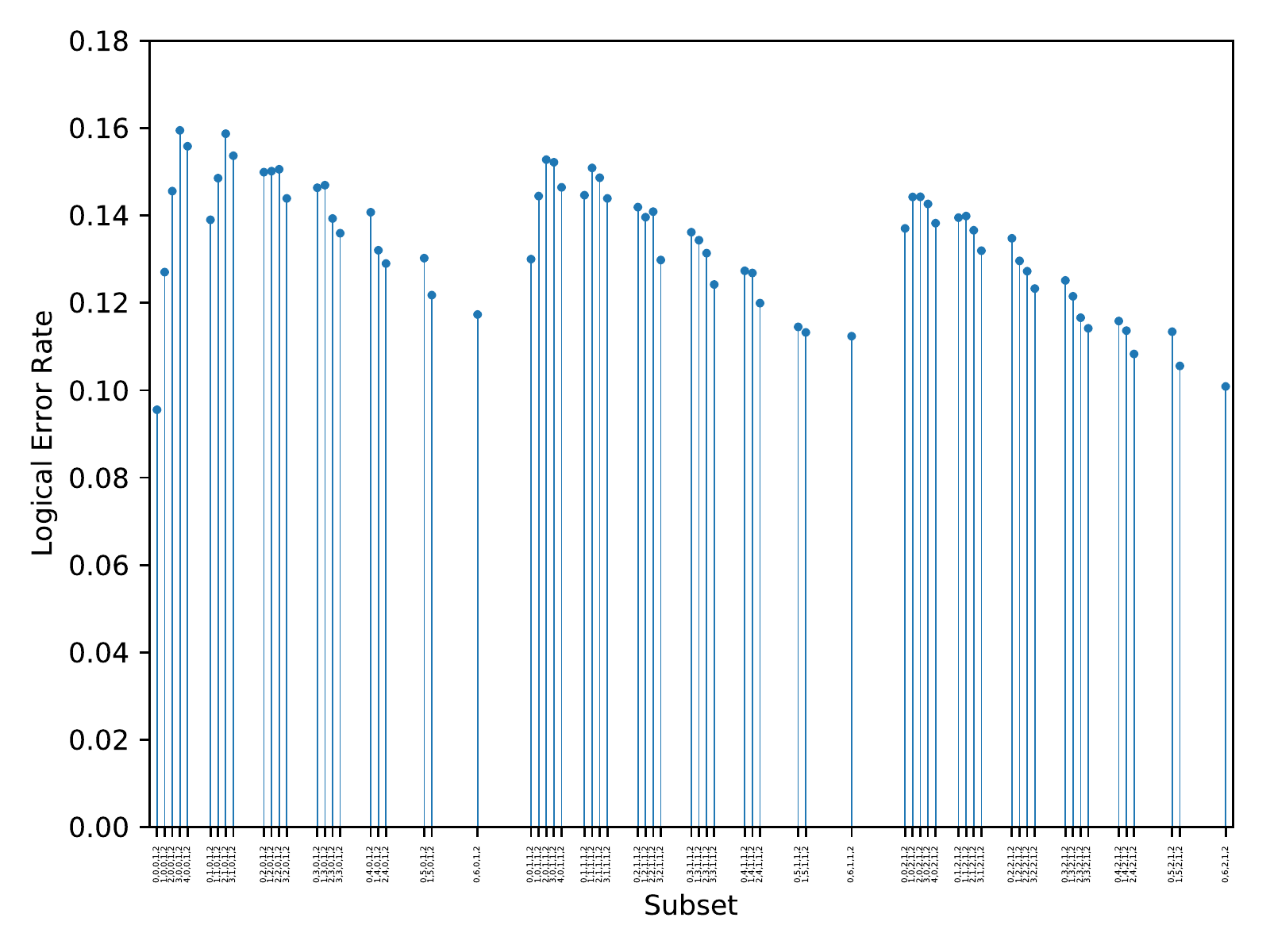}
\end{subfigure}

\begin{subfigure}[t]{\textwidth}
\includegraphics[width=\textwidth]{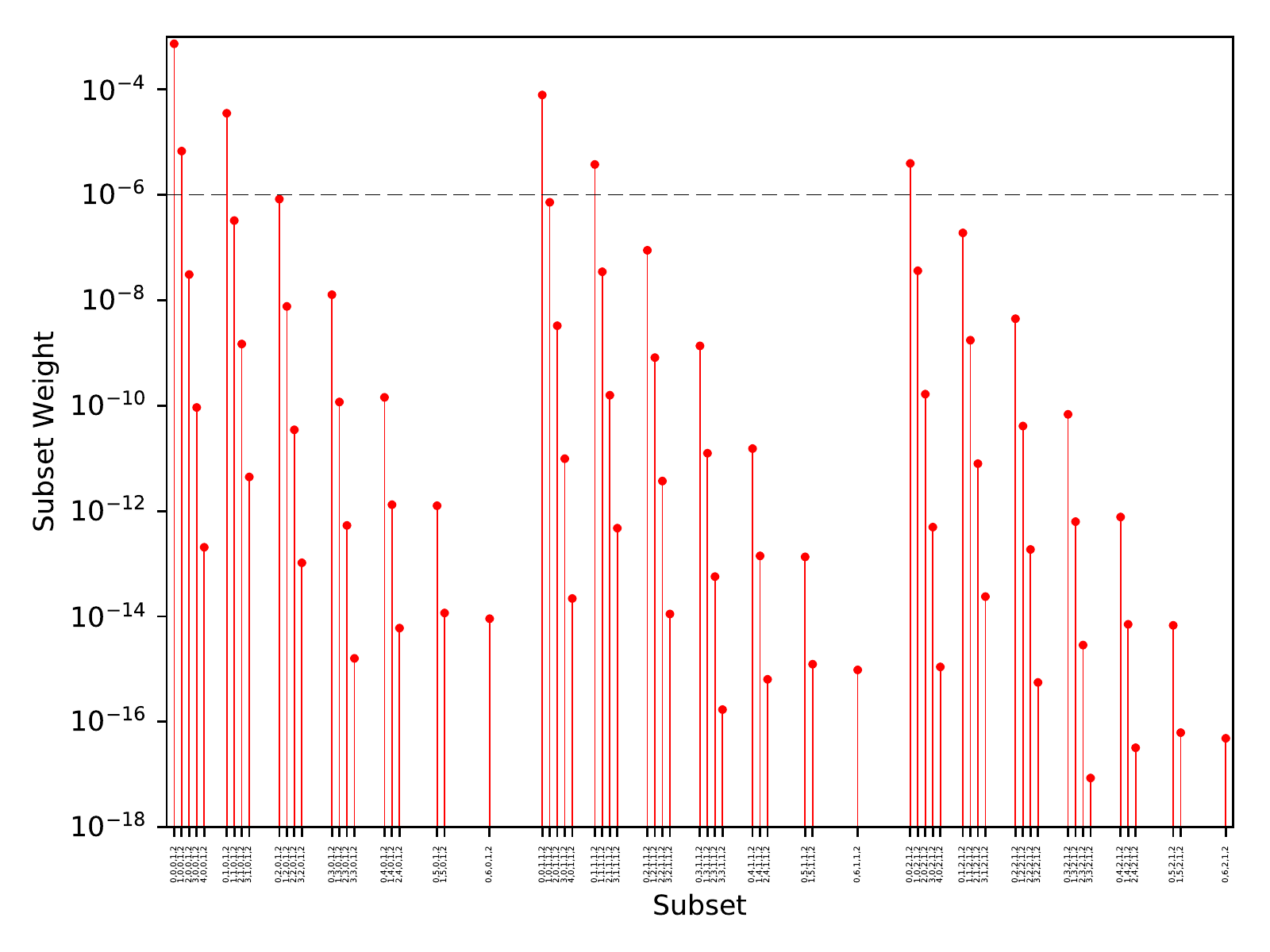}
\end{subfigure}
\end{figure}
\begin{figure}[b!]\ContinuedFloat
\begin{subfigure}[t]{\textwidth}
\includegraphics[width=\textwidth,trim={0 0.5cm 0 0.5cm}]{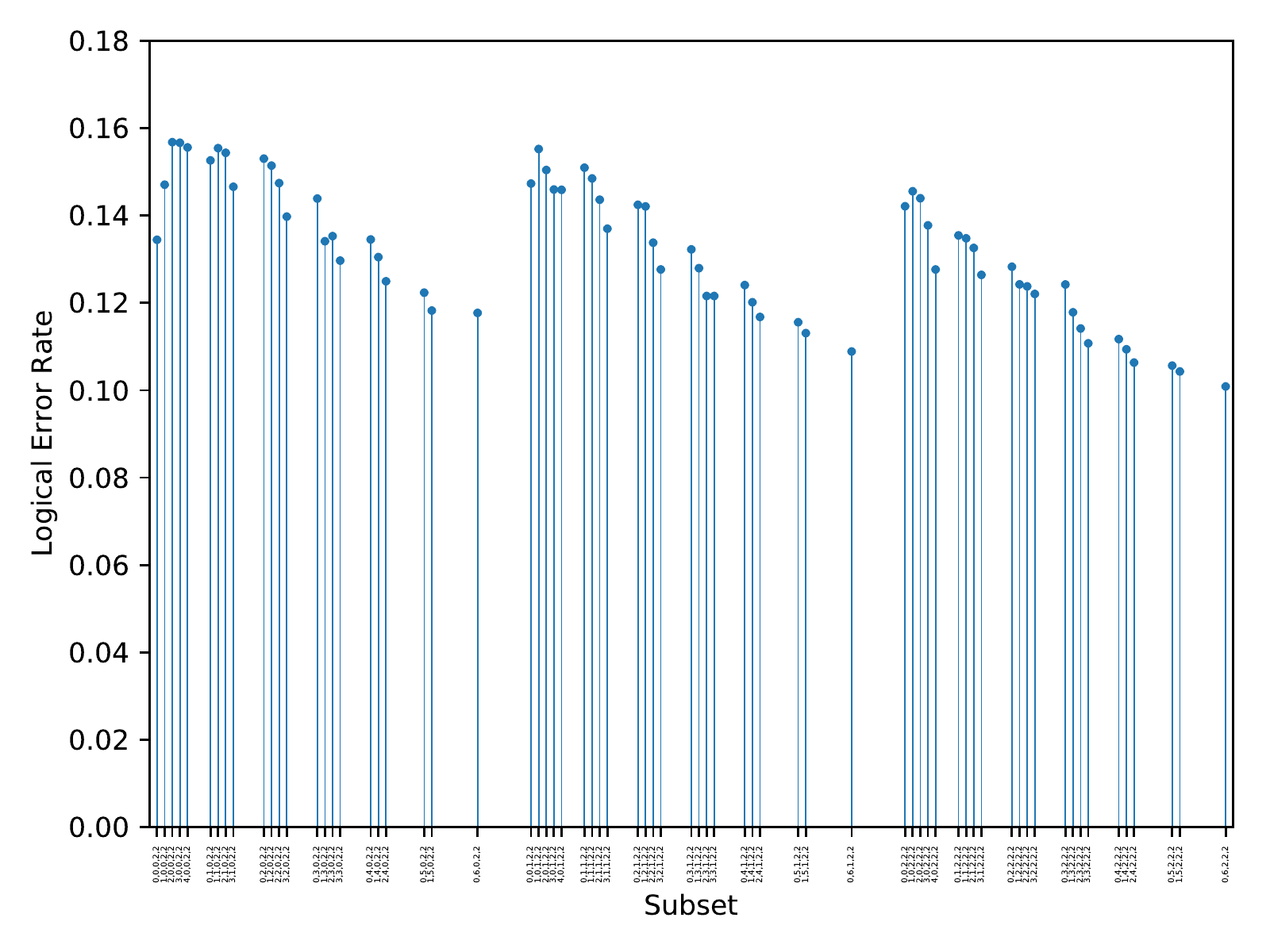}
\end{subfigure}

\begin{subfigure}[t]{\textwidth}
\includegraphics[width=\textwidth]{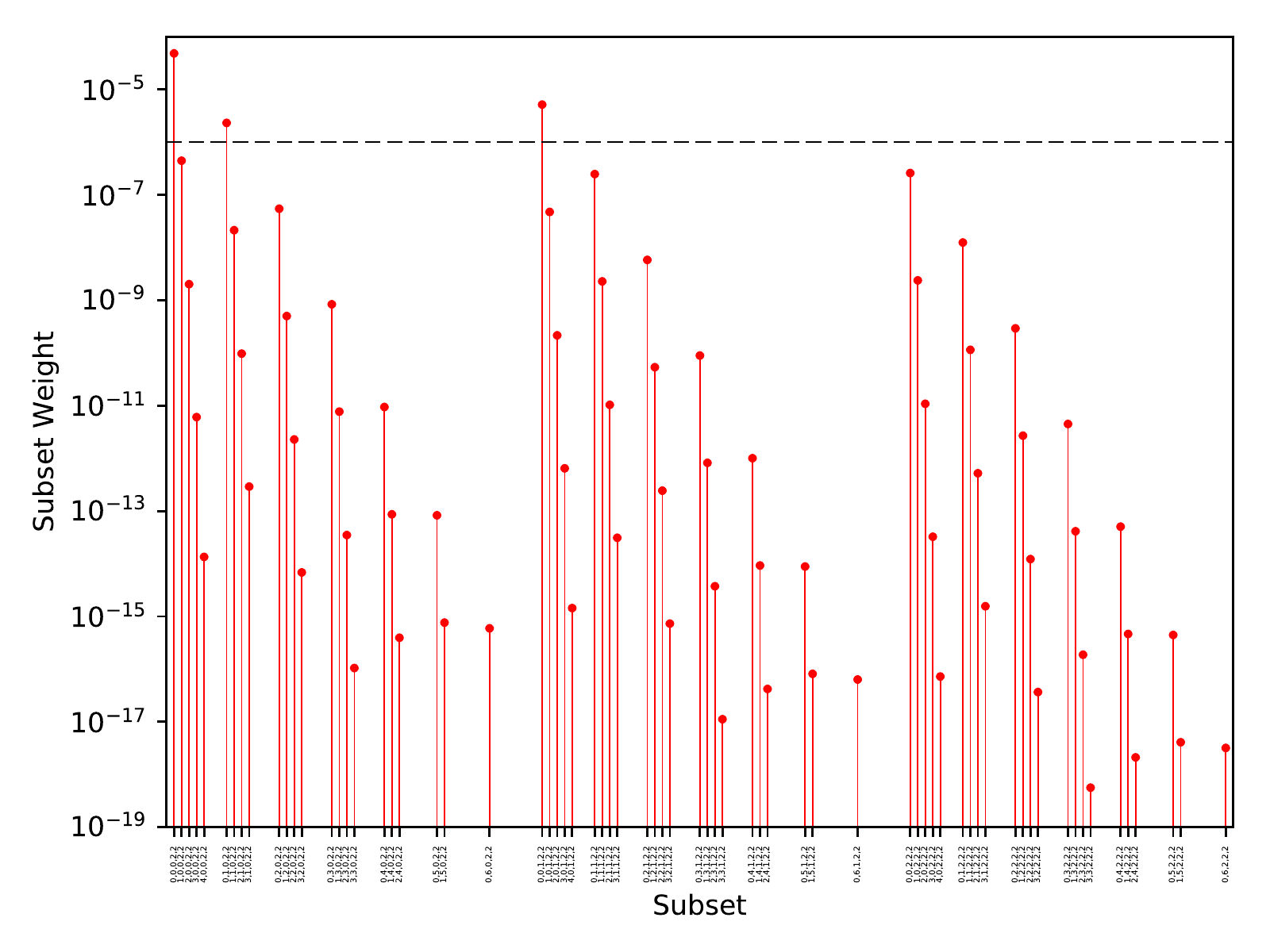}
\end{subfigure}
\end{figure}

\end{document}